\documentclass{article}
\usepackage{jheppub} 

\usepackage[T1]{fontenc} 
\usepackage{graphicx,amssymb,amstext,amsmath}
\usepackage{subcaption}
\usepackage{caption}
\usepackage{appendix}
\usepackage{float}
\usepackage{empheq}
\usepackage{mathtools}
\makeatletter
\allowdisplaybreaks

\usepackage[parfill]{parskip}
\newcommand{\be}{\begin{equation}}
\newcommand{\ee}{\end{equation}}
\newcommand{\p}{\partial}
\newcommand{\vast}{\bBigg@{4}}
\newcommand{\Vast}{\bBigg@{5}}

\preprint{CCTP-2021-6 \\
\hspace*{0cm} \hfill ITCP-IPP 2021/3 \\}

\title{On matched asymptotic expansions of backreacting
metastable anti-branes}

\author{Nam Nguyen$^{\dagger}$, Vasilis Niarchos$^{\natural}$}

\affiliation{$^{\dagger}$Department of Mathematical Sciences and Centre for Particle Theory,\\Durham University, Durham DH1 3LE, United Kingdom\\
$^{\natural}$CCTP and ITCP, Department of Physics,\\University of Crete, 71003 Heraklion, Greece}

\emailAdd{nam.h.nguyen@durham.ac.uk} 
\emailAdd{niarchos@physics.uoc.gr}

\abstract{We construct analytically a perturbative supergravity solution that captures the backreaction of a metastable state of anti-branes in the background of a particular modification of the Klebanov-Strassler throat in a long-wavelength approximation.
Our solution, which has no unphysical singularities, describes how non-supersymmetric spherical NS5-branes with dissolved anti-D3 brane charge backreact in a fluxed throat geometry.
It supports previous claims that there is a well-behaved supergravity description of the metastable state of wrapped NS5-branes proposed years ago by Kachru, Pearson, and Verlinde.}

\begin{document}
\maketitle
\flushbottom

\section{Introduction \& Summary of results}
\subsection{Introduction}
An understanding of controlled supersymmetry (SUSY) breaking in string theory is among the principal goals in string phenomenology for establishing a connection between string theory and our reality. It also plays a significant role in holographic descriptions of non-SUSY quantum field theories (QFTs) and within the general debate of Swampland conjectures.

One of the canonical methods for breaking SUSY in string theory involves balancing anti-branes in warped throats \cite{Maldacena:2001pb}. Despite their consequential applications in holographic QFT \cite{Kachru:2002gs, Klebanov:2010qs}, string cosmology \cite{Kachru:2003aw, Kachru:2003sx}, and black hole physics \cite{Bena:2012zi}, some aspects of these anti-branes configurations are not yet fully understood. Over the last decade, many works have been dedicated to exploring their properties. In the subsequent paragraphs, we review the Kachru-Pearson-Verlinde (KPV) \cite{Kachru:2002gs} configuration, a frequently-discussed exemplar metastable configuration of anti-branes, and briefly summarise the discussions on some of its properties. 

Our starting point is the Klebanov-Strassler (KS) throat \cite{Klebanov:2000hb} in ten-dimensional type IIB supergravity. The KS throat involves a six-dimensional deformed conifold, a four-dimensional Minkowski space, and non-trivial $F_3$, $F_5$, and $H_3$ fluxes. One can intuitively think of the KS throat as the solution resulting from placing D3 and D5 brane charges at the tip of a Ricci-flat (deformed) conifold. The D3 and D5 brane charges induce non-trivial $F_3$, $F_5$, and $H_3$ fluxes, resulting in a warped, fluxed throat geometry. 

\begin{figure}[t]
\begin{subfigure}{0.45\textwidth}
\includegraphics[width=0.95\linewidth]{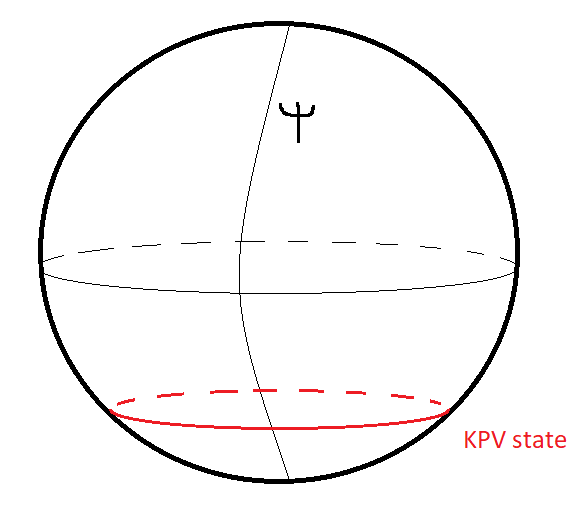} 
\end{subfigure}
\begin{subfigure}{0.55\textwidth}
\includegraphics[width=0.95\linewidth]{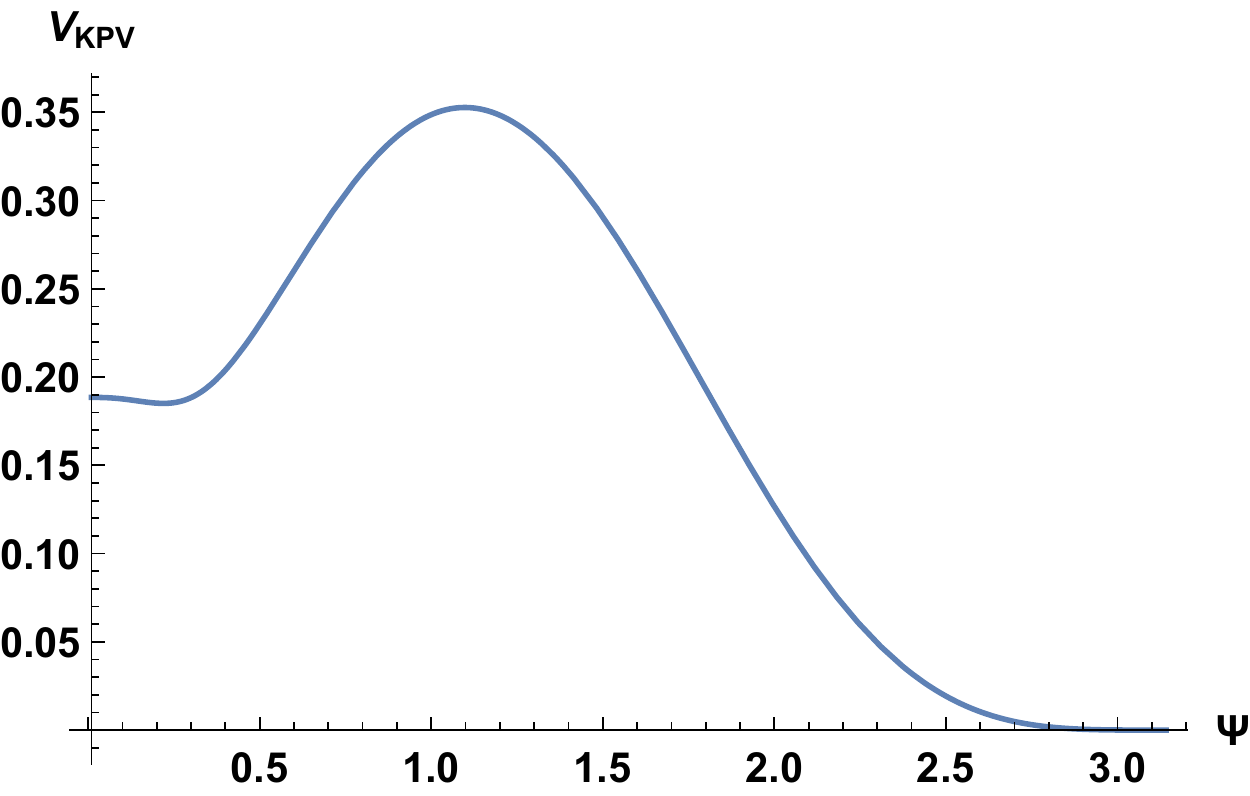}
\end{subfigure}
\caption{On the left, we present an illustrative picture of the tip of the KS throat, which is an $S^3$.  The coordinate $\psi \in (0, \pi)$ is the azimuthal angle of this $S^3$. The red circle illustrates the KPV state, which is a spherical NS5 state (wrapping an $S^2$ of the $S^3$) that experiences a balance of force along the azimuthal angle $\psi$. On the right, we present the effective potential of a spherical NS5 state at the tip of the KS throat for $p/M = 0.03$. As we can see from the plot, the potential has a metastable minimum at $\psi \approx 0.3$. This is the KPV state.}
\label{extra1}
\end{figure}

Anti-D3 branes near the tip of the KS throat are attracted to the tip via both gravitational and ``electromagnetic'' forces. By analysing the non-abelian action of the stack of anti-D3 branes in the probe approximation, Ref.\ \cite{Kachru:2002gs} observed that these anti-D3 branes polarise via the Myers effects \cite{Myers:1999ps} to a spherical NS5 brane with dissolved anti-D3 brane charge wrapping an $S^2$ inside the $S^3$ of the geometry at the tip of the KS throat. In some regime of parameters, this spherical NS5 brane stabilises along the azimuthal angle direction $\psi$. More specifically, let $p$ denote the number of anti-D3 branes and $M$ the strength of the background KS fluxes. Ref.\ \cite{Kachru:2002gs} demonstrated that when $p/M$ is between 0 and $(p/M)_{crit}\approx 0.080488$, the effective potential of the NS5 brane has a metastable minimum, see Fig.\ \ref{extra1}. This metastable state of spherical NS5 brane with dissolved anti-D3 brane charge at the tip of the KS throat is commonly referred to as the Kachru-Pearson-Verlinde (KPV) state. Note that this spherical NS5 state is classically stable,\footnote{The claim that the spherical NS5 brane is classically stable is not yet fully settled and requires further investigations. The observation in \cite{Kachru:2002gs} is simply that the spherical NS5 feels a balance of gravitational and ``electromagnetic'' forces in the azimuth direction $\psi$ of the $S^3$ that it wraps. One might be worried that there exist classical perturbations that make the NS5 state unstable. There have been studies on whether this happens, e.g. \cite{Bena:2014jaa, Nguyen:2019syc}, which we will discuss further in the Outlook section.} but can tunnel quantum mechanically through the classical barrier to the true minimum at the South pole. However, the rate of this quantum tunnelling effect can be exponentially suppressed by choosing a suitably large $M$ \cite{Kachru:2002gs}. 

As we mentioned already, the initial discovery of the KPV state was based on a brane probe analysis. Its existence beyond the probe limit became controversial when unphysical singularities were found in the backreacted supergravity description of anti-D3 branes in the KS throat. By studying the linearised backreaction of smeared\footnote{The anti-branes are smeared homogeneously across the $S^3$ tip of the KS throat.} anti-D3 branes, Ref.\ \cite{Bena:2009xk} observed that the backreacted supergravity description for the smeared anti-D3 branes must have unphysical\footnote{Initially, these divergences were deemed unphysical because they were divergences without an obvious physical origin. However, using Gubser's criterion for identifying good versus bad singularities \cite{Gubser:2000nd}, later works, e.g. \cite{Blaback:2014tfa}, argued that these singularities are indeed unphysical and cannot be resolved using string theory.} singularities in the 3-form flux. Later works \cite{Massai:2012jn, Bena:2012bk, Gautason:2013zw, Blaback:2014tfa} showed that the unphysical flux singularities in the anti-D3 branes supergravity description persist even when one gets rid of all the approximations (e.g., linearisation of the anti-branes backreaction, smearing). However, \cite{Cohen-Maldonado:2015ssa} observed that the arguments for unphysical singularities in the supergravity description of smeared/localised anti-D3 branes do not extend to the case of spherical NS5 branes. In particular, the IR and UV gluing conditions, which is the key to previous observations of unphysical singularities, can be expressed in terms of a Smarr relation \cite{Cohen-Maldonado:2015ssa, Cohen-Maldonado:2016cjh}. While localised/smeared anti-D3 branes cannot satisfy the Smarr relation with a regular horizon, spherical NS5 branes can. As the KPV state is the polarised state of anti-D3 branes and should be considered in supergravity as spherical NS5 branes with dissolved anti-D3 brane charge, \cite{Cohen-Maldonado:2015ssa, Cohen-Maldonado:2016cjh} effectively found for the KPV state a possible way out of the unphysical singularities that plagued backreacting anti-branes.

As the result of \cite{Cohen-Maldonado:2015ssa} is a negative of a negative, i.e. it is an argument that polarised branes do not run into the same problems that smeared/localised branes do, \cite{Cohen-Maldonado:2015ssa} had only provided the necessary (but not sufficient) conditions for the existence of a well-behaved supergravity solution of backreacted anti-branes. As such, supplementary direct evidence for the existence of the KPV state is needed. Using the blackfold approach \cite{Emparan:2009cs, Emparan:2009at, Armas:2016mes}, Ref.\ \cite{Armas:2018rsy} provided further positive evidence. The result of \cite{Armas:2018rsy} fits perfectly with the lifting of the no-go theorem in \cite{Cohen-Maldonado:2015ssa}. In the extremal limit, \cite{Armas:2018rsy} recovered the KPV spherical metastable NS5 state. Away from extremality, it uncovered a metastable black NS5 state and observed, in agreement with expectations, that its metastability is lost when its horizon geometry resembles that of a localised black anti-D3 state. As these works were done using very different methods, they together constitute a strong argument for the existence of the KPV state.

We note in passing that there are many works relevant to the existence of the KPV state that are not exhaustively covered in this introduction. Such works include, for example, a string theory resolution for the singularities in the case of a single anti-D3 brane   \cite{Michel:2014lva}, investigations on the effect of temperature on the anti-D3 branes singularities \cite{Bena:2012ek, Bena:2013hr, Blaback:2014tfa, Hartnett:2015oda}, discussions on the resolution of anti-D6-brane singularities in the supergravity regime via brane polarisation \cite{Blaback:2019ucp}, and studies of analogous anti-brane configurations in M-theory \cite{Klebanov:2010qs, Bena:2010gs, Cohen-Maldonado:2015ssa, M2M5brane}.

The results in \cite{Cohen-Maldonado:2015ssa} and \cite{Armas:2018rsy} are indirect. In both cases, an explicit construction of the wrapped NS5-brane state at the tip of the KS throat is technically very demanding and still missing. The strategy behind the blackfold approach of Ref.\ \cite{Armas:2018rsy} is based on a perturbative, long-wavelength analysis of the supergravity equations. In this context, one would like to construct a perturbative supergravity description of the KPV configuration using the technique of matched asymptotic expansions (MAE), where the solution is approximated in the far zone by the background solution of interest (here the KS throat) and in the near zone by a uniform flat-space $p$-brane solution (here the D3-NS5 bound state). By studying a subset of the matched asymptotic equations for D3-NS5 branes in the KS throat, \cite{Armas:2018rsy} was able to distil useful information about the backreacted description of the KPV state in the leading order of the perturbative expansion, but did not construct the full, leading-order perturbative solution of wrapped NS5 branes in supergravity. 

The idea of using a subset of the matched asymptotic equations (constraint equations) to learn about the backreacted description of a brane configuration in some background lies at the ``soul'' of the blackfold approach. In the blackfold approach, the constraint equations, which are dubbed blackfold equations, provide a $(p+1)$-dimensional effective worldvolume description of the dynamics of a $p$-brane solution. When viewed in isolation, the blackfold equations are only the necessary conditions for a matched asymptotic solution. As such, statements made using the blackfold approach are only truly conclusive when a regular perturbative solution in MAE can be constructed. The claim that the blackfold equations guarantee a regular matched asymptotic solution is dubbed the blackfold conjecture and this conjecture is baked into many of the blackfold applications including that of \cite{Armas:2018rsy}. The conjecture is correct in the fluid-gravity correspondence \cite{Bhattacharyya:2007vjd} in AdS/CFT and has been proven at leading order in the MAE expasion in pure Einstein gravity in flat space \cite{Camps:2012hw}. It is not known, however, if it is generally valid in supergravity with generic asymptotics. In particular, it has not been proved to be valid in the context of the highly-nontrivial configurations of anti-branes in the warped, fluxed KS throat. It is the lack of this proof that motivates the study in this paper.

This work aims to construct the (leading order) perturbative supergravity description for the KPV state by suitably employing the procedure of MAE. We will see that, even within the leading order MAE, the construction is demanding. A particularly challenging aspect that we were unable to address fully has to do with the expression of the KS fluxes in suitable adapted coordinates. We noticed, however, that there is a convenient modification of the KS solution (in the long-wavelength expansion of interest) that satisfies the SUGRA equations and simplifies the problem. We present a solution of the corresponding MAE that describes the backreaction of the KPV anti-brane state in the background of this modified KS solution. This result, taken in conjunction with previously known results \cite{Armas:2018rsy, Cohen-Maldonado:2015ssa}, serves as further, more direct, evidence in favor of the existence of a backreacted KPV state without unphysical singularities. It also serves as a non-trivial example for the validity of the blackfold conjecture in complicated charged $p$-brane configurations in supergravity with warped/fluxed asymptotics. An explicit perturbative supergravity solution is also expected to be useful in studies of other properties of the KPV state that are currently inaccessible. Some of these properties are discussed at the end of the paper, in the Outlook section.

\subsection{Method outline \& Summary of results}

Let us assume that we want to construct a solution describing a configuration of a (black) $p$-brane with two characteristic length scales ---$r_h$ and $\mathcal{R}$--- in an asymptotic background with characteristic length scale $L$. $r_h$ is a near-horizon scale (e.g.\ a horizon radius or a charge radius for charged solutions) and $\mathcal{R}$ is a characteristic scale of worldvolume inhomogeneities that parametrises the deviations of the configuration from the homogeneous, planar $p$-brane solution. A Matched Asymptotic Expansion (MAE), see e.g. \cite{Gorbonos:2004uc, Harmark:2003yz}, can be understood in this context as a general procedure where the (super)gravity profile of the above configuration is analysed in two asymptotic regions: the {\it far-zone} ($r \gg r_h$), where the profile can be approximated by small deformations of the asymptotic background solution and the {\it near-zone} ($r \ll \mathcal{R}$, $L$), where the profile can be approximated by small deformations of a seed near-horizon solution. $r$ is a radial coordinate transverse to the $p$-brane. In cases of a large scale separation, $r_h \ll \mathcal{R}$, $L$, one can obtain information on the (super)gravity profile by matching the two asymptotic regions across a large overlap zone. Using this information, one can construct an ansatz for the (super)gravity fields, based on long-wavelength deformations of the near-zone seed, where the worldvolume inhomogeneities/background effects on the seed are taken into account perturbatively in powers of $r/\mathcal{R}$, $r/L$. Plugging this ansatz into the (super)gravity equations, solving the resulting differential equations at appropriate order of $r/\mathcal{R}$ and $r/L$, one arrives at an approximate description for the desired configuration when $\mathcal{R}$ and $L$ are large.

In the context of the KPV metastable state, we are interested in the supergravity description of spherical NS5 branes with dissolved anti-D3 brane charge wrapping an $S^2$ inside an $S^3$ at the tip of the KS throat. In this paper, we will focus exclusively on extremal solutions, but we note that similar methods could be applied to study also non-extremal solutions with a finite black hole horizon. In this problem, the length scale $r_h$ is the characteristic near-horizon scale associated to the planar D3-NS5 bound state. This is the extremal horizon of the bound state, which will be denoted as $\rho_c$ in the main text. The length scale $\mathcal{R}$ is controlled by the size of the $S^2$ that the NS5 wraps, while the background length scale $L$ is set by the size of the $S^3$ at tip of the KS throat. As discussed in \cite{Armas:2018rsy}, as long as the spherical NS5 branes are not located at the North or South poles ($\sin \psi \neq 0$), we can always tune our configuration to have a large scale separation, i.e. \ $r_h \ll \mathcal{R}$, $L$, by choosing suitable large background fluxes, that is a suitably large parameter $M$.

\begin{figure}[h]\centering
\includegraphics[width= 0.8\textwidth]{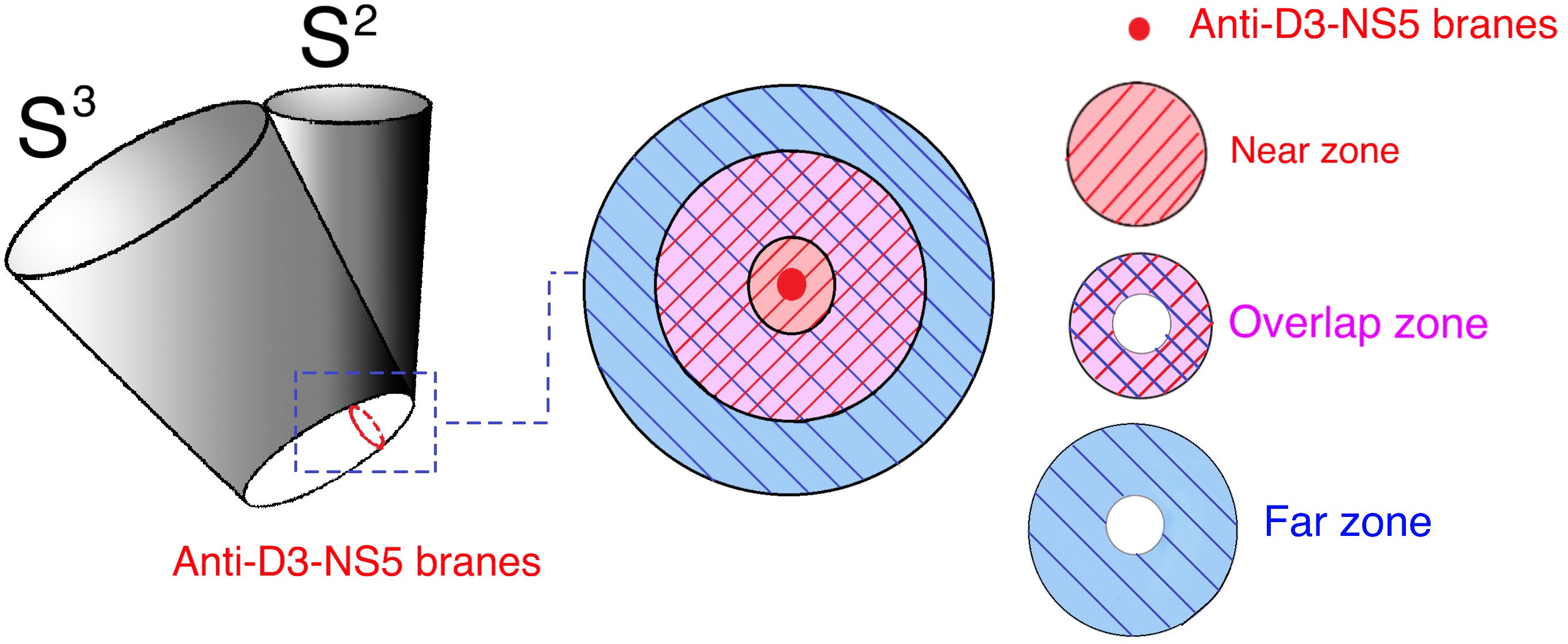}
\caption{Pictorial depiction of the MAE for KPV metastable states in the Klebanov-Strassler geometry.}
\label{MAEPicture}
\end{figure}

A pictorial depiction of the near, far, and overlap zone in the MAE of KPV metastable states appears in Fig \ref{MAEPicture}. In Fig \ref{MAEPicture}, we depict as a red circle the metastable state of spherical anti-D3-NS5 branes warping an $S^2$ inside the $S^3$ at the tip of the Klebanov-Strassler throat. The picture on the right is a pictorial description of the matched asymptotic solution. It is a zoomed-in description of the backreacted metastable anti-D3-NS5 branes at the tip of a large-$M$ KS throat. In the picture, we depict the metastable anti-D3-NS5 branes as a dot. The four transverse dimensions are represented as the two dimensions of the graph. To describe the different zones of the matched asymptotic expansion, we have drawn 3 concentric circles. The near zone is inside of the middle circle. The far zone is the area between the inner-most and the outer-most circle. The overlap zone is the area between the inner circle and the middle circle, the purple region.

The seed of our MAE is the extremal planar D3-NS5 bound state. The relevant asymptotic background is the KS throat expanded/truncated to some order in $\lambda \equiv 1/\sqrt{M}$\footnote{We use $\lambda \equiv 1/\sqrt{M}$ as our small parameter because the radius of the wrapped $S^2$ (i.e. the $\mathcal{R}$ scale) and the radius of the $S^3$ at the tip (i.e. the $L$ scale) both go like $\sqrt{M}$.}. The reason why we only care about the KS throat in the large-$M$ limit, instead of its exact description, is because the aim of our MAE procedure is to produce a perturbative description that approximates the metastable state when $\mathcal{R}$ and $L$ are large.
As such, we are really matching the seed D3-NS5 bound state to a far-zone asymptotic background approximated by the KS throat appropriately expanded/truncated to some order in $\lambda$. 

In this paper, we work up to first order in the $\lambda$-expansion. However, as we were unable to express all the necessary forms of the 3-form fluxes in adapted coordinates, we could not straightforwardly expand the fluxes in $\lambda$ and obtain their first-order description. Nevertheless, we noticed that there is a modification of the KS solution at first order in $\lambda$ which simplifies the profile of the metric and 3-form fluxes but (crucially) retains the validity of the supergravity equations. For this modified perturbative KS solution, we managed to solve the leading order MAE completely. We report results based on this modification.\footnote{It would be useful to understand better the origin of the 3-form flux modification, e.g.\ if it can be extended to higher orders in the $\lambda$-expansion and how it is related to the KS solution. In this paper, we will leave this important question aside and focus on the leading-order MAE of the modified perturbative solution as an instructive example of an explicit leading-order MAE in ten-dimensional type-IIB supergravity that describes the backreaction of anti-3-branes in a non-trivial fluxed, warped background.} The modified description ignores cross-angles components in the metric and 3-form fluxes (see App.\ \ref{KSLargeM}). For example, for the asymptotic background $F_3$ flux, we use $F_3 \sim \lambda \omega_3$, where $\omega_3$ is the volume form of the $S^3$ at the tip. If we expand and truncate the $F_3$ flux of the KS solution to first order in $\lambda$, we will find not only a $\lambda \omega_3$ component but also some cross-angles components. However, we note that if we scale the $\lambda \omega_3$ component of this $F_3$ by a constant factor of $2/\sqrt{3}$, we can reproduce without the cross-angles components the $F_3$ energy-momentum tensor $T^{\mu \nu}_{(F_3)}$ to leading order in $\lambda$. By taking this scaled $\lambda \omega_3$ component as the modified $F_3$, we guarantee that the leading order contribution of the $F_3$ to the metric equation is unchanged. A similar logic is applied to obtain the modified $H_3$ flux. Then, together with the metric and $F_5$ flux, which are left untouched, we can check explicitly that these fluxes satisfy all the supergravity equations to the relevant order in $\lambda$, i.e. to order $\lambda$ in the flux equations and to order $\lambda^2$ in the dilaton and metric equations.

With the form of the asymptotic solution fixed, our first task is to construct the overlap-zone solution. This solution is a matched asymptotic solution of the linearised D3-NS5 bound state, i.e.\ the D3-NS5 solution to leading order in $\rho_c$ (the extremal horizon radius of the D3-NS5 branes), and the modified leading order in $\lambda$ KS background. We require that the overlap-zone profile recovers the asymptotic background in the $\rho_c \rightarrow 0$ limit, the linearised seed in the $\lambda \rightarrow 0$ limit, and flat space in the simultaneous $\lambda, \rho_c \rightarrow 0$ limit. As such, in compact notation we can write the overlap-zone description as
\be
X_{overlap} = X_{flat} + \lambda \, X_{background} + \rho_c^2 \, X_{brane} + \lambda \rho_c^2 \, X_{corrections} ~.
\ee
The extremal horizon radius $\rho_c$ appears in this expression as $\rho_c^2$ because the leading non-trivial order of $\rho_c$ in the far-zone description of D3-NS5 bound state is $\rho_c^2$.
$X_{overlap}$ denotes the overlap-zone profile. $X_{flat}$ denotes the flat space profile. The combination $X_{flat} + \lambda \, X_{background}$ gives the description of the KS solution to leading order in $\lambda$, and the combination $X_{flat} + \rho_c^2 \, X_{brane}$ gives the description of the far-zone, flat-space D3-NS5 solution to leading order in $\rho_c$. The $\lambda \rho_c^2 \, X_{corrections}$ are the needed corrections to ensure that the overlap-zone profile $X_{overlap}$ is a SUGRA solution to leading order in $\lambda$ and $\rho_c$. By studying the SUGRA equations to leading order in $\lambda$ and $\rho_c$, together with the regularity conditions, the correction terms $\lambda \rho_c^2 \, X_{corrections}$ can be explicitly determined, giving us the overlap-zone solution. One interesting point in the overlap-zone discussion is the recovery of the effective worldvolume equations\footnote{This is the version of the worldvolume blackfold equations in \cite{Armas:2018rsy} when one considers the extremal, time-independent case. This is the relevant case for us because we are constructing a metastable state, which is a time-independent configuration of extremal D3-NS5 branes. In \cite{Armas:2018rsy}, the effective worldvolume equations were used to explore also the behaviour of the KPV state under time-dependence and non-zero thermal effects.} derived using the blackfold approach in \cite{Armas:2018rsy}. As the blackfold equations are the necessary conditions for the construction of an overlap-zone solution, we should be able to recover the above equations from our overlap-zone analysis; indeed, we show explicitly that we can. The only difference with the equations reported in \cite{Armas:2018rsy} are certain factors of $2/\sqrt{3}$ which can be traced back to the rescaled components of the $F_3$ and $H_3$ fluxes in the modified asymptotic KS solution.

Armed with an explicit overlap-zone solution, we proceed to set up an ansatz for the leading-order solution across the full spacetime that describes how the KS asymptotics connects to the near-zone deformation of the D3-NS5 seed. This ansatz contains 8 unknown functions, namely $g_{M}(\rho)$, $g_{\omega \omega} (\rho)$, $g_{\rho \rho} (\rho)$, $g_{\Phi} (\rho)$, $g_{\omega \varphi} (\rho)$, $g_{\theta_2 \phi_2} (\rho)$, $g_{x0123} (\rho)$, and $g_{\omega \varphi \theta_2 \phi_2} (\rho)$, which depend only on the adapted radial coordinate $\rho$. Plugging this ansatz into the SUGRA equations yields a set of coupled ordinary differential equations (ODEs) for these 8 unknown functions. 
We present an analytic solution of these equations. The analytic solution involves undetermined integration constants that can be fixed completely by further imposing the regularity conditions at the origin and by requiring that the solution recovers correctly the far-zone modified KS throat asymptotics. In this manner, we verify that there is an explicit leading-order MAE construction of a polarised anti-brane solution without unphysical singularities that interpolates between the D3-NS5 near-zone seed and the far-zone modified KS asymptotics.

\subsection{Outline of paper}

The main results of the paper are contained in section \ref{MAE}. In subsection \ref{MAE2} we setup the near-zone D3-NS5 bound state seed and elaborate on the far-zone Klebanov-Strassler asymptotics. In this subsection we explain the long-wavelength limit of interest and describe how we modify the expanded KS solution to evade the technical issues that arise when one attempts to express the KS 3-form fluxes in adapted coordinates.

In subsection \ref{MAE3} we present the results of the overlap-zone analysis relegating many of the details of the computation to App.\ \ref{OverlapKPV}. An aspect of the analysis that is highlighted here is the derivation of the blackfold effective worldvolume equations (in agreement with previous results in Ref.\ \cite{Armas:2018rsy}).

In subsections \ref{MAE4} and \ref{MAE5} we setup an ansatz for the full leading order MAE, insert it in the supergravity equations and solve the resulting ODEs. In the process we demonstrate how the regularity conditions fix the undetermined integration constants and exhibit the absence of unphysical singularities in the final solution. 
We conclude in section \ref{conclusion} with a summary of our results and a discussion of open questions and potential further computations. A summary of our conventions and useful details of our computations are collected in the appendices.

\section{The Matched Asymptotic Expansion}
\label{MAE}
In this section, we construct a perturbative description of a metastable NS5 state using the procedure of matched asymptotic expansion (MAE). We begin in subsection \ref{MAE2} with the description of the near and far-zone asymptotics in adapted coordinates. For the near-zone, this is the D3-NS5 bound state. For the far-zone, this is the KS throat to leading order in $\lambda$ (with a modification that will be specified). In subsection \ref{MAE3}, we derive the overlap-zone solution. In subsection \ref{MAE4}, we set up an ansatz for the leading order matched asymptotic solution and describe the set of ODEs that arise from inserting the ansatz into the SUGRA equations. We perform a preliminary check of our matched asymptotic equations by using them to recover the overlap-zone solution, and find the sub-leading order corrections required for an overlap-zone solution to higher order in $\rho_c$. In subsection \ref{MAE5}, we solve the matched asymptotic equations analytically with the aid of appropriate computer software. The general solution contains undetermined integration constants. However, all these constants can be fixed by imposing suitable regularity conditions and the far-zone asymptotics.

\subsection{Specifics of the near-zone and far-zone asymptotics}
\label{MAE2}
\subsubsection{The extremal D3-NS5 bound state}
In this subsection, we provide a description of the extremal D3-NS5 bound state in adapted coordinates. This solution serves as the seed (i.e., the near-zone asymptotics) for our matched asymptotic construction. Here, we simply quote the solution and refer the reader to appendix \ref{D3NS5Appen} for details on how it is obtained. 

The metric of the D3-NS5 bound state is
\begin{multline}
\label{mae36}
ds^2 = b_0^2\,  D^{-1/2} \Bigg( - (d x^0)^2 + (d x^1)^2 + (d x^2)^2 + (d x^3)^2 + D \, (d \omega^2 + \omega^2 d \varphi^2 ) \\
+ H \Big( d \rho^2 + \rho^2 \left( d \zeta^2 + \sin^2 \zeta \left( d \theta_2^2 + \sin^2 \theta_2 d \phi_2^2  \right) \right) \Big) \Bigg)
\end{multline}
with
\begin{align}
&H= 1 + \frac{\rho_c^2}{b_0^2 \, \rho^2} ~ , & &D = \left( \sin^2 \theta H^{-1} + \cos^2 \theta \right) ^{-1} ~.
\end{align}
The dilaton field is 
\be
\label{new1}
e^{2 \phi} = H D^{-1} ~,
\ee
and the gauge fields are 
\be
C_2 = \frac{\omega}{\sin^2 \psi_0} \left( \psi_0 - \frac{1}{2} \sin 2 \psi_0  \right) d\omega \wedge d \varphi + b_0^2 \left( 1 - D H^{-1} \right)\, \tan \theta \, \omega d \omega \wedge d \varphi  ~ ,
\ee
\be
\label{new2}
B_2 = - 2 \, \rho_c^2 \cos \theta \sin^2 \zeta \cos \theta_2 \, d \zeta \wedge d \phi_2 ~ ,
\ee
\begin{multline}
\label{mae17}
C_4 = - b_0^4 \, (1 - H^{-1}) \sin \theta \, d x^0 \wedge d x^1 \wedge d x^2 \wedge dx^3 - 2 b_0^2 \, \rho_c^2 \sin \theta \, \sin^2 \zeta \cos \theta_2 \, \omega \, d\zeta \wedge d \phi_2 \wedge d \omega \wedge d \varphi\\
+  \frac{\omega}{\sin^2 \psi_0} \left( \psi_0 - \frac{1}{2} \sin 2 \psi_0  \right) d\omega \wedge d \varphi   \wedge B_2 ~.
\end{multline}

$b_0$ is a constant given by $b_0^2 \approx 0.93266$. In the above description, the coordinates $x^i$, $\omega$, and $\rho$ have dimensions of length, the coordinates $\varphi$, $\zeta$, $\theta_2$, and $\phi_2$ are angles and, are, thus, dimensionless. The only dimension-full parameter is $\rho_c$, which has units of length. The rest, e.g., $\psi_0$ and $\theta$, are dimensionless.

\subsubsection{The Klebanov-Strassler throat in the large-$M$ limit}
\label{mae18}
In this subsection, we collect the far-zone asymptotics of our matched asymptotic solution. Recall that the method of MAE only allows one to describe configurations where the effects of the bending/background can be formulated perturbatively as long-wavelength deformations of a known seed solution. This is possible when the configuration of interest possesses a large separation of scales, i.e.\ the characteristic length scale of the seed is much smaller than the characteristic length scale of the bending/background. As discussed in \cite{Armas:2018rsy}, this happens when $M$ is sufficiently large. Thus, technically, we are not discussing a generic KPV configuration but KPV configurations with a sufficiently large $M$. Accordingly, for our far-zone asymptotics, we want to consider a perturbative expansion of the KS throat in the large-$M$ limit. 

In the subsequent paragraphs, we summarise the properties of a modified version of the KS throat solution at leading order in $\lambda \equiv 1/\sqrt{M}$. We will explain how we come to such a description, but relegate most details to the appendices \ref{KS} and \ref{KSLargeM}. In appendix \ref{KS}, we discuss aspects of the KS throat that are immediately relevant for its role as a background for metastable anti-branes. This includes a derivation of the KS metric in adapted coordinates.\footnote{A similar appendix has already appeared in \cite{Nguyen:2019syc}, but, for convenience of the reader, we present the relevant details again here.} In appendix \ref{KSLargeM}, we derive our modified leading order in $\lambda$ description of the KS throat.

\paragraph{Coordinates in the large-$M$ limit} 
When we take the limit $M \rightarrow \infty$, we have to appropriately rescale our coordinates to make sense of the KS metric and gauge fields in such limit. As an illustrative example, let us consider the metric of an $S^2$ with radius $\sqrt{M}$:
\be
\label{mae11}
ds^2 = M (d \psi^2 + \sin^2 \psi d \omega^2) ~.
\ee
In the limit $M \rightarrow \infty$, the metric blows up and the inverse metric becomes degenerate. To obtain finite expressions, we define a rescaled coordinate:
\be
\label{mae10}
\tilde{\psi} = \sqrt{M} \psi
\ee
with $\tilde{\psi} \in (0, \infty)$. Then, in the $M \rightarrow \infty$ limit, the metric becomes the well-defined metric of flat space in polar coordinates:
\be
\label{flat}
ds^2 = d \tilde{\psi}^2 + \tilde{\psi}^2 d \omega^2 ~.
\ee

Defining the small parameter $\lambda \equiv 1/\sqrt{M}$, we can easily find the expansion of the metric \eqref{mae11}
\be
ds^2 = d \tilde{\psi}^2 + \left( \tilde{\psi}^2 - \lambda^2 \, \frac{\tilde{\psi}^4}{3} + \lambda^4 \, \frac{2 \, \tilde{\psi}^6}{45} \right) d \omega^2 + \mathcal{O} \left( \lambda^6 \right) ~,
\ee
which corrects the leading order flat-space result \eqref{flat}.

By using the rescaled coordinate $\tilde{\psi}$ as (\ref{mae10}), we have effectively zoomed into the North pole of the $S^2$ where $\psi = 0$. For the study of the KPV state, we would like to zoom into the local patch around a local equilibrium at some azimuthal angle $\psi_0 \neq 0$. In that case, the required coordinate scaling is
\be
\tilde{\psi} = \lambda^{-1} \, (\psi - \psi_0) 
~.\ee
Expressing $\psi$ in terms of $\tilde{\psi}$, the $S^2$ metric \eqref{mae11} becomes
\begin{align}
ds^2 &= d \tilde{\psi}^2 + \lambda^{-2} \sin^2 \left( \lambda \tilde{\psi} + \psi_0 \right) d \omega^2 \nonumber\\
&=  d \tilde{\psi}^2 + \lambda^{-2} \left( \sin \lambda \tilde{\psi} \, \cos \psi_0 + \cos \lambda \tilde{\psi} \, \sin \psi_0 \right)^2 d \omega^2 \nonumber\\
&= d \tilde{\psi}^2 + \lambda^{-2} \left( \sin^2 \psi_0  + 2 \lambda \tilde{\psi} \cos \psi_0 \sin \psi_0  + \lambda^2 \tilde{\psi}^2 \cos^2 \psi_0 - \lambda^2 \tilde{\psi}^2 \sin^2 \psi_0 + \mathcal{O} (\lambda^3) \right) \, d \omega^2 ~.
\end{align}
In this case, we also need to rescale the $\omega$ coordinate:
\be
\tilde{\omega} = \lambda^{-1} \sin \psi_0 \, \omega  ~.
\ee
Altogether, in this limit the $S^2$ metric can be written as:
\be
ds^2 = d \tilde{\psi}^2 + d \tilde{\omega}^2 + 2 \lambda \cot \psi_0 \, \tilde{\psi}\, d \omega^2 + \lambda^2  \left( \cot^2 \psi_0 - 1 \right) \tilde{\psi} \, d \omega^2 + \mathcal{O} (\lambda^3) ~.
\ee

We can now return to the KS geometry. Near the tip, the KS metric can be written as
\begin{multline}
g_{\mu \nu} d x^\mu d x^\nu = M b_0^2  \Big ( - (dx^0)^2 + (dx^1)^2 + (dx^2)^2 + (dx^3)^2  + dr^2 \\
+ d \psi ^2 + \sin^2 \psi \left( d \omega^2 + \sin^2 \omega d \varphi^2 \right)  + r^2 (d \theta_2^2 +  \sin^2 \theta_2 d \phi_2^2) \Big) + \mathcal{O} (r^2)
\end{multline}
where $b_0^2 \approx 0.93266$. Completely analogous to the $S^2$ case, we need to implement in the large-$M$ limit the coordinate transformations:
\begin{align}
&\tilde{x}^i =  \lambda^{-1} x^i ~,& &\tilde{r} =  \lambda^{-1} r ~, \nonumber\\
\label{mae13}
&\tilde{\psi} = \lambda^{-1} \, (\psi - \psi_0)  ~, & &\tilde{\omega} = \lambda^{-1} \sin \psi_0 \, \omega  ~,
\end{align}
where $i$ runs from $0$ to $3$.

\paragraph{Modified KS throat solution at leading order in $\lambda$.}

Expanding the KS metric at any order of $\lambda$ is straightforward. One only needs to apply the coordinate scaling \eqref{mae13} to the KS metric in adapted coordinates and truncate to the desired order. As an example, we present the explicit form of the KS metric to order $\lambda^2$ in Eq.\ \eqref{mae28}. Similarly, one can obtain the expansion of the $\tilde{F}_5$ flux to any order in $\lambda$ in the same way. In Eq.\ \eqref{mae30}, we present $\tilde{F}_5$ to order $\lambda^2$.

Unfortunately, the 3-form fluxes are considerably more involved. In this case, we could not find a full description of the $H_3$ and $F_3$ fluxes in adapted coordinates, and did not manage to implement a straightforward expansion in powers of $\lambda$. However, we noticed two things. First, the components of the leading-order $H_3$ and $F_3$ fluxes along the two spheres of the background are easily expressed in adapted coordinates
\begin{align}
\label{SH1}
H_3 &= - 2 \lambda \, r^2 \sin \theta_2 \, d r \wedge d \theta_2 \wedge d \phi_2  + ... ~, \\ 
\label{SH2}
F_3 &= 2 \lambda \,  \omega \, d \psi \wedge d \omega  \wedge d \varphi + ...
~.
\end{align}
These components alone do not yield a leading-order SUGRA solution.
Nevertheless, we observed that if we rescale these components by a factor of $2/\sqrt{3}$, we arrive at a profile that satisfies all the SUGRA equations to the relevant order of $\lambda$. In particular, with this simple rescaling (and by dropping the complicated cross-angle components of the 3-form fluxes), one can satisfy all the flux SUGRA equations to order $\lambda$ and the dilaton/metric SUGRA equations to order $\lambda^2$ (the leading order where the $H_3$, $F_3$ fluxes are relevant). The resulting background is an on-shell modification of the leading-order expansion of the KS solution near the tip. It contains all of the key ingredients of the KS solution (3- and 5-form fluxes and warped metric), hence, we will proceed to consider the MAE of anti-branes in this context.

Let us now collect the modified KS throat solution at first order in $\lambda$. This profile will serve as the far-zone asymptotics for our MAE. We have the metric is 
\begin{multline}
\label{mae45}
ds_{10}^2 = b_0^2 \Big(  - (d x^0)^2 + (dx^1)^2 + (dx^2)^2 + (dx^3)^2 + \left(1 + 2 \lambda \rho \cos \zeta \cot \psi_0 \right)  (d\omega^2 + \omega^2 d \varphi^2) \\
d \rho^2 + \rho^2 \left( d \zeta^2 + \sin^2 \zeta (d \theta_2^2 + \sin^2 \theta_2 d \phi_2^2) \right) 
\Big) ~.
\end{multline}
The non-trivial gauge fields are
\be
B_2 = - \frac{4}{\sqrt{3}} \lambda \rho^3 \sin^2 \zeta \cos \zeta \cos \theta_2 \, d \zeta \wedge d \phi_2 - \frac{4}{\sqrt{3}} \lambda \rho^2 \sin^3 \zeta \cos \theta_2 \, d \rho \wedge d \phi_2 ~ ,
\ee
\be
C_2 = \frac{4}{\sqrt{3}} \lambda \, \rho  \cos \zeta  \, \omega \, d \omega \wedge d \varphi +  \frac{\omega}{\sin^2 \psi_0} \left( \psi_0 - \frac{1}{2} \sin 2 \psi_0  \right) d\omega \wedge d \varphi ~ ,
\ee
\begin{multline}
\label{mae46}
C_4 = - \frac{4}{\sqrt{3}} \lambda \rho^3 \sin^2 \zeta \cos \zeta \cos \theta_2\, \frac{\omega}{\sin^2 \psi_0} \left( \psi_0 - \frac{1}{2} \sin 2 \psi_0  \right)  \, d \zeta \wedge d \phi_2  \wedge  d\omega \wedge d \varphi \\
- \frac{4}{\sqrt{3}} \lambda \rho^2 \sin^3 \zeta \cos \theta_2 \, \frac{\omega}{\sin^2 \psi_0} \left( \psi_0 - \frac{1}{2} \sin 2 \psi_0  \right) d \rho \wedge d \phi_2 \wedge d\omega \wedge d \varphi ~ .
\end{multline}
Through direct substitution into the SUGRA equations \eqref{mae6}-\eqref{mae7}, one can check that this profile indeed satisfies all SUGRA equations to first order in $\lambda$.

\subsection{Overlap-zone analysis}
\label{MAE3}
In this subsection, we discuss the overlap-zone description of our matched asymptotic solution. This is obtained by matching the linearised D3-NS5 bound state and the modified leading order KS throat in such a way that all SUGRA equations are satisfied to leading order in $\lambda$ and $\rho_c^2$. We relegate the details of the derivation to appendix \ref{OverlapKPV}. Here, we present the final result and, subsequently, make connections to the results obtained in \cite{Armas:2018rsy}.

\subsubsection{The overlap-zone solution}

The overlap-zone metric takes the form 
\begin{multline}
\label{mae47}
g_{\mu \nu} dx^\mu dx^\nu = b_0^2 \left( 1 - \frac{\rho_c^2 \sin^2 \theta}{2 \, b_0^2 \rho^2}  \right) \Big( - (d x^0)^2 + (d x^1)^2 + (d x^2)^2 + (d x^3)^2 \Big) \\
+  b_0^2 \left( 1 + \frac{\rho_c^2 \sin^2 \theta}{2 \, b_0^2 \rho^2} + 2 \, \lambda \rho \cos \zeta \cot \psi_0  \right) \Big(d \omega^2 + \omega^2 d \varphi^2 \Big) \\
+ b_0^2 \left( 1 + \frac{\rho_c^2}{ b_0^2 \rho^2} - \frac{\rho_c^2 \sin^2 \theta}{2 \, b_0^2  \rho^2}  \right) \left( d \rho^2 + \rho^2 \left( d \zeta^2 + \sin^2 \zeta \left( d \theta_2^2 +  \sin^2 \theta_2 \, d \phi_2^2  \right) \right) \right) \\
+ \lambda \rho_c^2 \cos \zeta \Bigg[ g_M(\rho) \, \Big( - (d x^0)^2 + (d x^1)^2 + (d x^2)^2 + (d x^3)^2 \Big) + g_{\omega \omega} (\rho) \, \Big( d \omega^2 + \omega^2 d \varphi^2 \Big) \\
+ g_{\rho \rho} (\rho) \, \Big( d \rho^2 + \rho^2 \left( d \zeta^2 + \sin^2 \zeta \left( d \theta_2^2 + \sin^2 \theta_2 d \phi_2^2  \right) \right) \Big) \Bigg]~,
\end{multline}
where the correction functions $g_{M} (\rho)$, $g_{\omega \omega} (\rho)$, and $g_{\rho \rho}(\rho)$ are given by\footnote{In the expressions below, we have used \eqref{Final1} to express $\cot \psi_0$ in term of $\theta$. 
}
\begin{align}
g_M (\rho) &=  \frac{\sin^2 \theta \left( 2 \cot \theta + \sec \theta + \tan \theta \right)}{\sqrt{3} \, b_0^2 \, \rho} ~ , \\
g_{\omega \omega} (\rho) &=  \frac{\sin^2 \theta \left( - 2 \cot \theta + \sec \theta + \tan \theta \right)}{\sqrt{3} \, b_0^2 \, \rho} ~, \\
g_{\rho \rho} (\rho) &= \frac{4 \cos \theta \left( 1 - 2 \sin \theta \right) + \left( 1 + 7 \cos 2 \theta \right) \left( \sec \theta + \tan \theta \right)}{6 \sqrt{3} \, b_0^2 \,\rho} ~ .
\end{align}
The dilaton sourced by the anti-branes is 
\be
\phi =  \frac
{\rho_c^2 \cos^2 \theta}{ 2 \, b_0^2 \, \rho^2}  + \lambda \rho_c^2 \,  \frac{\cos \theta \left(  1+ \sin \theta \right)}{\sqrt{3} \, b_0^4 \, \rho} ~.
\ee
The $C_2$ gauge field is 
\begin{multline}
C_2 = \frac{\omega}{\sin^2 \psi_0} \left( \psi_0 - \frac{1}{2} \sin 2 \psi_0  \right) d\omega \wedge d \varphi + \frac{4}{\sqrt{3}} \lambda \, \rho  \cos \zeta  \, \omega d \omega \wedge d \varphi \\
+ \frac{\rho_c^2}{\rho^2} \sin \theta \cos \theta \, \omega d \omega \wedge d \varphi + \lambda \rho_c^2 \, \frac{\mathcal{A} \cos \zeta}{\rho} \, \omega d \omega \wedge d \varphi
\end{multline}
with
\be
\mathcal{A} = \frac{2}{\sqrt{3} \, b_0^2} \Big( 1+ \sin \theta \Big) \Big( 3 \sin \theta - 1 \Big) ~. 
\ee
The $B_2$ gauge field is 
\begin{multline}
\label{KPV6}
B_2 = - \frac{4}{\sqrt{3}} \lambda \rho^3 \sin^2 \zeta \cos \zeta \cos \theta_2 \, d \zeta \wedge d \phi_2 - \frac{4}{\sqrt{3}} \lambda \rho^2 \sin^3 \zeta \cos \theta_2 \, d \rho \wedge d \phi_2 \\
- 2 \, \rho_c^2 \cos \theta \sin^2 \zeta \cos \theta_2 \, d \zeta \wedge d \phi_2 \\
+ \lambda \rho_c^2 \,  \mathcal{B}  \, \rho \sin^3 \zeta \, \sin \theta_2 d \theta_2 \wedge d \phi_2 
\end{multline}
with
\be
\mathcal{B} = - \frac{2 }{\sqrt{3} \, b_0^2} \cos^2 \theta ~.
\ee

Finally, the $C_4$ gauge field is 
\begin{multline}
\label{KPV7}
C_4 = - b_0^2 \sin \theta \, \frac{\rho_c^2}{\rho^2}  \, d x^0 \wedge d x^1 \wedge d x^2 \wedge dx^3 - 2 \, b_0^2 \,  \rho_c^2 \sin \theta \sin^2 \zeta  \, \omega  \cos \theta_2 \, d \omega \wedge d \varphi  \wedge d \zeta \wedge d \phi_2 \\ 
+ \lambda \rho_c^2 \,  \frac{\mathcal{C}_1 \cos \zeta}{\rho}  d x^0 \wedge d x^1 \wedge d x^2 \wedge d x^3 + \lambda \rho_c^2 \, \mathcal{C}_2 \, \rho \sin^3 \zeta  \, \omega \sin \theta_2 \, d \omega \wedge d \varphi \wedge d \theta_2 \wedge d \phi_2 \\
+ B_2 \wedge (C_2)_{0} 
\end{multline}
with 
\be
C_0 = \frac{\omega}{\sin^2 \psi_0} \left( \psi_0 - \frac{1}{2} \sin 2 \psi_0  \right) d\omega \wedge d \varphi ~,
\ee
and
\begin{align}
\mathcal{C}_1 &= \frac{2}{\sqrt{3}} \Big( \cos \theta \Big( 1 - \sin \theta \Big) + \tan \theta \Big(1 + \sin \theta \Big) \Big) ~, ~~
\mathcal{C}_2 = - \frac{2}{\sqrt{3}} \left(  \cos \theta \Big(  1 +  \sin \theta \Big) + \tan \theta \Big(1 + \sin \theta \Big) \right) ~.
\end{align}

\subsubsection{Recovery of the blackfold equations} 
In \cite{Armas:2018rsy}, the conditions for the existence of the metastable state were obtained by solving the effective worldvolume equations of the blackfold approach. For a static configuration at extremality, these equations are 
\be
\label{KPV1}
\cot \psi_0 = \frac{1}{b_0^2} \sqrt{1  + \tan^2 \theta}  + \frac{1}{b_0^2} \tan \theta
\ee
with 
\be
\label{KPV2}
\tan \theta = \frac{1}{b_0^2 \sin^2 \psi_0} \left( \frac{\pi p}{M} - \left( \psi_0 - \frac{1}{2} \sin 2 \psi_0 \right) \right)  ~.
\ee
Here, $\psi_0$ is the value of the azimuthal angle of the $S^3$ around which the spherical anti-D3-NS5 branes wraps and $p$ is defined as 
\be
p = \frac{\mathbb{Q}_3}{4 \pi^2 \mathbb{Q}_5} 
\ee
with $\mathbb{Q}_3$ and $\mathbb{Q}_5$ the conserved charges carried by the anti-D3-NS5 branes. As the blackfold equations form the necessary conditions for the construction of a matched asymptotic solution, we should be able to recover them from our overlap-zone solution. 

In our matched asymptotic construction, we have introduced two independent parameters: $\psi_0$, which is the azimuthal angle of the $S^3$ at the tip of the KS throat that we zoom into, and $\tan \theta$, which is the ratio of the D3 and NS5 brane charges in the D3-NS5 bound state. We observe that the overlap-zone SUGRA equations (SUGRA equations to leading order in $\lambda$ and $\rho_c^2$) force these two parameters to obey a relationship that is independent of the correction terms:
\be
\label{Final1}
\cot \psi_0 = \frac{2}{\sqrt{3}} \left( \frac{1}{b_0^2} \sec \theta + \frac{1}{b_0^2} \tan \theta \right) ~.
\ee
This equation is identical to equation \eqref{KPV1} with the exception of a $2/\sqrt{3}$ factor in front of the RHS. The origin  of this factor can be traced to the fact that we have rescaled the relevant component of the KS $F_3$ and $H_3$ fluxes by a factor of $2/\sqrt{3}$ in order to obtain a simplified leading order in $\lambda$ description of them. 

Recall that equation \eqref{KPV2} is derived in \cite{Armas:2018rsy} by, first, computing the conserved charges $\mathbb{Q}_3$ and $\mathbb{Q}_5$ from current conservation equations. In particular, the conserved charges are given by:
\begin{align}
\mathbb{Q}_5 &= - \int * j_6 = - \mathcal{C} \rho_c^2 \cos \theta ~, \\
\label{KPV8}
\mathbb{Q}_3 &= - \int * (J_4 - *( * j_6 \wedge C_2)) = - \, \Omega_{\text{Wrapped $S^2$}} \times \mathcal{C} \rho_c^2 \left(b^2_0  \sin \theta     +  \frac{\cos \theta}{\sin^2 \psi_0} \, \left( \psi_0 - \frac{1}{2} \sin 2 \psi_0 \right) \right)~,
\end{align}
where $\Omega_{\text{Wrapped $S^2$}} = 4 \pi M \sin^2 \psi_0$ is the surface area of the $S^2$ wrapped by the anti-D3-NS5 branes. The constant $\mathcal{C}$ is given by $\mathcal{C} = \Omega_3/ 8 \pi G$ with $\Omega_3 = 2 \pi^2$ the surface area of a unit $S^3$. Then, by introducing the parameter $p/M$ via the identification $p = \mathbb{Q}_3/(4 \pi^2 \mathbb{Q}_5) $, one obtains equation \eqref{KPV2}.

Before continuing, let us stress that the parameter $p/M$ is finite in the large-$M$ regime. As $\mathbb{Q}_3$ is a total charge that includes the surface area of the spherical NS5 ($4 \pi  M \sin^2 \psi_0$), the charge ratio $p$ actually scales as $M$. Therefore, the parameter $p/M$ has its $M$ factors cancelled and, thus, is finite in the large-$M$ limit. For convenience, we define
\be
{\mathfrak p}^* \equiv p/M  = \frac{\mathbb{Q}_3}{4 \pi^2 M \, \mathbb{Q}_5 } = \frac{\Omega_{\text{Wrapped $S^2$}} \times \mathcal{Q}_3}{4 \pi^2  M \mathbb{Q}_5} =  \frac{4 \pi M \sin^2 \psi_0 \mathcal{Q}_3}{4 \pi^2 M \mathbb{Q}_5} = \frac{ \sin^2 \psi_0 \mathcal{Q}_3}{ \pi \mathbb{Q}_5} 
\ee
with $\mathcal{Q}_3$ a charge density over the wrapped 2-sphere. Using $\mathfrak p^*$, Eq.\ \eqref{KPV2} becomes:
\be
\label{KPV5}
\tan \theta = \frac{1}{b_0^2 \sin^2 \psi_0} \left( \pi \mathfrak p^* - \left( \psi_0 - \frac{1}{2} \sin 2 \psi_0 \right) \right)  ~.
\ee

We can similarly derive Eq.\ \eqref{KPV5} from the overlap-zone solution. First, we compute the conserved charges $\mathbb{Q}_3$ and $\mathbb{Q}_5$ of the overlap-zone solution from the fluxes (using Gauss' law), and then, parametrise the charge ratio $\mathbb{Q}_3/ \mathbb{Q}_5$ using $\mathfrak p*$. We obtain: 
\begin{align}
\mathbb{Q}_5 &= \frac{1}{16 \pi G} \lim_{\rho \rightarrow \infty}  \int_{\zeta, \theta_2, \phi_2} H_3 = - \mathcal{C} \rho_c^2 \cos \theta ~ ,\\
\mathbb{Q}_3 &= \frac{1}{16 \pi G} \lim_{\rho \rightarrow \infty} \int_{\zeta, \theta_2, \phi_2, \omega, \varphi} F_5 = - Vol_2 \times \mathcal{C} \rho_c^2 \left( b_0^2 \sin \theta + \frac{\cos \theta}{\sin^2 \psi_0} \left( \psi_0 - \frac{1}{2} \sin 2 \psi_0 \right)  \right)~,
\end{align}

where $\mathcal{C}$ is again $\mathcal{C} = \Omega_3/ 8 \pi G$ with $\Omega_3 = 2 \pi^2$ the surface area of a unit $S^3$, and $Vol_2$ the (regularized) volume of an $\mathbb{R}^2$. In the above computations, we have used $H_3 = d B_2$ with $B_2$ given in \eqref{KPV6}, and $F_5 = d C_4$ with $C_4$ given in \eqref{KPV7}. Comparing the $\mathbb{Q}_3$ charge computed here and the one computed in \cite{Armas:2018rsy}, i.e. \eqref{KPV8}, we note that there is a difference in volume factors ---$\Omega_{\text{Wrapped $S^2$}}$ versus $Vol_2$. This difference is a result of blowing up the wrapped $S^2$ ($\sqrt{M} \sin \psi_0 \rightarrow \infty$). On the other hand, we note that the charge densities $\mathcal{Q}_3$ are identical in both computations. Therefore, defining the parameter $\mathfrak p^*$ as
\be
\mathfrak p^* = \frac{\sin^2 \psi_0 \mathcal{Q}_3}{\pi \mathbb{Q}_5}
\ee
and plugging in $\mathcal{Q}_3 = - \mathcal{C} \rho_c^2 \left( b_0^2 \sin \theta + \frac{\cos \theta}{\sin^2 \psi_0} \left( \psi_0 - \frac{1}{2} \sin 2 \psi_0 \right)  \right)$ and $\mathbb{Q}_5 = - \mathcal{C} \rho_c^2 \cos \theta$ yields equation \eqref{KPV5}.

To summarise, we have obtained directly from the overlap-zone solution the ``blackfold equations'' 
\be
\label{Final12}
\cot \psi_0 = \frac{2}{\sqrt{3}} \left( \frac{1}{b_0^2} \sec \theta + \frac{1}{b_0^2} \tan \theta \right) ~,
\ee
\be
\label{Final13}
\tan \theta = \frac{1}{b_0^2 \sin^2 \psi_0} \left( \pi p^* - \left( \psi_0 - \frac{1}{2} \sin 2 \psi_0 \right) \right)  ~.
\ee

\subsection{ODE reduction of the SUGRA equations} 
\label{MAE4}

\subsubsection{Ansatz and resulting ODEs}

Drawing inspiration from the overlap-zone solution \eqref{mae47}-\eqref{KPV7}, we postulate the following ansatz for the leading-order matched asymptotic solution at all orders of $\rho_c$. For the metric we set\footnote{Note that all unknown functions are functions of $\rho_c^2$. The $\lambda \rho_c^2$ scaling in front is simply for convenience.}
\begin{multline}
\label{KPV9}
ds^2_{10} = b_0^2\,  D^{-1/2} \Bigg[ - (d x^0)^2 + (d x^1)^2 + (d x^2)^2 + (d x^3)^2 + D \, \left(d \omega^2 + \omega^2 d \varphi^2 \right) \\
+ H \Big( d \rho^2 + \rho^2 \left( d \zeta^2 + \sin^2 \zeta \left( d \theta_2^2 + \sin^2 \theta_2 d \phi_2^2  \right) \right) \Big) \Bigg]  \\
+ 2 b_0^2 \lambda \,  \rho \cos \zeta \cot \psi_0 ( d \omega^2 + \omega^2 d \varphi^2) \\
+ \lambda \rho_c^2 \cos \zeta \Bigg[ g_M(\rho) \, \Big( - (d x^0)^2 + (d x^1)^2 + (d x^2)^2 + (d x^3)^2 \Big) + g_{\omega \omega} (\rho) \, \Big( d \omega^2 + \omega^2 d \varphi^2 \Big) \\
+ g_{\rho \rho} (\rho) \, \Big( d \rho^2 + \rho^2 \left( d \zeta^2 + \sin^2 \zeta \left( d \theta_2^2 + \sin^2 \theta_2 d \phi_2^2  \right) \right) \Big) \Bigg]
\end{multline}
with
\begin{align}
&H= 1 + \frac{\rho_c^2}{b_0^2 \, \rho^2} ~ , & &D = \left( \sin^2 \theta H^{-1} + \cos^2 \theta \right) ^{-1} ~.
\end{align}
For the dilaton we set
\be
\phi = \frac{1}{2} \ln (H D^{-1}) + \lambda \rho_c^2 \cos \zeta \, g_{\Phi} (\rho) ~,
\ee
and for the gauge fields 
\begin{multline}
C_2 = \frac{\omega}{\sin^2 \psi_0} \left( \psi_0 - \frac{1}{2} \sin 2 \psi_0  \right) d\omega \wedge d \varphi + b_0^2 \left( 1 - D H^{-1} \right)\, \tan \theta \, \omega d \omega \wedge d \varphi  \\
+ \frac{4}{\sqrt{3}} \lambda \, \rho  \cos \zeta  \, \omega \, d \omega \wedge d \varphi + \lambda \rho_c^2 \cos \zeta \, g_{\omega \varphi} (\rho) \, \omega d \omega \wedge d \varphi ~,
\end{multline}
\begin{multline}
B_2 = - \frac{4}{\sqrt{3}} \lambda \rho^3 \sin^2 \zeta \cos \zeta \cos \theta_2 \, d \zeta \wedge d \phi_2 - \frac{4}{\sqrt{3}} \lambda \rho^2 \sin^3 \zeta \cos \theta_2 \, d \rho \wedge d \phi_2 \\
- 2 \, \rho_c^2 \cos \theta \sin^2 \zeta \cos \theta_2 \, d \zeta \wedge d \phi_2 \\
+ \lambda \rho_c^2 \, \sin^3 \zeta \, g_{\theta_2 \phi_2} (\rho) \, \sin \theta_2 d \theta_2 \wedge d \phi_2 ~,
\end{multline}
\begin{multline}
\label{KPV10}
C_4 = - b_0^4 \, (1 - H^{-1}) \sin \theta \, d x^0 \wedge d x^1 \wedge d x^2 \wedge dx^3 - 2 b_0^2 \, \rho_c^2 \sin \theta \, \sin^2 \zeta \cos \theta_2 \, \omega \, d\zeta \wedge d \phi_2 \wedge d \omega \wedge d \varphi\\
+ \lambda \rho_c^2 \,  \cos \zeta \, g_{x0123} (\rho) \,  d x^0 \wedge d x^1 \wedge d x^2 \wedge d x^3 + \lambda \rho_c^2 \, \sin^3 \zeta  \, g_{\omega \varphi \theta_2 \phi_2} (\rho) \, \omega \sin \theta_2 \, d \omega \wedge d \varphi \wedge d \theta_2 \wedge d \phi_2 \\
+ B_2 \wedge (C_2)_0  ~.
\end{multline}
We have denoted 
\be
(C_2)_0 =  \frac{\omega}{\sin^2 \psi_0} \left( \psi_0 - \frac{1}{2} \sin 2 \psi_0  \right) d\omega \wedge d \varphi ~.
\ee
The $\psi_0$ and $\theta$ parameters are related by the equation:
\be
\label{Near1}
\cot \psi_0 = \frac{2}{\sqrt{3}} \left( \frac{1}{b_0^2} \sec \theta + \frac{1}{b_0^2} \tan \theta \right) ~.
\ee

In this ansatz, the functions $g_{M}(\rho)$, $g_{\omega \omega} (\rho)$, $g_{\rho \rho} (\rho)$, $g_{\Phi} (\rho)$, $g_{\omega \varphi} (\rho)$, $g_{\theta_2 \phi_2} (\rho)$, $g_{x0123} (\rho)$, and $g_{\omega \varphi \theta_2 \phi_2} (\rho)$ are unknown functions of the adapted radial coordinate $\rho$ only. They should also depend parametrically on the parameters $\rho_c$ and $\theta$ (assuming we have expressed $\psi_0$ in terms of $\theta$ via Eq.\ \eqref{Near1}). Inserting the ansatz \eqref{KPV9}-\eqref{KPV10} into the SUGRA equations yields a set of ODEs on the $\rho$-dependence of the 8 unknown functions. The reduction of the full set of PDEs to a system of ODEs is encouraging. We emphasise that our ansatz is at least cohomogeneity-two and depends on $\rho$ and $\zeta$. We have explicitly identified the $\zeta$ dependence and the remaining $\rho$-dependence is captured by the aforementioned functions.

As a preliminary check of the ansatz, we expand and solve the resulting ODEs order by order in the $\rho_c$ expansion to recover the overlap-zone solution in \eqref{mae47}-\eqref{KPV7} and produce the sub-leading order corrections needed for an overlap-zone solution to higher order in $\rho_c$.

\paragraph{Overlap-zone solution to sub-leading orders}
To find the overlap-zone solution beyond the leading order in $\rho_c$ we insert the expansions
\begin{align}
\label{Near2}
g_{M}(\rho) &= \frac{\mathtt{a}_1}{\rho} + \frac{\mathtt{b}_1 \rho_c^2}{\rho^3} + \frac{\mathtt{c}_1 \rho_c^4}{\rho^5} + \frac{\mathtt{d}_1 \rho_c^6}{\rho^7} + ...  ~,\\ 
g_{\omega \omega} (\rho) &= \frac{\mathtt{a}_2}{\rho} + \frac{\mathtt{b}_2 \rho_c^2}{\rho^3} + \frac{\mathtt{c}_2 \rho_c^4}{\rho^5} + \frac{\mathtt{d}_2 \rho_c^6}{\rho^7} + ...  ~,\\ 
g_{\rho \rho} (\rho) &= \frac{\mathtt{a}_3}{\rho} + \frac{\mathtt{b}_3 \rho_c^2}{\rho^3} + \frac{\mathtt{c}_3 \rho_c^4}{\rho^5} + \frac{\mathtt{d}_3 \rho_c^6}{\rho^7} + ...  ~,\\ 
g_{\Phi} (\rho) &= \frac{\mathtt{a}_4}{\rho} + \frac{\mathtt{b}_4 \rho_c^2}{\rho^3} + \frac{\mathtt{c}_4 \rho_c^4}{\rho^5} + \frac{\mathtt{d}_4 \rho_c^6}{\rho^7} + ...  ~,\\ 
g_{\theta_2 \phi_2} (\rho) &= \mathtt{a}_5 \rho + \frac{\mathtt{b}_5 \rho_c^2}{\rho} + \frac{\mathtt{c}_5 \rho_c^4}{\rho^3} + \frac{\mathtt{d}_5 \rho_c^6}{\rho^5} + ...  ~,\\ 
g_{\omega \varphi} (\rho) &= \frac{\mathtt{a}_6}{\rho} + \frac{\mathtt{b}_6 \rho_c^2}{\rho^3} + \frac{\mathtt{c}_6 \rho_c^4}{\rho^5} + \frac{\mathtt{d}_6 \rho_c^6}{\rho^7} + ...  ~,\\ 
g_{x0123} (\rho) &=  \frac{\mathtt{a}_7}{\rho} + \frac{\mathtt{b}_7 \rho_c^2}{\rho^3} + \frac{\mathtt{c}_7 \rho_c^4}{\rho^5} + \frac{\mathtt{d}_7 \rho_c^6}{\rho^7} + ...  ~,\\ 
g_{\omega \varphi \theta_2 \phi_2} (\rho) &= \mathtt{a}_8 \rho + \frac{\mathtt{b}_8 \rho_c^2}{\rho} + \frac{\mathtt{c}_8 \rho_c^4}{\rho^3} + \frac{\mathtt{d}_8 \rho_c^6}{\rho^5} + ...  ~.
\end{align}
into the matched asymptotic ODEs and truncate the resulting equations to some order of $\rho_c$ to obtain a set of algebraic equations for the coefficients: $a$, $b$, $c$, etc. In particular, to recover the overlap-zone solution, we solve for the coefficients $a_i$ ($i=1,\ldots,8)$ by considering the order $\rho_c^2$ algebraic equations. In this manner, we obtain 
\begin{align}
\label{Final4}
\mathtt{a}_1 &= \frac{\sin^2 \theta (2 \cot \theta + \sec \theta + \tan \theta)}{\sqrt{3} \,b_0^2} ~,\\
\mathtt{a}_2 &= \frac{\sin^2 \theta (- 2 \cot \theta + \sec \theta + \tan \theta)}{\sqrt{3} \, b_0^2} ~,\\
\mathtt{a}_3 &= \frac{4 \cos \theta (1 - 2 \sin \theta) + (1 + 7 \cos 2 \theta)(\sec \theta + \tan \theta)}{6 \sqrt{3}\,  b_0^2}~, \\
\mathtt{a}_4 &= \frac{\cos \theta (1 + \sin \theta)}{\sqrt{3} \, b_0^4} ~, \\
\mathtt{a}_5 &= \frac{- 2 }{\sqrt{3} \, b_0^2} \cos^2 \theta ~, \\
\mathtt{a}_6 &= \frac{2}{\sqrt{3} \, b_0^2} \Big( 1+ \sin \theta \Big) \Big( 3 \sin \theta - 1 \Big) ~, \\
\mathtt{a}_7 &=  \frac{2}{\sqrt{3}} \Big( \cos \theta \Big( 1- \sin \theta \Big) + \tan \theta \Big( 1+ \sin \theta \Big)  \Big)~, \\ 
\label{Final5}
\mathtt{a}_8 &=  - \frac{2}{\sqrt{3}} \Big( \cos \theta \Big( 1 + \sin \theta \Big) + \tan \theta \Big( 1+ \sin \theta \Big)  \Big) ~ .
\end{align}
As one can easily check, these results are indeed consistent with the overlap-zone solution in Eqs.\ \eqref{mae47}-\eqref{KPV7}. 

With all the $a_i$ coefficients determined, we can proceed to consider the order $\rho_c^4$ algebraic equations and solve for the $b_i$ coefficients. At this point we notice that the order $\rho_c^4$ algebraic equations can be reduced to only two constraints. This means that 6 of the $b_i$ are left undetermined. We choose to set 
\be
\label{bchoice}
b_5 = b_8 = 0 ~.
\ee
This choice is motivated by the observation that in the seed D3-NS5 solution \eqref{mae36}-\eqref{mae17}, the $B_2$ gauge field \eqref{new2} and the components of the $C_4$ gauge field \eqref{mae17} in the angular directions truncate to order $\rho_c^2$ (as opposed to being an infinite sum of $\rho_c^k$ terms). We will see momentarily that this truncation can be implemented consistently at arbitrarily high order (and in the next subsection that this choice leads to a well-behaved solution). At this point, Eqs.\ \eqref{bchoice} leave 4 undetermined $b$-coefficients. 

Considering the order-$\rho_c^6$ algebraic equations, we can solve for the $c_i$ coefficients, which are expressed in terms of the 4 undetermined $b_i$ coefficients. We can choose the latter to set $c_5 = c_8 = 0$. For the higher-order coefficients we can similarly use the allowed freedom to set the corresponding corrections to the $g_{\theta_2 \phi_2}$ and $g_{\omega \varphi \theta_2 \phi_2}$ unknown functions to zero. We will implement this observation to motivate a concrete ansatz for the unknown functions $g_{\theta_2 \phi_2}$ and $g_{\omega \varphi \theta_2 \phi_2}$ in the full solution of the next subsection.

\subsection{Solving the matched asymptotic ODEs}
\label{MAE5}

The expectation that the unknown functions in the $B_2$ gauge field and the components of the $C_4$ gauge field along the angular directions truncate to order $\rho_c^2$ means that we have already obtained their full profile in the above perturbative expansion:
\begin{align}
\label{new3}
g_{\theta_2 \phi_2} (\rho) &= - \frac{2 \cos^2 \theta\, \rho}{\sqrt{3} \, b_0^2}  ~, ~~
g_{\omega \varphi \theta_2 \phi_2} (\rho) = - \frac{2 \left( 1 + \sin \theta \right) \left( \tan \theta + \cos \theta \right)}{\sqrt{3}}  \rho ~.
\end{align}
Inserting these functions into the $\tilde{F}_5$ duality equation, we obtain the $g_{x0123} (\rho)$ function for free 
\be
\label{new4}
g_{x0123} (\rho) = \frac{2 \, H \sec \theta \Big( \rho_c^2 + \left( -2 b_0^2 \, \rho^2 + \rho_c^2 \right) \sin \theta \Big) + D \Big( \rho_c^2 \sin 2 \theta + 4 b_0^2 \rho^2 \tan \theta \Big)}{\sqrt{3} \rho_c^2 \, \rho  \, D \, H^2} ~.
\ee
Expanding $g_{x0123} (\rho)$ in terms of $\rho_c$, we verify that it reproduces the sub-leading order overlap-zone results (namely, the coefficients $\mathtt{a}_7$, $\mathtt{b}_7$, $\mathtt{c}_7$, $\mathtt{d}_7$, etc).

With 3 functions in Eqs.\ \eqref{new3}-\eqref{new4} determined, our task reduces to finding only 5 unknown functions. These remaining unknown functions are: $g_\phi (\rho)$, $g_M (\rho)$, $g_{\omega \omega} (\rho)$, $g_{\rho \rho} (\rho)$, and $g_{\omega \varphi} (\rho)$. To make progress with our system of ODEs, we first note that the $B_2$ equation and the $\tilde{F}_5$ duality equation\footnote{Above, we mentioned already the $\tilde{F}_5$ duality condition. Notice that the $\tilde{F}_5$ duality condition yields 2 independent equations, one from the $d x^0 \wedge ... \wedge d x^3 \wedge d \rho$ component and the other from the angular component $d \omega \wedge d \varphi \wedge d \rho \wedge d \theta_2 \wedge d \phi_2$. The determination of $g_{x0123} (\rho)$ arises from the angular component $d \omega \wedge d \varphi \wedge d \rho \wedge d \theta_2 \wedge d \phi_2$ of the duality equation. Here we are interested in the equation that comes from the $d x^0 \wedge ... \wedge d x^3 \wedge d \rho$ component.} are algebraic equations of the functions $g_\phi (\rho)$, $g_M (\rho)$, $g_{\omega \omega} (\rho)$, $g_{\rho \rho} (\rho)$, and $g_{\omega \varphi} (\rho)$. Using these two equations, one can write 2 unknowns in terms of the others. In particular, we can express $g_{\omega \varphi} (\rho)$ and $g_{M} (\rho)$ in terms of $g_\phi (\rho)$, $g_{\omega \omega} (\rho)$, and $g_{\rho \rho} (\rho)$. After the above step, we are left with three unknowns: $g_\phi (\rho)$, $g_{\rho \rho} (\rho)$ and $g_{\omega \omega} (\rho)$. Naively, we still have a list of ODEs to satisfy. However, by substituting in all the known corrections functions, we easily see that the $g_{\rho \rho} (\rho)$, $g_\phi (\rho)$, and $g_{\omega \omega} (\rho)$ also share an algebraic relationship. This allows us to write $g_{\rho \rho} (\rho)$ in terms of $g_\phi (\rho)$ and $g_{\omega \omega} (\rho)$. After a few trivial simplifications, we are left with only two unknowns functions: $g_{\omega \omega} (\rho)$ and $g_\phi (\rho)$, and two independent ODEs. With the help of appropriate computer software (Mathematica), these two second order ODEs can be solved analytically to give $g_{\omega \omega} (\rho)$ and $g_\phi (\rho)$. The solution has 4 unknown integration constants. However, we note that these 4 integration constants can be uniquely determined by requiring the absence of all unphysical singularity at the origin and by imposing that the solution recovers the correct far-zone asymptotics in the $\rho_c \rightarrow 0$ limit. Finally, as a check, we take all our unknown functions, e.g. $g_{M}(\rho)$, $g_{\omega \omega} (\rho)$, $g_{\rho \rho} (\rho)$, $g_{\Phi} (\rho)$, $g_{\omega \varphi} (\rho)$, $g_{\theta_2 \phi_2} (\rho)$, $g_{x0123} (\rho)$, and $g_{\omega \varphi \theta_2 \phi_2} (\rho)$, substitute them into the ansatz \eqref{KPV9}-\eqref{KPV10}, and plug everything into the SUGRA equations \eqref{mae6}-\eqref{mae7} to verify that we indeed have a valid solution to first order in $\lambda$.

Before writing down the explicit description of the $g_M (\rho)$, $g_{\omega \omega} (\rho)$, $g_{\rho \rho} (\rho)$, and $g_{\omega \varphi} (\rho)$ unknown functions, let us mention a computational caveat. As the matched asymptotic ODEs are rather complicated, analysing them with generic $\theta$ causes Mathematica to be extremely slow. As such, it is computationally convenient to set the parameter $\theta$  to  specific values, e.g.\ $\theta = \pi/6$.  The choice of $\theta$ specifies the exact metastable state we are interested in. For example, by looking at the blackfold equations \eqref{Final12}-\eqref{Final13}, we see that the choice $\theta = \pi/6$ corresponds to a metastable state at $\psi_0 \approx 0.436351$ with $p/M \equiv {\mathfrak p}^* \approx 0.0475869$. While computationally a choice like $\theta = \pi/6$ is convenient, physically such a choice is not special. If we set $\theta$ to any other value, given that such a value of $\theta$ corresponds to a metastable state, we expect everything to work in the same way. The $g_M (\rho)$, $g_{\omega \omega} (\rho)$, $g_{\rho \rho} (\rho)$, and $g_{\omega \varphi} (\rho)$ correction functions in the case $\theta = \pi/6$ are given by:
\begin{align}
g_{\phi} (\rho) &= \frac{3 \rho }{4 b_0^4 \rho^2 + 3 b_0^2 \rho^2_c} ~,~~~~
g_M (\rho) = \frac{3 \rho}{2 \sqrt{4 b_0^4 \rho^4 + 7 b_0^2 \rho^2 \rho_c^2 +3 \rho_c^4}} ~, \\
\begin{split}
g_{\omega \omega} (\rho) &= \frac{2 \rho}{\rho_c \left(\frac{4 b_0^2 \rho ^2}{\rho_c^2}+3\right)^{3/2} \sqrt{b_0^2 \rho ^2+\rho_c^2}}  \vast\{ -6 \sqrt{\frac{4 b_0^4 \rho ^4}{\rho_c^4}+\frac{7 b_0^2 \rho ^2}{\rho_c^2}+3}+9 \\
   &+  \frac{b_0^2 \rho ^2 \left(16 b_0^2 \rho ^2-8 \sqrt{4 b_0^4 \rho ^4+7 b_0^2 \rho ^2 \rho_c^2+3 \rho_c^4}+25 \rho_c^2\right)}{\rho_c^4} \vast\} ~, 
\end{split} \\
g_{\rho \rho} (\rho) &= \frac{3}{4 b_0^2 \, \rho} \sqrt{\frac{\rho_c^2}{4 b_0^2 \rho^2+3 \rho_c^2}+1} ~, \\
\begin{split}
g_{\omega \varphi} (\rho) &= \frac{8 \sqrt{3} b_0^2 \rho^3}{\left(2 b_0^2 \rho ^2+3 \rho_c^2\right) \left(4 b_0^2 \rho ^2 \rho_c + 3 \rho_c^3\right)^2} \vast\{\rho_c^4 \left(3-18 \sqrt{\frac{b_0^2 \rho ^2+\rho_c^2}{4 b_0^2 \rho ^2+3 \rho_c^2}}\right) \\
   &- 32 b_0^4 \rho ^4 \sqrt{\frac{b_0^2 \rho ^2+\rho_c^2}{4 b_0^2 \rho ^2+3 \rho_c^2}} +6 \rho_c^2 \sqrt{4 b_0^4 \rho ^4+7 b_0^2 \rho ^2 \rho_c^2+3 \rho_c^4} \\
   &+2 b_0^2 \rho ^2
   \left[\rho_c^2 \left(1-24 \sqrt{\frac{b_0^2 \rho ^2+\rho_c^2}{4 b_0^2 \rho ^2+3 \rho_c^2}}\right)+4 \sqrt{4 b_0^4
   \rho ^4+7 b_0^2 \rho ^2 \rho_c^2+3 \rho_c^4}\right] \vast\} ~ .
\end{split}
\end{align}

As one can easily check by taking the leading order terms in the large-$\rho$ expansion, our unknown functions recreate the overlap-zone results \eqref{Final4}-\eqref{Final5}. We note that all our unknown functions do not exhibit any unphysical singularities at $\rho = 0$. They are either explicitly regular or exhibit a divergence that is already present in the seed D3-NS5 solution. In particular, the function $g_{\rho \rho} (\rho)$ diverges at $\rho =0$, but the $\rho \rho$ component of the D3-NS5 solution is already singular at $\rho =0$, since it involves the prefactor $D^{-1/2} H \sim 1/\rho^2$. Consequently, we can express $g_{\rho \rho} (\rho)$ as $g_{\rho \rho} (\rho) = D^{-1/2} H \, g_{\rho} (\rho)$. Then, one can use $g_{\rho} (\rho) = \left( D^{-1/2} H \right)^{-1} g_{\rho \rho} (\rho)$ as the unknown function in the ansatz, instead of $g_{\rho \rho} (\rho)$. We observe that $g_{\rho} (\rho)$ is regular at $\rho = 0$.

\begin{center}
\begin{figure}[H]
\begin{subfigure}{0.45\textwidth}
\includegraphics[width=0.9\linewidth]{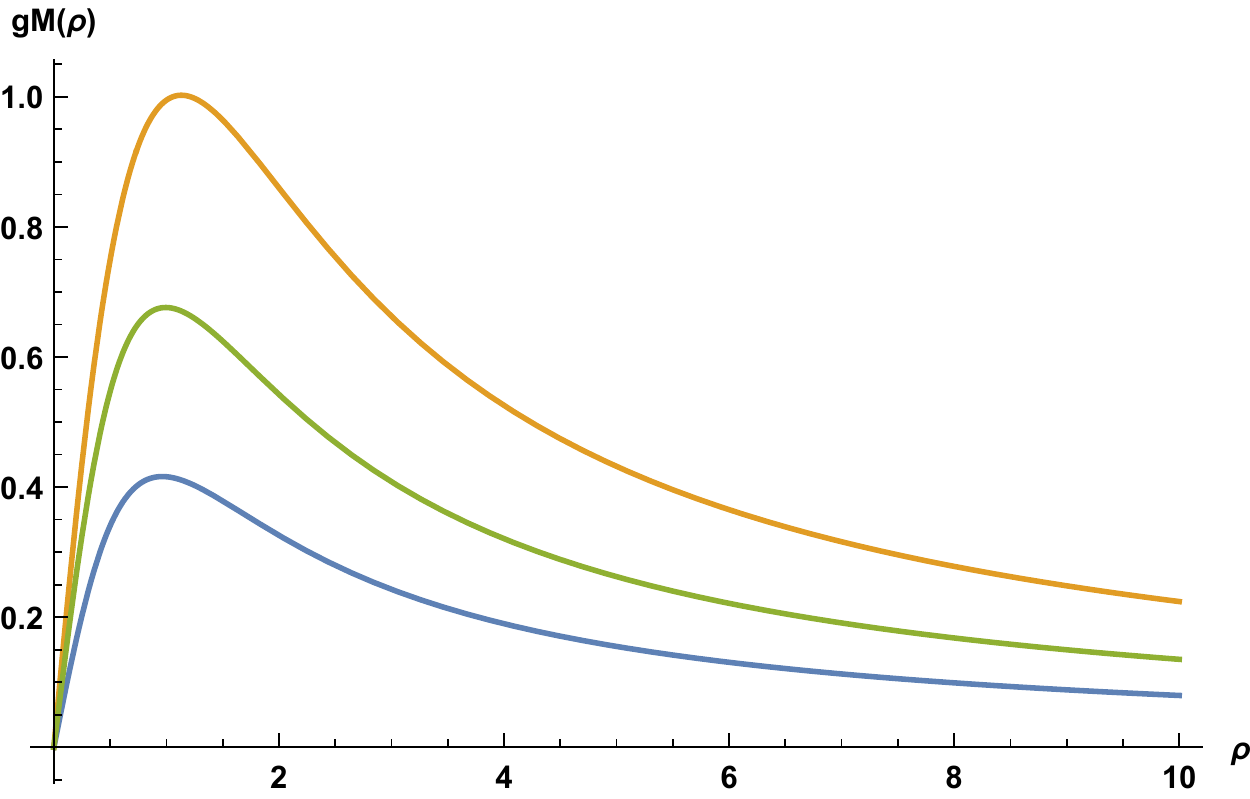} 
\caption{$g_M (\rho)$ for $\rho \in (0, 10)$ }
\end{subfigure}
\begin{subfigure}{0.45\textwidth}
\includegraphics[width=0.9\linewidth]{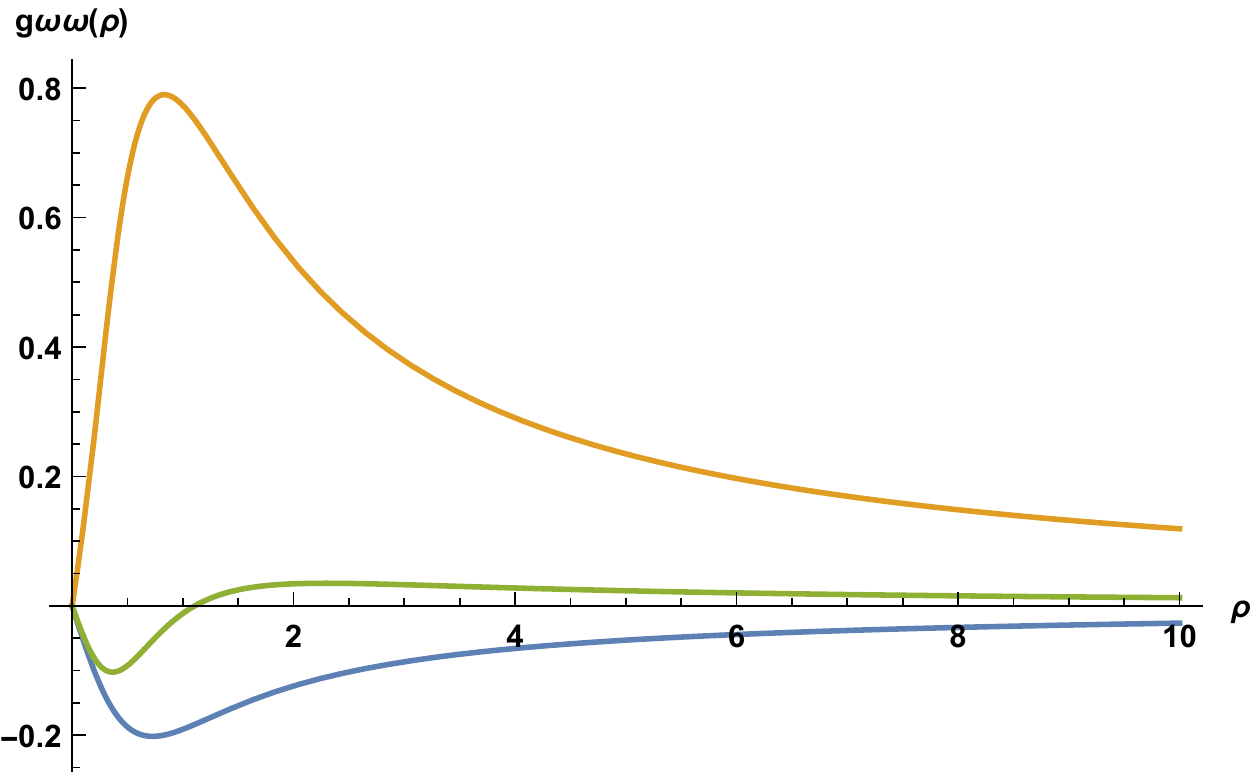}
\caption{$g_{\omega \omega} (\rho)$ for $\rho \in (0, 10)$}
\end{subfigure} \\
\begin{subfigure}{0.45\textwidth}
\includegraphics[width=0.9\linewidth]{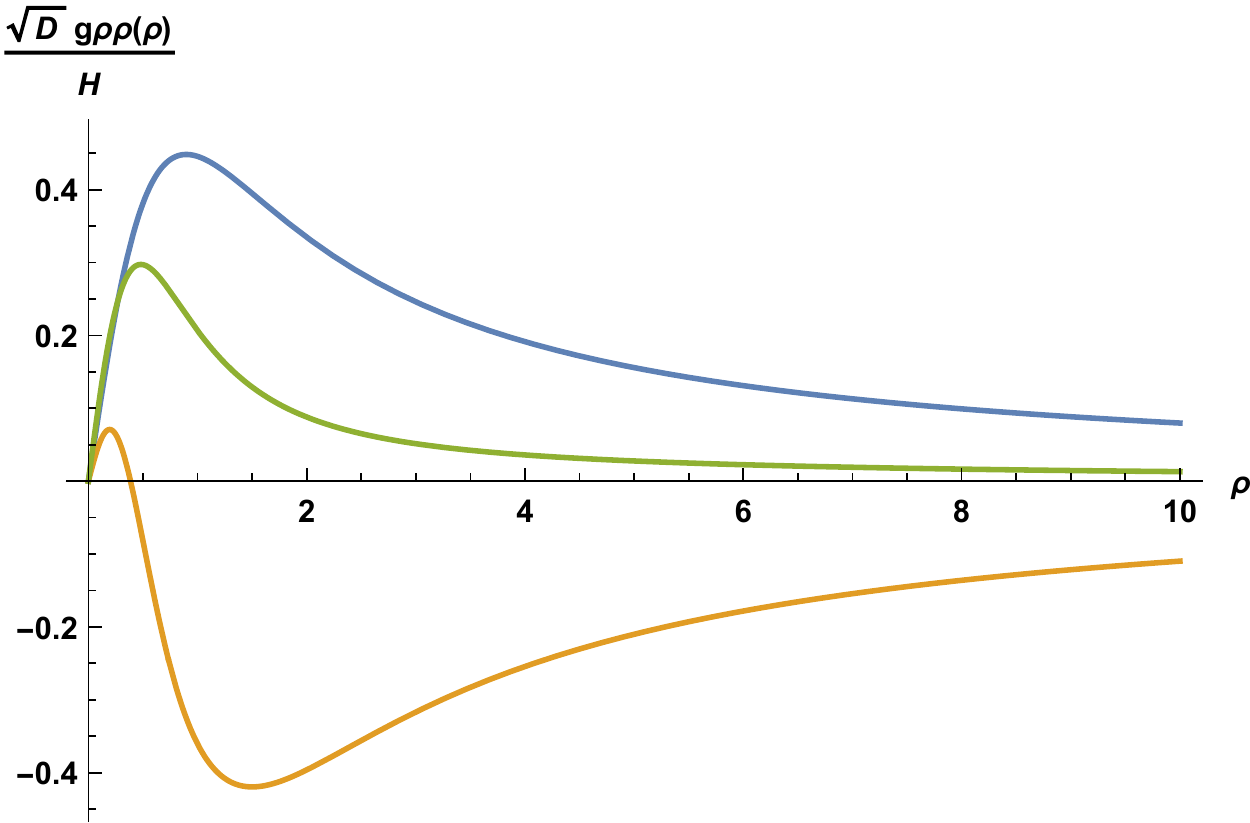} 
\caption{$ \left(D^{-1/2} H \right)^{-1} g_{\rho \rho} (\rho)$  for $\rho \in (0, 10)$ }
\end{subfigure}
\begin{subfigure}{0.45\textwidth}
\includegraphics[width=0.9\linewidth]{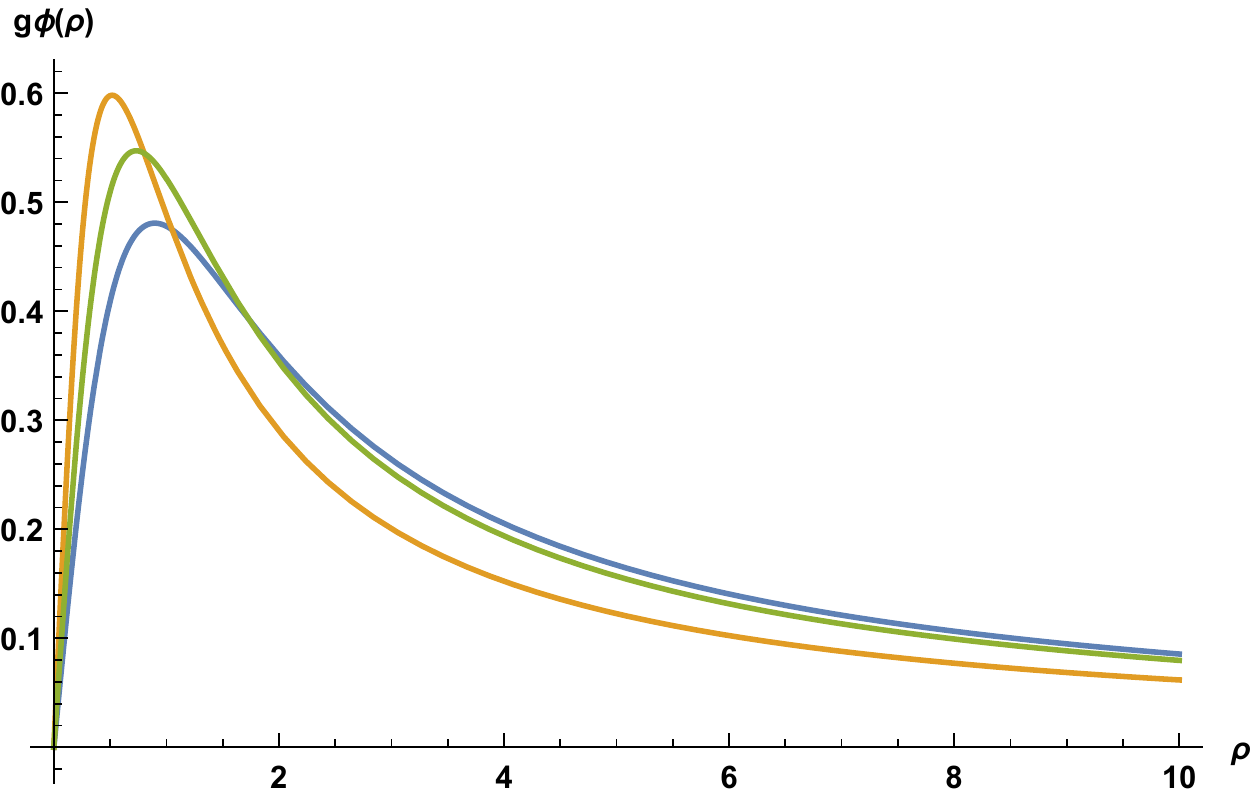}
\caption{$g_{\phi} (\rho)$ for $\rho \in (0, 10)$}
\end{subfigure}
\\
\begin{subfigure}{0.45\textwidth}
\includegraphics[width=0.9\linewidth]{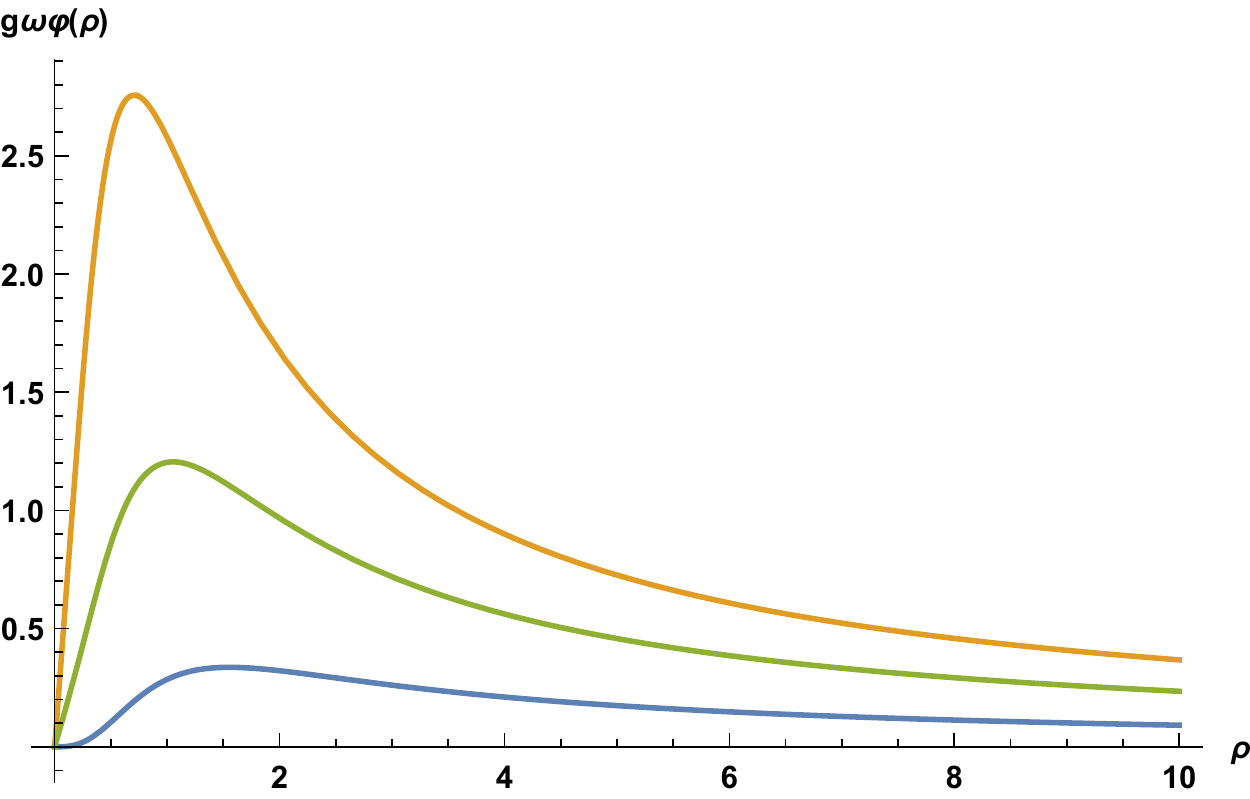} 
\caption{$g_{\omega \varphi} (\rho)$ for $\rho \in (0, 10)$ }
\end{subfigure}
\begin{subfigure}{0.45\textwidth}
\includegraphics[width=0.9\linewidth]{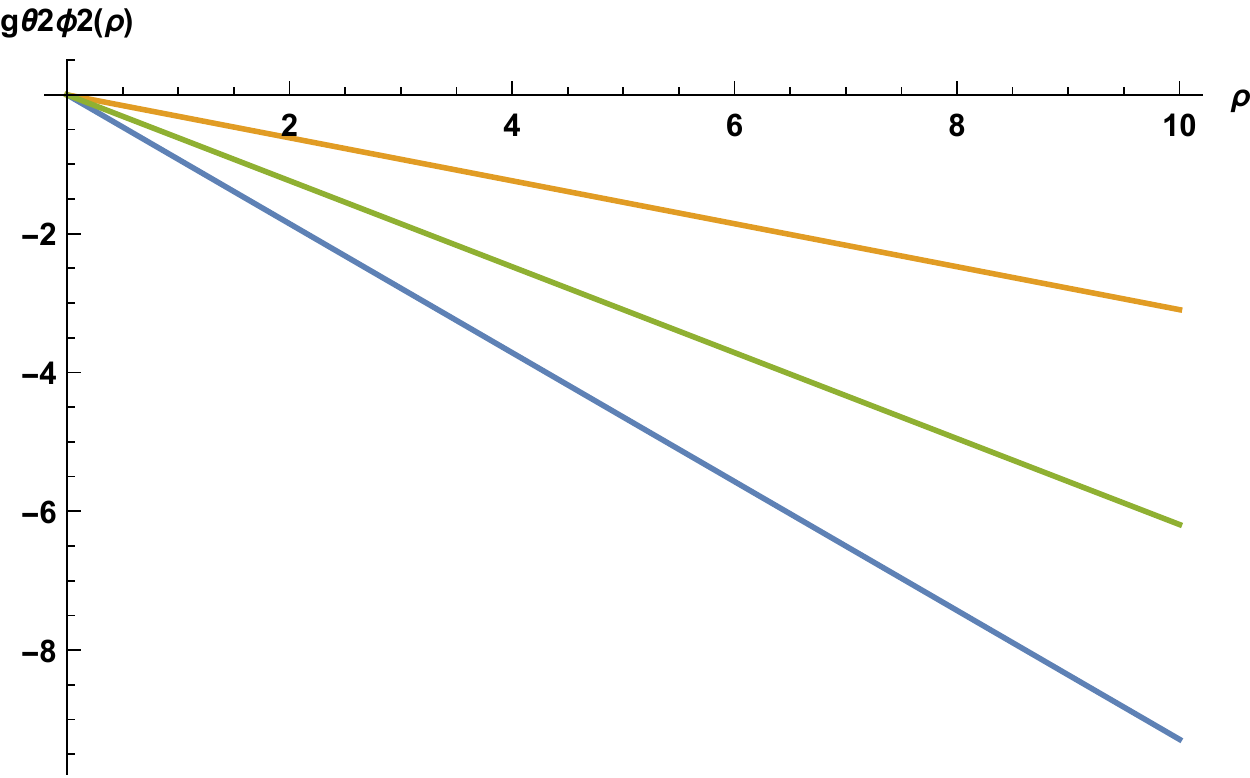}
\caption{$g_{\theta_2 \phi_2} (\rho)$ for $\rho \in (0, 10)$}
\end{subfigure}
\\
\begin{subfigure}{0.45\textwidth}
\includegraphics[width=0.9\linewidth]{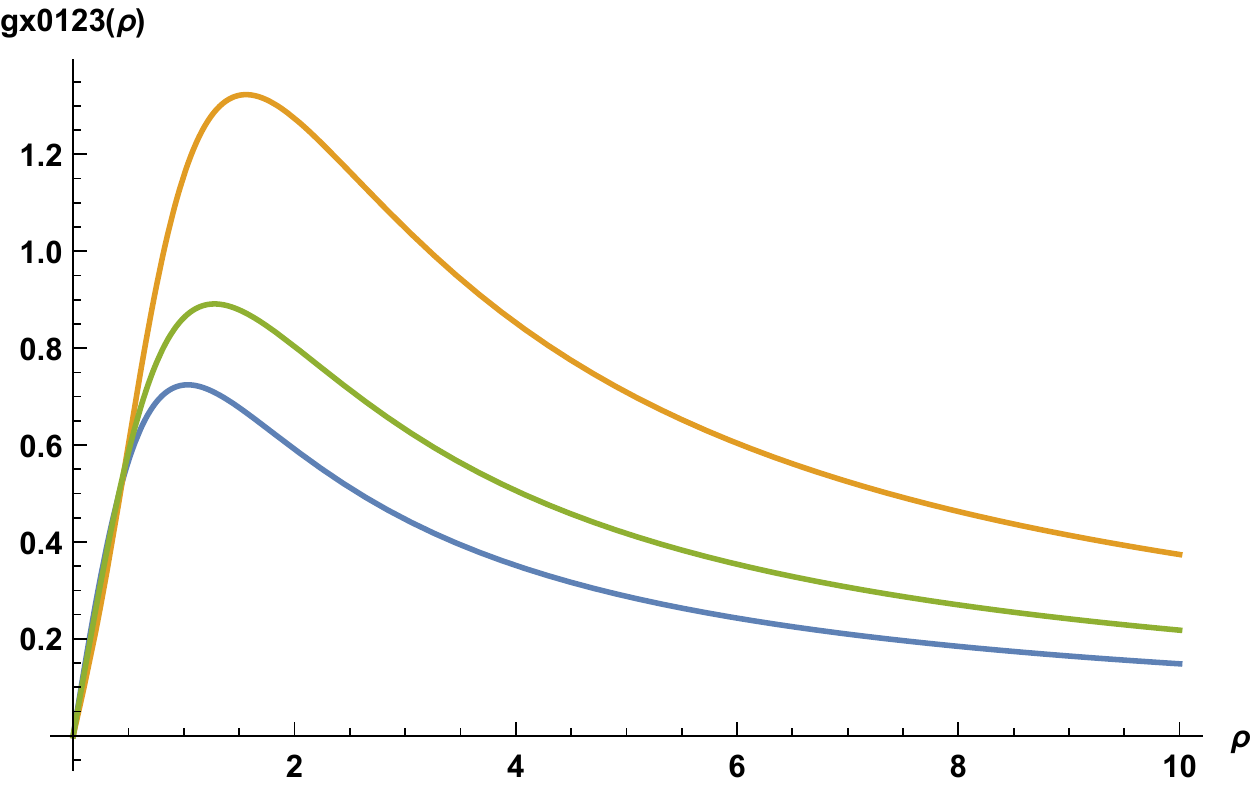} 
\caption{$g_{x0123} (\rho)$ for $\rho \in (0, 10)$ }
\end{subfigure}
\begin{subfigure}{0.45\textwidth}
\includegraphics[width=0.9\linewidth]{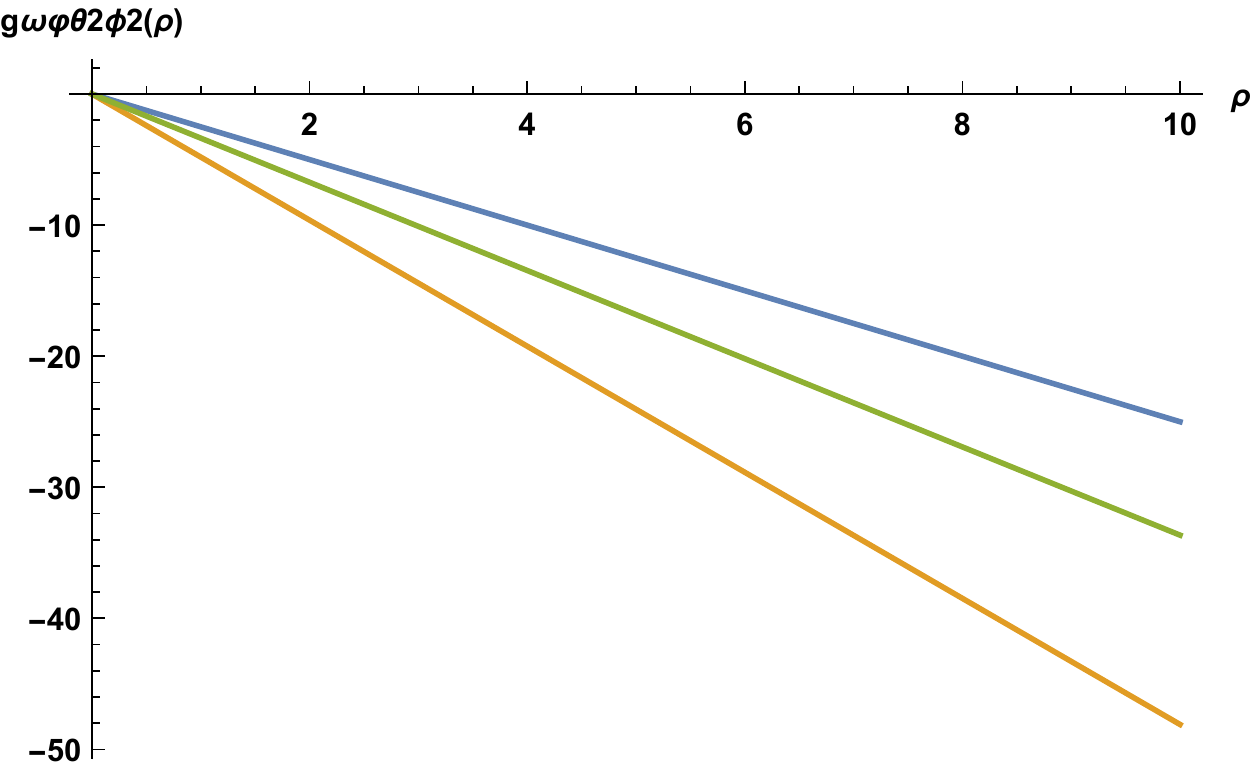}
\caption{$g_{\omega \varphi \theta_2 \phi_2} (\rho)$ for $\rho \in (0, 10)$}
\end{subfigure}
\caption{Plot of the 8 correction functions, namely $g_{M}(\rho)$, $g_{\omega \omega} (\rho)$, $g_{\rho \rho} (\rho)$, $g_{\Phi} (\rho)$, $g_{\omega \varphi} (\rho)$, $g_{\theta_2 \phi_2} (\rho)$, $g_{x0123} (\rho)$, and $g_{\omega \varphi \theta_2 \phi_2} (\rho)$, in the domain $\rho \in (0,10)$ with $\rho_c$ set to 1 for the case $\theta = \pi/6$ (blue curves), $\theta = \pi/4$ (green curves), and $\theta = \pi/3$ (yellow curves).}
\label{FigBig}
\end{figure}
\end{center}

In figure \ref{FigBig}, we plot the 8 correction functions for the case $\theta = \pi/6$ (blue curves), $\theta = \pi/4$ (green curves), and $\theta = \pi/3$ (yellow curves) with $\rho_c$ set to 1 \footnote{The choice $\rho_c = 1$ can be understood as a particular choice of scaling of the radial coordinate $\rho$ that measures distances in units of $\rho_c$.} in the domain $\rho \in (0,10)$. The key take away from these plots is that there is no unexpected divergence, all corrections are indeed regular throughout their domain. In addition, as one can explicitly calculate, the potentially worrisome 3-form field strength $H_3$, $F_3$ and their squares are all regular. 

\section{Conclusions \& Outlook}
\label{conclusion}

\subsection{Conclusions}

In this paper, we constructed a perturbative solution of wrapped five-branes with 3-brane charge in ten-dimensional type IIB supergravity. More specifically, we constructed the leading-order matched asymptotic description of a metastable state of NS5 branes in a warped background with fluxes that is closely related to a specific long-wavelength limit of the Klebanov-Strassler solution. Our solution interpolates between a near-zone wrapped D3-NS5 bound state with negative 3-brane charge and a far-zone modification of the KS background with positive 3-brane charge. The lack of a convenient description of the KS 3-form fluxes in suitable adapted coordinates prevented us from constructing the solution in the background of the exact long-wavelength limit of the KS solution. As a compromise, we were able to find a simplifying modification of the long-wavelength expansion of the KS solution that solves the supergravity equations and allows a straightforward use of adapted coordinates. In this simplified context, we could address concretely the backreaction properties of a metastable anti-brane state that shares many of the characteristic features of the KPV state in the KS background.

We established the following key results:
\begin{itemize}
    \item [$(a)$] From the constraint equations of the full supergravity analysis we recovered the conditions for the existence of the wrapped NS5 metastable state that are already present in \cite{Kachru:2002gs} (from a DBI perspective), or in \cite{Armas:2018rsy} (from the perspective of the blackfold equations). Our result includes an extra factor of $2/\sqrt{3}$ that can be traced back to the modification of the KS asymptotics mentioned above.
    
    \item[$(b)$] We have established the existence of a leading-order perturbative solution that does not exhibit any unphysical singularities. This is the first example of an explicit construction of a MAE for multi-charge black branes in a background with non-trivial geometry and fluxes in supergravity. The construction of a regular solution gives further evidence in favor of the blackfold conjecture reviewed in the Introduction. In addition, the close relation of our construction to the physics of the metastable KPV state is further supportive evidence to the claim that the KPV state is described by a well-behaved solution in the supergravity regime. In this sense, our results are a useful addition to the effective worldvolume analysis of Ref.\ \cite{Armas:2018rsy}.
\end{itemize}

\subsection{Outlook}

As we mentioned in the previous paragraphs, in this paper, we considered the backreaction of the polarised state of anti-3 branes in a perturbative modification of the KS background. It would be very interesting to determine if this modification can be extended to an exact supergravity solution and to understand its properties in supergravity and holography. One can also examine analogous questions in M-theory in the context of the CGLP background \cite{Cvetic:2000db}.

Another interesting direction has to do with the stability properties of the backreacted metastable solutions. Claims regarding the metastability of the KPV state from the DBI analysis in \cite{Kachru:2002gs} and subsequently from the blackfold approach in \cite{Armas:2018rsy} only refer to the balance of force felt by the spherical NS5 branes in the azimuthal angle $\psi$ of the $S^3$ at the tip of the KS throat. They are not complete statements about the stability properties of the state under generic perturbations. In the context of their numerous applications, particularly the cosmological string de Sitter construction of Ref.\  \cite{Kachru:2003aw}, it is important to determine whether these anti-branes are truly metastable under generic perturbations. By studying the backreacted profile of localised anti-D3 branes at the tip of the KS throat, \cite{Bena:2014jaa} observed that a test anti-D3 brane in such a background would feel a repulsive force away from the localised branes. This tachyonic behaviour is then used to argued that the polarised state of anti-D3 branes, i.e. the spherical NS5 branes with dissolved anti-D3 brane charge, has classical instabilities  \cite{Bena:2015kia}. On the other hand, by studying generic deformations of the KPV state using the blackfold approach, \cite{Nguyen:2019syc} observed that classical instabilities are not allowed. The disparity of the two results, in \cite{Bena:2014jaa,Bena:2015kia} and \cite{Nguyen:2019syc}, could be due to the fact that these works are applicable in complementary regimes of validity. In particular, while the technique employed in \cite{Bena:2014jaa} is reliable for studying situations where the spherical polarised state has a small/finite radius and is located near the North pole, i.e.\ when $M$ is small, the results in \cite{Nguyen:2019syc} are reliable when the spherical polarised state has a very large radius and is located away from the North pole, i.e. when $M$ is large. Furthermore, it is worth noting that the setup in \cite{Nguyen:2019syc} may not be able to observe directly\footnote{In certain cases, the leading order blackfold equations can detect the onset of a fragmentation instability. For example, the onset of the Gregory-Laflamme (GL) instability in black strings \cite{Gregory:1993vy, Hovdebo:2006jy} and black rings \cite{Santos:2015iua} whose end point is fragmentation \cite{Lehner:2010pn, Figueras:2015hkb} can be observed by an analogous blackfold stability analysis \cite{Emparan:2009at,Armas:2019iqs}. } potential fragmentation instabilities. Therefore, it could be that the instabilities observed in \cite{Bena:2014jaa} simply went undetected. 

It would be interesting to explore further the stability properties of the perturbative metastable solution derived in this paper. In particular, one can compute the dynamics of a probe anti-brane in such background and see if the tachyonic modes observed in the description of localised anti-D3 branes persist in the description of spherical NS5 branes.\footnote{It is possible that instabilities observed for the polarised sphere in the small/finite radius regime (small $M$) do not persist in the large radius regime (large $M$). An analogous example of this picture can be found in the study of black rings where an instability (elastic mode instability) is observed for fat rings, i.e. rings with small curvature radius, but is not observed for thin rings, i.e. rings with large curvature radius, \cite{Figueras:2015hkb, Armas:2019iqs}.}
It would also be interesting to determine the supergravity solution of the non-extremal metastable states of wrapped D3-NS5 branes by solving the corresponding MAE to leading order. Ref.\ \cite{Armas:2018rsy} demonstrated (using the blackfold equations) that such configurations should exist up to a critical value of the non-extremality parameter. In this context one can also perform a further probe analysis to examine the existence of potential fragmentation instabilities in the presence of thermal effects. Similar exercises can be performed in the context of metastable states of anti-branes in M-theory \cite{Klebanov:2010qs} supplementing the results in \cite{M2M5brane}. 

\acknowledgments
We would like to thank Thomas Van Riet for useful discussions and suggestions. NN is supported by the EPSRC Additional Funding Programme for Mathematical Sciences. 

\begin{appendices}

\section{Notations and conventions}

For the benefit of the reader we collect here some of the basic conventions that we use throughout the paper:

\begin{itemize}

\item The signature is mostly plus $(- + + + ...)$.

\item Greek letters ($\alpha, \beta, \, ...$) are used for the indices of the ten-dimensional spacetime. Latin letters ($a, b, \, ...$) are used for the worldvolume indices. 

\item The Hodge star operator of a $p$-form on an $n$-dimensional manifold is defined as 
\be
(* A)_{\mu_1 ... \mu_{n-p}} = \frac{1}{p!} \epsilon_{\nu_1 ... \nu_p \mu_1 .... \mu_{n-p}} A^{\nu_1 ... \nu_p}
\ee
with $\epsilon_{\nu_1 ... \nu_p \mu_1 .... \mu_{n-p}}$ the Levi-Civita tensor. 

\item The type IIB supergravity action is given by 
\begin{multline}
\label{sugraaction}
\mathcal{I}_{IIB} = \frac{1}{16 \pi G} \int_{\mathcal{M}_{10}} d^{10} x \Bigg\{ \sqrt{- g} \left[ e^{- 2 \phi} \left( R + 4 \p_\mu \phi \p^\mu \phi - \frac{1}{2} | H_3|^2 \right) - \frac{1}{2} |\tilde{F}_1|^2 - \frac{1}{2} | \tilde{F}_3|^2 - \frac{1}{4} |\tilde{F}_5|^2 \right] \\
- \frac{1}{2} C_4 \wedge H_3 \wedge F_3 + \frac{1}{2} B_2 \wedge C_2 \wedge H_3 \wedge F_3
\Bigg\}
\end{multline}
where the gauge invariant field strengths are defined as
\be
\tilde{F}_{q+2} = F_{q+2} - H_3 \wedge C_{q-1} 
\ee
with $F_{q+2} \equiv d C_{q+1}$. There is an extra Chern-Simon term in the action \eqref{sugraaction} compared to the one in most string theory books, e.g. \cite{Polchinski:1998rr}, because we use a different convention of $\tilde{F}_5$. The corresponding type IIB SUGRA equations, that we are solving throught the paper, are:

{\it $\phi$ equation:}
\be
\label{mae6}
4 e^{2 \phi} \nabla^\mu \left( e^{-2 \phi} \p_\mu \phi \right) + R + 4 \p_\mu \phi \p^\mu \phi - \frac{1}{12} (H_3)_{\mu_1 \mu_2 \mu_3} (H_3)^{\mu_1 \mu_2 \mu_3}  = 0 ~ .
\ee
{\it $B_2$ equation:}
\be
d \left( e^{- 2\phi} \star H_3 - \star \tilde{F}_3 \wedge C_0 - \frac{1}{2} \tilde{F}_5 \wedge C_2 + \frac{1}{2} C_4 \wedge F_3 - \frac{1}{2} B_2 \wedge C_2 \wedge F_3 \right) = 0 ~.
\ee
{\it $C_0$ equation:}
\be
d (\star F_1) + H_3 \wedge \star \tilde{F}_3 = 0 ~ .
\ee
{\it $C_2$ equation:}
\be
d \left( \star \tilde{F}_3 \right) + H_3 \wedge \star \tilde{F}_5 = 0 ~ .
\ee
{\it $C_4$ equation:}
\be
d \left( \star \tilde{F}_5 \right) - H_3 \wedge F_3 = 0 ~ .
\ee
{\it $g_{\mu \nu}$ equation:}
\be
\label{mae7}
e^{- 2 \phi} G^{\mu \nu} +  2 \nabla^\mu \left( e^{- 2 \phi} \p^\nu \phi \right) - 2 \nabla_\rho \left( e^{- 2 \phi} \p^\rho \phi \right) g^{\mu \nu} = T^{\mu \nu}_{(\phi)} + T^{\mu \nu}_{(H_3)} + T^{\mu \nu}_{(F_1)} + T^{\mu \nu}_{(F_3)} + T^{\mu \nu}_{(F_5)} 
\ee
with 
\begin{align}
T^{\mu \nu}_{(\phi)} &= 4 e^{- 2 \phi} \left( \p^\mu \phi \p^\nu \phi - \frac{1}{2} g^{\mu \nu} \p_\lambda \phi \p^\lambda \phi \right) ~ , \\
T^{\mu \nu}_{(H_3)} &= \frac{e^{-2 \phi}}{4} \left( H_3^{\mu \mu_1 \mu_2} H^{\ \nu}_{3 \ \mu_1 \mu_2}  - \frac{1}{6} g^{\mu \nu} |H_3|^2\right) ~ , \\
T^{\mu \nu}_{(F_1)} &= \frac{1}{2} \left( F_1^\mu F_1^\nu - \frac{1}{2} g^{\mu\nu} |F_1|^2 \right) ~ , \\
T^{\mu \nu}_{(F_3)} &= \frac{1}{4} \left( \tilde{F}_3^{\mu \mu_1 \mu_2} \tilde{F}^{ \ \nu}_{3 \  \mu_1 \mu_2} - \frac{1}{6} g^{\mu \nu} |\tilde{F}_3|^2 \right) ~ ,  \\
T^{\mu \nu}_{(F_5)} &= \frac{1}{2} \frac{1}{48} \left( \tilde{F}_5^{\mu \mu_1 ... \mu_4} \tilde{F}^{\ \nu}_{5 \ \mu_1 ... \mu_4} - \frac{1}{10} g^{\mu \nu} |\tilde{F}_5|^2 \right) 
\end{align}
where $|F_{p}|^2 = \frac{1}{p!} (F_p)_{\mu_1 ... \mu_p} (F_p)^{\mu_1 ... \mu_p}$.

\end{itemize}

\section{The extremal D3-NS5 bound state}\label{D3NS5Appen}

The type IIB supergravity solution of the extremal D3-NS5 bound state is well known. In this appendix, we remind readers of the pertinent details and, for later convenience, rewrite it in adapted coordinates. We also make some convenient gauge and convention choices.

\paragraph{Extremal D3-NS5 solution in a different convention and coordinates system} From \cite{Nguyen:2019syc} and references therein, we have the supergravity description of the extremal D3-NS5 bound state:
\begin{equation}\label{3}
ds^2 = D^{-1/2} \left( - dt^2 + D \left( (dx^1)^2 + (dx^2)^2 \right)+ \sum_{i=3}^5 (dx^i)^2 \right)+ H D^{-1/2} \left( dr^2 + r^2 d\Omega_3^2 \right)
\end{equation}
with
\begin{align}
& D = \left( \sin^2 \theta H^{-1} + \cos^2 \theta \right) ^{-1} ~ , & H = 1 + \frac{r_h^2}{r^2} 
\end{align}
where $d \Omega_3^2$ is the standard $S^3$ metric $d \Omega_3^2 = d \psi^2 + \sin^2 \psi \left( d \omega^2 + \sin^2 \omega d \varphi^2 \right)$. The dilaton field is given by
\be
e^{2 \phi} = H D^{-1} ~,
\ee
and the gauge fields are given by
\begin{align} 
C_2 &= - \tan \theta (H^{-1} D - 1) \, d x^1 \wedge d x^2 ~ , \\
B_2 &= - 2 \, r_h^2  \cos \theta \ \varphi \sin^2 \psi \sin \omega \, d \psi \wedge d \omega  ~ , \\
\label{4}
C_4 &= (H^{-1} - 1) \sin \theta \, d t \wedge d x^3 \wedge d x^4 \wedge dx^5 + \frac{r^2}{r_h^2  \cos^2 \theta} B_2 \wedge C_2 ~ .
\end{align}

This solution is expressed in the convention where 
\be
\tilde{F}_5 = F_5 + B_2 \wedge F_3
\ee
with $F_{q+2} \equiv d C_{q+1}$. Note that this is \textit{not} the convention we will follow. Our convention will be
\be
\tilde{F}_5 = F_5 - H_3 \wedge C_{2} ~. 
\ee
More details about how the $C_4$ gauge field  in the two conventions are related  will be presented momentarily.

\paragraph{Converting to adapted coordinates} Firstly, let us transform the above description of the D3-NS5 bound state to coordinates adapted to the geometry of our problem. In our case, this involves expressing the $x_1, x_2$ directions in spherical coordinates together with some trivial scaling and renaming. In particular, we do:
\begin{align}
&t \rightarrow b_0 \, x_0 ~ , & &x^3 \rightarrow b_0 \, x^3 ~ , & &x^4 \rightarrow b_0 \, x^4~ , & &x^5 \rightarrow b_0  \, x^5 ~ , & &r \rightarrow b_0  \, \rho ~ ,  \\
&x^1 \rightarrow b_0 \, \omega \cos \varphi ~, & &x^2 \rightarrow b_0 \, \omega \sin \varphi ~, & &\psi \rightarrow \zeta ~, & &\omega \rightarrow \theta_2 ~, & &\varphi \rightarrow \phi_2 
\end{align}
with $b_0$ a constant given by $b_0^2 \approx 0.93266$.\footnote{At this point, the $b_0$ scaling looks arbitrary. Later on when we discuss the KS throat, we will write KS metric with the constant factor $b_0^2  \equiv \frac{2^{2/3} \sqrt{a_0}}{\sqrt[3]{3}}$ in front. The $b_0$ factors here will make things more convenient when doing the MAE.} For convenience, we shall also rename the extremal horizon radius $r_h $ as $\rho_c$. Altogether, the D3-NS5 metric in adapted coordinates is
\begin{multline}
ds^2 = b_0^2\,  D^{-1/2} \Bigg( - (d x^0)^2 + (d x^1)^2 + (d x^2)^2 + (d x^3)^2 + D \, (d \omega^2 + \omega^2 d \varphi^2 ) \\
+ H \Big( d \rho^2 + \rho^2 \left( d \zeta^2 + \sin^2 \zeta \left( d \theta_2 + \sin^2 \theta_2 d \phi_2^2  \right) \right) \Big) \Bigg)
\end{multline}
with
\begin{align}
&H= 1 + \frac{\rho_c^2}{b_0^2 \, \rho^2} ~ , & &D = \left( \sin^2 \theta H^{-1} + \cos^2 \theta \right) ^{-1} ~.
\end{align}
The dilaton field is
\be
e^{2 \phi} = H D^{-1} ~,
\ee
and the gauge fields are  
\begin{align} 
C_2 &= b_0^2 \left( 1 - D H^{-1} \right)\, \tan \theta \, \omega d \omega \wedge d \varphi  ~ , \\
\label{mae4}
B_2 &= - 2 \, \rho_c^2  \cos \theta \, \phi_2 \sin^2 \zeta \sin \theta_2 \, d \zeta \wedge d \theta_2  ~ , \\
\label{mae5}
C_4 &=  b_0^4 (H^{-1} - 1) \sin \theta \, d x^0 \wedge d x^1 \wedge d x^2 \wedge dx^3 + \frac{b_0^2 \, \rho^2}{\rho_c^2  \cos^2 \theta} B_2 \wedge C_2 ~ .
\end{align}

\paragraph{Relating different conventions of $\tilde{F}_5$} Different conventions for $\tilde{F}_5$ translate to different conventions for $C_4$. They are related by redefining $C_4$ as
\be
(C_4)_{new} = (C_4)_{old} + \xi B_2 \wedge C_2
\ee
with a constant $\xi$. In particular, we can move from the $\tilde{F}_5 = F_5 + B_2 \wedge F_3$ convention to the $\tilde{F}_5 = F_5 - H_3 \wedge C_{2}$ convention by a shift of $C_4$ with $\xi = 1$:
\be
(\tilde{F}_5)_{old} = d (C_4)_{old} + B_2 \wedge F_3 = d (C_4)_{new} - H_3 \wedge C_2 = (\tilde{F}_5)_{new}
\ee
It should be clear that the value of $\tilde{F}_5$ is unchanged under the  shifting of conventions. They are really just equivalent ways of collecting the gauge fields.    

In our convention of $\tilde{F}_5 = F_5 - H_3 \wedge C_2$, the $C_4$ gauge field is
\be
(C_4)_{new} = b_0^4 (H^{-1} - 1) \sin \theta \, d x^0 \wedge d x^1 \wedge d x^2 \wedge dx^3 + \left( \frac{b_0^2 \, \rho^2}{\rho_c^2  \cos^2 \theta} + 1 \right) B_2 \wedge C_2 ~ .
\ee

\paragraph{Some gauge choices}
With retrospect, let us make some gauge choices that will make things simpler in subsequent manipulations. Let us add to the $C_2$ gauge field a pure gauge term:
\be
\label{mae39}
\frac{\omega}{\sin^2 \psi_0} \left( \psi_0 - \frac{1}{2} \sin 2 \psi_0  \right) d\omega \wedge d \varphi
\ee
where $\psi_0$ is a constant. Note that, in our convention of $\tilde{F}_5 = d C_4 - H_3 \wedge C_2$, a pure gauge shift in the $C_2$ field also induces a gauge shift in the $C_4$ field. As we require the gauge invariant field strength $\tilde{F}_5$ to be gauge invariant, the shift in $C_4$ is given by
\be
\label{mae40}
\frac{\omega}{\sin^2 \psi_0} \left( \psi_0 - \frac{1}{2} \sin 2 \psi_0  \right) d\omega \wedge d \varphi \wedge B_2 ~ .
\ee

Let us also express the $B_2$ field in a new gauge:
\be
B_2 = - 2 \, \rho_c^2 \cos \theta \sin^2 \zeta \cos \theta_2 \, d \zeta \wedge d \phi_2 ~ .
\ee
We can easily check that this expression of $B_2$ and \eqref{mae4} are equivalent as they give the same $H_3 = d B_2$. 

\paragraph{Extremal D3-NS5 solution in adapted coordinates and the right convention} With all the massaging done, we come to our description of the extremal D3-NS5 solution in adapted coordinates. The metric takes the form
\begin{multline}
ds^2 = b_0^2\,  D^{-1/2} \Bigg( - (d x^0)^2 + (d x^1)^2 + (d x^2)^2 + (d x^3)^2 + D \, (d \omega^2 + \omega^2 d \varphi^2 ) \\
+ H \Big( d \rho^2 + \rho^2 \left( d \zeta^2 + \sin^2 \zeta \left( d \theta_2^2 + \sin^2 \theta_2 d \phi_2^2  \right) \right) \Big) \Bigg)
\end{multline}
with
\begin{align}
&H= 1 + \frac{\rho_c^2}{b_0^2 \, \rho^2} ~ , & &D = \left( \sin^2 \theta H^{-1} + \cos^2 \theta \right) ^{-1} ~.
\end{align}
The dilaton field is 
\be
e^{2 \phi} = H D^{-1} ~,
\ee
and the gauge fields are 
\be
C_2 = \frac{\omega}{\sin^2 \psi_0} \left( \psi_0 - \frac{1}{2} \sin 2 \psi_0  \right) d\omega \wedge d \varphi + b_0^2 \left( 1 - D H^{-1} \right)\, \tan \theta \, \omega d \omega \wedge d \varphi  ~ ,
\ee
\be
B_2 = - 2 \, \rho_c^2 \cos \theta \sin^2 \zeta \cos \theta_2 \, d \zeta \wedge d \phi_2 ~ ,
\ee
\begin{multline}
C_4 = - b_0^4 \, (1 - H^{-1}) \sin \theta \, d x^0 \wedge d x^1 \wedge d x^2 \wedge dx^3 - 2 b_0^2 \, \rho_c^2 \sin \theta \, \sin^2 \zeta \cos \theta_2 \, \omega \, d\zeta \wedge d \phi_2 \wedge d \omega \wedge d \varphi\\
+  \frac{\omega}{\sin^2 \psi_0} \left( \psi_0 - \frac{1}{2} \sin 2 \psi_0  \right) d\omega \wedge d \varphi   \wedge B_2 ~.
\end{multline}

$b_0$ is a constant given by $b_0^2 \approx 0.93266$. In the above description, the coordinates $x^i$, $\omega$, and $\rho$ have dimensions of length, the coordinates $\varphi$, $\zeta$, $\theta_2$, and $\phi_2$ are angles and, are, thus, dimensionless. The only dimension-full parameter is $\rho_c$, which has units of length. The rest, e.g., $\psi_0$ and $\theta$, are dimensionless.

As a check, one can take the above description of the D3-NS5 bound state and plug it into the SUGRA equations \eqref{mae6}-\eqref{mae7} to verify that, indeed, it is a solution.

\section{The Klebanov-Strassler throat}\label{KS}
In this appendix, we discuss the Klebanov-Strassler (KS) throat in 10-dimensional type IIB supergravity. The throat involves a 6-dimensional deformed conifold, a 4-dimensional Minkowski space, and non-trial $F_3, F_5, H_3$ fluxes which in turn induce warping effects on the flat space and the conifold. For further information on the KS throat, we refer readers to the original paper \cite{Klebanov:2000hb} or the review \cite{Herzog:2001xk}.

\subsection{The 6-dimensional deformed conifold}
In this section, we review the parametrisation of the 6-dimensional deformed conifold. The conifold is defined by the equation
\be
\sum^4_{i = 1} z_i^2 = \varepsilon^2
\ee
where $z_i$ are complex numbers and $\varepsilon$ characterises the degree of deformation, i.e. if $\varepsilon = 0$, we have a normal cone. In order to obtain a parametrisation of the space, a clever trick is to define the matrix
\be
W = \left( \begin{matrix}
z_3 + i z_4 & z_1 - i z_2 \\
z_1 + i z_2 & - z_3 + i z_4 
\end{matrix}\right) ~.
\ee
Then, the defining equation becomes
\be
\det W = - \varepsilon^2 ~.
\ee
It is easy to see that 
\be
W_0 = \left( \begin{matrix}
0 & \varepsilon e^{\tau/2} \\
\varepsilon e^{- \tau/2} & 0
\end{matrix}\right)
\ee
is one possible solution. Furthermore, if we define two $SU(2)$ matrices $L_j$ with $j = 1,2$ then 
\be
W = L_1 . W_0 . L_2^{\dagger}
\ee
also satisfies the equation $\det W = - \varepsilon^2$. As argued in \cite{Minasian:1999tt}, the metric of the deformed conifold is then
\be
\label{a}
ds^2_{6} = \mathcal{F} tr \left( d W^{\dagger} d W \right) + \mathcal{G} | tr (W^\dagger d W) |^2
\ee
where
\begin{gather}
\mathcal{F} (\tau) =  \frac{(\sinh 2 \tau  - 2 \tau)^{1/3}}{2 \times 2^{1/3} \times   \varepsilon^{2/3} \sinh \tau} ~,\\
\mathcal{G}(\tau) = \frac{2 - 3 \coth^2 \tau + 3 \tau (\cosh \tau/ \sinh^3 \tau) }{12  \times   \varepsilon^{8/3} (\cosh \tau \sinh \tau - \tau)^{2/3}} ~.
\end{gather}

\paragraph{Euler angles parametrisation of the deformed conifold}
One can parametrise the $L_j$ matrices using Euler angles as
\be
\label{EulerPara}
L_j = \left( \begin{matrix}
\cos \frac{\theta_j}{2} e^{i (\psi_j + \phi_j)/2} & - \sin \frac{\theta_j}{2} e^{-i (\psi_j - \phi_j)/2} \\
\sin \frac{\theta_j}{2} e^{i (\psi_j - \phi_j)/2} & \cos \frac{\theta_j}{2} e^{-i (\psi_j + \phi_j)/2} 
\end{matrix}\right)
\ee
with $(\psi_j, \phi_j)$ range from $0$ to $ 2 \pi$ and $\theta$ ranges from $0$ to $\pi$. Plugging the parametrised expression of $W = L_1 . W_0 . L_2^{\dagger}$ into (\ref{a}) yields the metric of the deformed conifold written in angular coordinates $\psi_j, \theta_j, \phi_j$. As the coordinates $\psi_1$ and $\psi_2$ only appear in $W$ as $\psi_1 + \psi_2$, we can define a new coordinate $\psi = \psi_1 + \psi_2$. The deformed conifold metric in these coordinates is 
\begin{equation}
\label{mae14}
ds_6^2 = \frac{1}{2} \varepsilon^{4/3} K(\tau) \Bigg[ \frac{1}{3 K^3 (\tau) } (d \tau^2 + (g^5)^2 ) + \cosh^2 \left ( \frac{\tau}{2} \right) [(g^3)^2 + (g^4)^2]  + \sinh^2 \left( \frac{\tau}{2} \right) [(g^1)^2 + (g^2)^2  ] 
\Bigg]
\end{equation}
where the function $K(\tau)$ is given by
\be
\label{mae15}
K (\tau) = \frac{(\sinh 2 \tau  - 2 \tau)^{1/3}}{2^{1/3} \sinh \tau}  ~,
\ee
and the $g^i$ forms are given by:
\begin{align}
g^1 &=  \frac{- \sin \theta_1 d \phi_1 - \cos \psi \sin \theta_2 d \phi_2 + \sin \psi d \theta_2}{\sqrt{2}} ~ ,\\
g^2 &=  \frac{d \theta_1 - \sin \psi \sin \theta_2 d \phi_2 - \cos \psi d \theta_2}{\sqrt{2}} ~ ,\\
g^{3} &= \frac{- \sin \theta_1 d \phi_1 + \cos \psi \sin \theta_2 d \phi_2 - \sin \psi d \theta_2}{\sqrt{2}} ~ ,\\
g^{4} &= \frac{d \theta_1 + \sin \psi \sin \theta_2 d \phi_2 + \cos \psi d \theta_2}{\sqrt{2}} ~ , \\
g^5 &= d \psi + \cos \theta_1 d \phi_1 + \cos \theta_2 d \phi_2 ~.
\end{align} 
Let us note further that, as argued in \cite{Minasian:1999tt}, the metric 
\be
ds^2 = \frac{1}{2} (g^5)^2 + (g^4)^2 + (g^3)^2 
\ee
and \be
ds^2 = (g^1)^2 + (g^2)^2
\ee
are the metric of respectively the standard $S^3$ sphere with radius $\sqrt{2}$ and the standard $S^2$ sphere with radius $\sqrt{2}$. By expanding \eqref{mae14} in $\tau$, we note that the $S^3$ described by the combination $ds^2 = \frac{1}{2} (g^5)^2 + (g^4)^2 + (g^3)^2$ is the $S^3$ at the tip and the $S^2$ described by the combination $ds^2 = (g^1)^2 + (g^2)^2$ is the $S^2$ away from the tip.

\paragraph{Spherical parametrisation of the deformed conifold} The Euler parametrisation of the deformed conifold described above is the parametrisation found in the original paper by Klebanov and Strassler \cite{Klebanov:2000hb}. However, for our purposes, it is more convenient to parametrise the conifold using spherical coordinates. 

In the Euler angles parametrisation of $W$, the coordinates $\psi_1$ and $\psi_2$ only appear as $\psi_1 + \psi_2$. So, instead of relabelling the final result, we can parametrise $L_2$ with only two angles $(\theta_2, \phi_2)$:
\be
L_2 = \left( \begin{matrix}
\cos \frac{\theta_2}{2} e^{i \phi_2/2} & - \sin \frac{\theta_2}{2} e^{i \phi_2/2} \\
\sin \frac{\theta_2}{2} e^{- i \phi_2/2} & \cos \frac{\theta_2}{2} e^{-i \phi_2/2} 
\end{matrix}\right) ~.
\ee
Previously, we inserted this $L_2$, and an $L_1$ that is parametrised as \eqref{EulerPara}, into $W = L_1 W_0 L_2^\dagger$ to obtain a parametrised expression for $W$. Now, before plugging in the explicit parametrisation for $L_1$ and $L_2$, we note that $W_0$ can be written as
\be
\label{1000}
W_0 = \varepsilon f(\tau) \sigma_1 + \varepsilon  g(\tau) \sigma_2
\ee
where
\begin{align}
&\sigma_{1} = \left( \begin{matrix}
0 & 1 \\
1 & 0 
\end{matrix}\right) ~, & & \sigma_2 = \left( \begin{matrix}
0 & 1 \\
- 1 & 0 
\end{matrix}\right) ~,
\end{align}
and
\begin{align}
&f(\tau) = \cos(i \, \tau/2)  ~, &g(\tau) = - i \sin (i \,  \tau /2) ~.
\end{align}
Thus, we can write
\begin{align}
W &= L_1 . \big( \varepsilon f(\tau) \sigma_1 + \varepsilon  g(\tau) \sigma_2 \big) . L_2^\dagger \\
&= \varepsilon f(\tau) L + \varepsilon g(\tau) L . \hat{L}   
\end{align}
where $L \equiv L_1 . \sigma_1 . L_2^\dagger$ and $\hat{L} \equiv L_2 . (\sigma_1)^{-1} . \sigma_2 . L_2^\dagger$.

As $L$ is a unitary complex matrix with $\det L = - 1$, we can parametrise $L$ using spherical coordinates as
\be
L = \left( \begin{matrix}
- \sin \psi \sin \omega \cos \varphi + i \sin \psi \sin \omega \sin \varphi & \cos \psi - i \sin \psi \cos \omega \\
\cos \psi + i \sin \psi \cos \omega & \sin \psi \sin \omega \cos \varphi + i \sin \psi \sin \omega \sin \varphi 
\end{matrix}\right) ~.
\ee
On the other hand, the parametrisation of $\hat{L}$ comes directly from the parametrisation of $L_2$:
\be
\hat{L} = \left( \begin{matrix}
- \cos \theta_2 & - e^{i \phi_2} \sin \theta_2 \\
- e^{- i \phi_2} \sin \theta_2 & \cos \theta_2
\end{matrix}\right) ~.
\ee
Inserting the spherically parametrised $W$ into (\ref{a}), we obtain the metric of the deformed conifold in spherical coordinates. As the description of the deformed conifold in spherical coordinates is complicated and not particularly illuminating, we will not write down its full expression here.

\subsection{The KS metric in adapted coordinates}
\label{ks2}
The original description of the KS throat is expressed in Euler angles coordinates: $x^0$, $x^1$, $x^2$, $x^3$,
$\tau$, $\psi$, $\theta_1$, $\phi_1$, $\theta_2$, $\phi_2$. For our purposes, it proves useful to express the KS metric in spherical angles coordinates: $x^0$, $x^1$, $x^2$, $x^3$, $r$, $\psi$, $\omega$, $\varphi$, $\theta_2$, $\phi_2$. These new coordinates are obtained by a trivial rescaling of the original Minkowski and radial coordinates, and a spherical parametrisation of the 6-dimensional deformed conifold. In particular, the coordinates $x^0$, $x^1$, $x^2$, $x^3$, and $\tau$ of the Euler angles coordinates system can be transformed to the coordinates $x^0$, $x^1$, $x^2$, $x^3$, and $r$ of the spherical angles coordinates system via the scaling
\begin{align}
\label{1002}
&x^i \rightarrow \frac{\sqrt{2} \sqrt{a_0} M}{3^{1/6} \times  \epsilon^{2/3}} \ x^i ~, &  &\tau \rightarrow 2 \ r  ~ ,
\end{align}
where $i$ runs from $0$ to $3$.

As discussed in \cite{Klebanov:2000hb}, the KS metric is given by
\be
\label{BB49}
ds_{10}^2 = h^{-1/2} (\tau) \left( - (d x^0)^2 + (d x^1)^2 + (d x^2)^2 + (d x^3)^2 \right) + h^{1/2} (\tau) ds_6^2
\ee
where $ds_6^2$ is the metric of the deformed conifold and the $h(\tau)$ is the warping effect induced by the non-trivial fluxes:
\begin{align}
\label{A44}
h (\tau) &= M^2 \, 2^{2/3} \epsilon^{-8/3} \int_\tau^\infty dx \frac{x \coth x - 1}{\sinh^2 x} (\sinh 2 x - 2 x)^{1/3}\\
&= M^2 2^{2/3} \epsilon^{-8/3} \ ( a_0 + a_2 \tau^2  + a_4 \tau^4)  + \mathcal{O} (\tau^6)
\end{align}
with $a_0 \approx 0.71805$, $a_2 = - (3 \times 6^{1/3})^{-1} $, and $a_4 = (18 \times 6^{1/3})^{-1}$. 

Applying to \eqref{BB49} the coordinate scaling (\ref{1002}) and substituting into $ds_6^2$ the description of the conifold metric in spherical coordinates, we arrive at the expression of the KS metric in our desired coordinates system. Again, as such a metric is complicated, we shall not write it down explicitly here. However, let us note that if we subdue terms of order $r^2$ or higher in all but the $d \Omega^2_2$ directions, we can write the KS metric as:
\begin{multline}
\label{bb51}
g_{\mu \nu} d x^\mu d x^\nu = M b_0^2  \Big ( - (dx^0)^2 + (dx^1)^2 + (dx^2)^2 + (dx^3)^2  + dr^2 \\
+ d \psi ^2 + \sin^2 \psi \left( d \omega^2 + \sin^2 \omega d \varphi^2 \right)  + r^2 (d \theta_2^2 +  \sin^2 \theta_2 d \phi_2^2) \Big) + \mathcal{O} (r^2)
\end{multline}
where $b_0^2= \frac{2^{2/3} \sqrt{a_0}}{3^{1/3}} \approx 0.93266$. This metric is particularly useful when we care only about the physics at the tip of the KS throat, e.g. \cite{Armas:2018rsy}.

\section{The Klebanov-Strassler throat in the large-$M$ limit}
\label{KSLargeM}

In this appendix, we discuss a long-wavelength approximate description for the KS throat in the large-$M$ limit. In particular, we shall derive a modified leading order in $\lambda \equiv 1/\sqrt{M}$ description of the KS throat. The modified description ignores cross-angles components in the 3-form fluxes and the metric but retains the validity of the supergravity equations to appropriate order of $\lambda$. 

\subsection{The KS metric in the large-$M$ limit}
\label{mae24}

As discussed in section \ref{mae18} in the main text, to make sense of the KS solution in the large-$M$ limit, one has to introduce the coordinates transformations:
\begin{align}
&\tilde{x}^i =  \lambda^{-1} x^i ~,& &\tilde{r} =  \lambda^{-1} r ~, \nonumber \\
\label{mae21}
&\tilde{\psi} = \lambda^{-1} \, (\psi - \psi_0)  ~, & &\tilde{\omega} = \lambda^{-1} \sin \psi_0 \, \omega  ~,
\end{align}
where $i$ runs from $0$ to $3$. Applying the above transformations to the description of the KS metric in adapted coordinates, and truncating to appropriate order of $\lambda$, we arrive at an approximate description for the KS metric in the large-$M$ limit. For MAE, it is most convenient to define an adapted radial coordinate that controls the normal distance from our branes. As our D3-NS5 branes source is located at $\tilde{\psi} = 0$, $\tilde{r} = 0$, such a radial coordinate is given by
\be
\tilde{\rho}^2 = \tilde{r}^2 + \tilde{\psi}^2 ~. 
\ee
Note that 
$\tilde{\psi} \in (-\infty, \infty)$.
Thus, instead of having $\tilde{r}, \tilde{\psi}$, we can parametrise the KS throat with $\tilde{\rho}, \tilde{\zeta}$:
\begin{align}
\label{mae22}
&\tilde{r} = \tilde{\rho} \sin \tilde{\zeta}  ~, ~~\tilde{\psi} = \tilde{\rho} \cos \tilde{\zeta} ~ .
\end{align}

The purpose of the tildes 
on the coordinates is to differentiate between the original coordinates and the $\lambda$-scaled ones. However, as we will only work with $\lambda$-scaled coordinates from here onwards, let us subdue all the tildes 
to simplify our syntax. To order $\lambda^2$, the KS metric in adapted coordinates is
\begin{footnotesize}
\begin{multline}
\label{mae28}
ds_{10}^2 = \frac{2^{2/3}
   \left(- (dx^0)^2 + (dx^1)^2 + (dx^2)^2 + (dx^3)^2\right) \left(a_0-2 a_2 \lambda ^2 \rho ^2 \sin ^2(\zeta )\right)}{\sqrt[3]{3} \sqrt{a_0}} \\
   -\frac{2^{2/3} (d\rho )^2 \left(-5 a_0 \lambda ^2 \rho ^2 \sin ^2(\zeta )-10
   a_2 \lambda ^2 \rho ^2 \sin ^2(\zeta )-a_0 \lambda ^2 \rho ^2 \cos (2 \zeta ) \sin ^2(\zeta )+2 a_0 \lambda ^2 \rho ^2 \cos ^2(\zeta ) \cos (2 \theta_2) \sin ^2(\zeta )-5 a_0\right)}{5 \sqrt[3]{3}
   \sqrt{a_0}}\\
   -\frac{2^{2/3} \rho ^2 (d\zeta )^2 \left(2 a_0 \lambda ^2 \rho ^2 \cos (2 \theta_2) \sin ^4(\zeta )-5 a_0 \lambda ^2 \rho ^2 \sin ^2(\zeta )-10 a_2 \lambda ^2 \rho ^2 \sin ^2(\zeta )+a_0 \lambda ^2 \rho ^2 \cos (2 \zeta ) \sin
   ^2(\zeta )-5 a_0\right) }{5 \sqrt[3]{3} \sqrt{a_0}}\\
   -\frac{2^{2/3} \rho ^2 \sin ^2(\zeta ) \left(d(\theta_2)^2+d(\phi_2)^2 \sin ^2(\theta_2)\right) \left((a_0-30 a_2) \lambda ^2 \rho ^2 \sin ^2(\zeta )-15 a_0\right) }{15 \sqrt[3]{3}
   \sqrt{a_0}}\\
   +\frac{2^{2/3}}{5 \sqrt[3]{3} \sqrt{a_0}} (d\omega )^2 \Bigg(-5 a_0 \lambda ^2 \rho ^2 \cos ^2(\zeta )+5 a_0 \lambda ^2 \rho ^2 \cot ^2(\psi_0) \cos ^2(\zeta )+10 a_0 \lambda  \rho  \cot (\psi_0) \cos (\zeta ) \\+7 a_0
   \lambda ^2 \rho ^2 \sin ^2(\zeta )+10 a_2 \lambda ^2 \rho ^2 \sin ^2(\zeta )+a_0 \lambda ^2 \rho ^2 \cos (2 \theta_2) \sin ^2(\zeta )+2 a_0 \lambda ^2 \rho ^2 \cos (2 \varphi -2 \phi_2-2 \psi_0) \sin
   ^2(\zeta ) \sin ^2(\theta_2)+5 a_0\Bigg)\\
   -\frac{2^{2/3}}{15 \sqrt[3]{3} \sqrt{a_0}} \omega ^2 (d\varphi )^2 \Bigg(15 a_0 \lambda ^2 \rho ^2 \cos ^2(\zeta )-15 a_0 \lambda ^2 \rho ^2 \cot ^2(\psi_0) \cos ^2(\zeta)
   - 30 a_0 \lambda  \rho  \cot (\psi_0) \cos (\zeta )+5 a_0 \lambda ^2 \omega ^2 \csc ^2(\psi_0)\\ 
   -21 a_0 \lambda ^2 \rho ^2 \sin ^2(\zeta ) -30 a_2 \lambda ^2 \rho ^2 \sin ^2(\zeta )-3 a_0 \lambda ^2 \rho ^2
   \cos (2 \theta_2) \sin ^2(\zeta )+6 a_0 \lambda ^2 \rho ^2 \cos (2 \varphi -2 \phi_2-2 \psi_0) \sin ^2(\zeta ) \sin ^2(\theta_2)-15 a_0\Bigg)\\
   +d(\zeta ) \Bigg[-\frac{8\ 2^{2/3} \sqrt{a_0} \lambda ^2 \rho ^3 \cos (\zeta ) d(\rho ) \sin ^2(\theta_2) \sin ^3(\zeta )}{5 \sqrt[3]{3}}+\frac{2\ 2^{2/3} \sqrt{a_0}
   \lambda ^2 \rho ^3 \omega  \cos (\varphi -\phi_2) \csc (\psi_0) d(\theta_2) \sin ^3(\zeta )}{\sqrt[3]{3}}\\+\frac{4\ 2^{2/3} \sqrt{a_0} \lambda ^2 \rho ^3 \omega  \cos (\varphi -\phi_2-\psi_0) d(\varphi
   ) \sin (2 \theta_2) \sin ^3(\zeta )}{5 \sqrt[3]{3}}+\frac{2\ 2^{2/3} \sqrt{a_0} \lambda  \rho ^3 d(\phi_2) \sin ^2(\theta_2) (\lambda  \omega  \cot (\theta_2) \csc (\psi_0) \sin (\varphi -\phi_2)-1) \sin ^3(\zeta )}{\sqrt[3]{3}}\\
   +\frac{4\ 2^{2/3} \sqrt{a_0} \lambda ^2 \rho ^3 d(\omega ) \sin (2 \theta_2) \sin (\varphi -\phi_2-\psi_0) \sin ^3(\zeta )}{5 \sqrt[3]{3}}\Bigg] \\
   -\frac{2^{2/3}
   \sqrt{a_0} \lambda  \rho ^2 d(\phi_2) d(\omega ) \sin ^2(\zeta ) \left(2 \lambda  \omega  \cot (\psi_0) \sin ^2(\theta_2)+\lambda  \rho  \cos (\zeta ) \csc (\psi_0) \sin (2 \theta_2) \sin (\varphi
   -\phi_2-2 \psi_0)+\sin (2 \theta_2) \sin (\varphi -\phi_2-\psi_0)\right)}{\sqrt[3]{3}}\\
   +d(\rho ) \Bigg[-\frac{4\ 2^{2/3} \sqrt{a_0} \lambda ^2 \rho ^2 \omega  \cos (\zeta ) \cos (\varphi -\phi_2-\psi_0) d(\varphi ) \sin (2 \theta_2) \sin ^2(\zeta )}{5 \sqrt[3]{3}}\\
   -\frac{2\ 2^{2/3} \sqrt{a_0} \lambda  \rho ^2 \cos (\zeta ) d(\phi_2) \sin (\theta_2) (\lambda  \omega  \cos (\theta_2) \csc
   (\psi_0) \sin (\varphi -\phi_2)-\sin (\theta_2)) \sin ^2(\zeta )}{\sqrt[3]{3}}\\
   -\frac{4\ 2^{2/3} \sqrt{a_0} \lambda ^2 \rho ^2 \cos (\zeta ) d(\omega ) \sin (2 \theta_2) \sin (\varphi -\phi_2-\psi_0) \sin ^2(\zeta )}{5 \sqrt[3]{3}}\Bigg]\\
   +d(\theta_2) \Bigg[-\frac{2\ 2^{2/3} \sqrt{a_0} \lambda ^2 \rho ^2 \omega  \cos (\zeta ) \cos (\varphi -\phi_2) \csc (\psi_0) d(\rho ) \sin ^2(\zeta
   )}{\sqrt[3]{3}}\\
   -\frac{2\ 2^{2/3} \sqrt{a_0} \lambda  \rho ^2 (\cos (\varphi -\phi_2-\psi_0)+\lambda  \rho  \cos (\zeta ) \cos (\varphi -\phi_2-2 \psi_0) \csc (\psi_0)) d(\omega ) \sin ^2(\zeta
   )}{\sqrt[3]{3}}\\
   +\frac{2\ 2^{2/3} \sqrt{a_0} \lambda  \rho ^2 \omega  d(\varphi ) (\lambda  \rho  \cos (\zeta ) \csc (\psi_0) \sin (\varphi -\phi_2-2 \psi_0)+\sin (\varphi -\phi_2-\psi_0)) \sin ^2(\zeta
   )}{\sqrt[3]{3}}\Bigg]\\
   +d(\varphi ) \Bigg[-\frac{4\ 2^{2/3} \sqrt{a_0} \lambda ^2 \rho ^2 \omega  d(\omega ) \sin ^2(\theta_2) \sin (2 \varphi -2 \phi_2-2 \psi_0) \sin ^2(\zeta )}{5 \sqrt[3]{3}}\\
   -\frac{2^{2/3}
   \sqrt{a_0}}{\sqrt[3]{3}} \lambda  \rho ^2 \omega \sin (\psi_0) \sin ^2(\zeta )   d(\phi_2) \Bigg(\lambda  \rho  \cos (\zeta ) \cos (\varphi -\phi_2) \cos (2 \psi_0) \sin (2 \theta_2) \csc ^2(\psi_0)+2 \lambda  \omega  \sin ^2(\theta_2)
   \csc (\psi_0)\\+\cos (\varphi -\phi_2) \cot (\psi_0) \sin (2 \theta_2)+2 \lambda  \rho  \cos (\zeta ) \cot (\psi_0) \sin (2 \theta_2) \sin (\varphi -\phi_2)+\sin (2 \theta_2) \sin
   (\varphi -\phi_2)\Bigg) \Bigg] \\
   + \mathcal{O} \left( \lambda^3 \right) ~.
\end{multline}
\end{footnotesize}
Note that, in the strict limit $\lambda \rightarrow 0$, the KS metric becomes flat:
\begin{multline}
\label{mae23}
ds_{10}^2 = \frac{2^{2/3} \sqrt{a_0}}{\sqrt[3]{3}} \Bigg[  - (d t)^2 + (dx^1)^2 + (dx^2)^2 + (dx^3)^2 + d\omega^2 + \omega^2 d \varphi^2 \\
d \rho^2 + \rho^2 \left( d \zeta^2 + \sin^2 \zeta \left(d \theta_2^2 + \sin^2 \theta_2 d \phi_2^2 \right) \right)
\Bigg] + \mathcal{O}\left( \lambda \right)
~.
\end{multline}

\subsection{The KS fluxes in the large-$M$ limit}

From \cite{Klebanov:2000hb}, we can write the $\tilde{F}_5$ flux in adapted coordinates:
\begin{multline}
\tilde{F}_5 =  - \frac{8 M^3 \sqrt{2} \sqrt{a_0} \, l(r)}{3^{1/6} \epsilon^{10/3} \, h^2(r) \, K^2(r)} \, d t \wedge d x^1 \wedge d x^2 \wedge d x^3 \wedge d r\\
+ 2 M^2 \, l(r) \sin^2 \psi \sin \omega \sin \theta_2 \, d \psi \wedge d \omega \wedge d \varphi \wedge \theta_2 \wedge d \phi_2
\end{multline}
with 
\begin{align}
l(r) &= \frac{2 r \coth 2r -1}{4 \sinh^2 2 r}   \left( \sinh 4r - 4r \right) ~,\\
h (r) &= M^2 \, 2^{2/3} \epsilon^{-8/3} \int_{2r}^\infty dx \frac{x \coth x - 1}{\sinh^2 x} (\sinh 2 x - 2 x)^{1/3} ~, \\
K (r) &= \frac{(\sinh 4r  - 4r)^{1/3}}{2^{1/3} \sinh 2 r} ~.
\end{align}
Applying the coordinate transformations \eqref{mae21} and the radial transformation \eqref{mae22}, and truncating to the appropriate order of $\lambda$, we arrive at an approximate description of the $\tilde{F}_5$ flux in the large-$M$ limit. In particular, the $\tilde{F}_5$ flux to order $\lambda^2$ is 
\begin{multline}
\label{mae30}
\tilde{F}_5 = - \frac{16}{9} \lambda^2 \rho \sin \zeta \Bigg[ \rho \cos \zeta \, d t \wedge d x^1 \wedge d x^2 \wedge d x^3 \wedge d \zeta + 
\sin \zeta \,  d t \wedge d x^1 \wedge d x^2 \wedge d x^3 \wedge d \rho \\
+ \rho^2 \sin^2 \zeta \, \omega \sin \theta_2 \Big( \rho \sin \zeta \, d \omega \wedge d \varphi \wedge d \zeta \wedge d \theta_2  \wedge d \phi_2 - \cos \zeta \,d \omega \wedge d \varphi \wedge d \rho \wedge d \theta_2 \wedge d \phi_2  \Big) 
\Bigg] \\
+ \mathcal{O} \left( \lambda^3 \right) ~.
\end{multline}

As we cannot easily write down the description of the $H_3$ and $F_3$ fluxes in adapted coordinates, it is trickier to obtain their approximate description in the large-$M$ limit. However, let us make a few comments. Firstly, as the KS metric becomes Minkowski in the strict $\lambda \rightarrow 0$ limit \eqref{mae23}, we expect all fluxes to disappear in this limit. Secondly, though we don't have the full description of the $H_3$ and $F_3$ fluxes in the adapted coordinates, we know that they have the components:\footnote{Here, the coordinates are the original coordinates and not the $\lambda$-scaled ones.}
\begin{align}
H_3 &= - 2 M  \tanh^2 r \left( \frac{1}{2} + \frac{r}{\sinh 2 r} \right)  \, \sin \theta_2 \, dr \wedge d \theta_2 \wedge d \phi_2 + ... \\ 
F_3 &= 2 M \, \left( \frac{1}{2} + \frac{r}{\sinh 2 r} \right) \, \sin^2 \psi \sin \omega \ d \psi \wedge d \omega \wedge d \varphi + ...
\end{align}
where the $...$ refers to components of the fluxes that involve cross terms between the angles of the $S^3$ at the tip of the KS throat and the $S^2$ away from the tip. Thus, we know that the leading order in $\lambda$ descriptions of the $H_3$ and $F_3$ fluxes contain:
\begin{align}
\label{mae26}
H_3 &= - 2 \lambda \, r^2 \sin \theta_2 \, d r \wedge d \theta_2 \wedge d \phi_2  + ... ~, \\ 
\label{mae27}
F_3 &= 2 \lambda \,  \omega \, d \psi \wedge d \omega  \wedge d \varphi + ... ~,
\end{align}
where $...$ refers to other terms that can already appear at order $\lambda$. 

As we expect the cross-angles components to contribute to the metric equations (via the energy-momentum tensor) and the dilaton equation (via the square $|H_3|^2$), simply taking the components described in \eqref{mae26}-\eqref{mae27} as our far-zone $H_3$ and $F_3$ fluxes will not yield a valid SUGRA solution. However,  we note that if we scale the \eqref{mae26}-\eqref{mae27} components by a factor of $2/\sqrt{3}$, we can reproduce without the cross-angles components the $H_3$ and $F_3$ energy-momentum tensors to leading order in $\lambda$. By taking these scaled components as the modified $H_3$ and $F_3$ fluxes, we guarantee that their leading order contribution to the metric/dilaton equations is unchanged. Together with the order-$\lambda^2$ KS metric \eqref{mae28} and $\tilde{F}_5$ flux \eqref{mae30}, one can check that we indeed have a valid SUGRA solution to the relevant order in $\lambda$. In particular, the $H_3$ and $F_3$ flux profile:
\begin{align}
\label{mae37}
H_3 &= - \frac{4}{\sqrt{3}} \lambda \, r^2 \sin \theta_2 \, d r \wedge d \theta_2 \wedge d \phi_2  + \mathcal{O} (\lambda^2) ~, \\ 
\label{mae38}
F_3 &= \frac{4}{\sqrt{3}} \lambda \,  \omega \, d \psi \wedge d \omega  \wedge d \varphi + \mathcal{O} (\lambda^2) 
\end{align}
satisfies the dilaton/metric equations to order $\lambda^2$ and the flux equations to order $\lambda$.

\subsection{The KS throat to leading order in the large-$M$ limit}

Let us now collect the modified description of the KS throat to leading order in $\lambda$. 

The leading, first-order contribution to the KS metric can easily be obtained by truncating away the order $\lambda^2$ terms from the expression in \eqref{mae28}. We further observe that truncating away also the cross-angles components of the metric at first order in $\lambda$ doesn't affect the SUGRA equations to the relevant order in $\lambda$. As a result, to simplify the problem, we will implement this extra truncation as an additional feature of the modified KS solution in the far-zone region. In particular, we will assume that the metric of our far-zone asymptotics is:
\begin{multline}
\label{mae33}
ds_{10}^2 =  \frac{2^{2/3} \sqrt{a_0}}{\sqrt[3]{3}}  \Big(  - (d x^0)^2 + (dx^1)^2 + (dx^2)^2 + (dx^3)^2 + \left(1 + 2 \lambda \rho \cos \zeta \cot \psi_0 \right)  (d\omega^2 + \omega^2 d \varphi^2) \\
d \rho^2 + \rho^2 \left( d \zeta^2 + \sin^2 \zeta (d \theta_2^2 + \sin^2 \theta_2 d \phi_2^2) \right)
\Big) ~.
\end{multline}
The fluxes of the far-zone asymptotics are the modified first order description of the KS $H_3$, $F_3$, and $F_5$ fluxes presented in the previous subsection. These can be described via the $B_2$, $C_2$, and $C_4$ gauge fields:
\be
B_2 = - \frac{4}{\sqrt{3}} \lambda \rho^3 \sin^2 \zeta \cos \zeta \cos \theta_2 \, d \zeta \wedge d \phi_2 - \frac{4}{\sqrt{3}} \lambda \rho^2 \sin^3 \zeta \cos \theta_2 \, d \rho \wedge d \phi_2 ~ ,
\ee
\be
C_2 = \frac{4}{\sqrt{3}} \lambda \, \rho  \cos \zeta  \, \omega \, d \omega \wedge d \varphi +  \frac{\omega}{\sin^2 \psi_0} \left( \psi_0 - \frac{1}{2} \sin 2 \psi_0  \right) d\omega \wedge d \varphi ~ ,
\ee
\begin{multline}
\label{mae34}
C_4 = - \frac{4}{\sqrt{3}} \lambda \rho^3 \sin^2 \zeta \cos \zeta \cos \theta_2\, \frac{\omega}{\sin^2 \psi_0} \left( \psi_0 - \frac{1}{2} \sin 2 \psi_0  \right)  \, d \zeta \wedge d \phi_2  \wedge  d\omega \wedge d \varphi \\
- \frac{4}{\sqrt{3}} \lambda \rho^2 \sin^3 \zeta \cos \theta_2 \, \frac{\omega}{\sin^2 \psi_0} \left( \psi_0 - \frac{1}{2} \sin 2 \psi_0  \right) d \rho \wedge d \phi_2 \wedge d\omega \wedge d \varphi ~ .
\end{multline}
Through direct substitution, one can verify that the SUGRA profile \eqref{mae33}-\eqref{mae34} is a valid SUGRA solution to linear order in $\lambda$.

\section{Derivation of the overlap-zone solution}
\label{OverlapKPV}
In this appendix, we derive the leading overlap-zone description of the matched asymptotic solution, i.e., a SUGRA solution to leading order in $\rho_c^2$ and $\lambda$ where $\rho_c$ is the charge radius of the extremal D3-NS5 bound state. 

\subsection{The far-zone description of the D3-NS5 bound state}
The order $\rho_c^2$ far-zone description of the D3-NS5 bound state can be easily obtained from its full description in \eqref{mae36}-\eqref{mae17}. The metric is 
\begin{multline}
g_{\mu \nu} dx^\mu dx^\nu = b_0^2 \left( 1 - \frac{\rho_c^2 \sin^2 \theta}{2 \, b_0^2 \rho^2}  \right) \Big( - (d x^0)^2 + (d x^1)^2 + (d x^2)^2 + (d x^3)^2 \Big) \\
+  b_0^2 \left( 1 + \frac{\rho_c^2 \sin^2 \theta}{2 \, b_0^2 \rho^2} + 2 \, \lambda \rho \cos \zeta \cot \psi_0  \right) \Big(d \omega^2 + \omega^2 d \varphi^2 \Big) \\
+ b_0^2 \left( 1 + \frac{\rho_c^2}{ b_0^2 \rho^2} - \frac{\rho_c^2 \sin^2 \theta}{2 \, b_0^2  \rho^2}  \right) \left( d \rho^2 + \rho^2 \left( d \zeta^2 + \sin^2 \zeta \left( d \theta_2^2 +  \sin^2 \theta_2 \, d \phi_2^2  \right) \right) \right) ~ .
\end{multline}
The dilaton is 
\be
\phi = \frac{\rho_c^2 \cos^2 \theta}{2 \, b_0^2 \rho^2} ~ ,
\ee
and the non-trivial gauge fields are
\be
C_2 = \frac{\omega}{\sin^2 \psi_0} \left( \psi_0 - \frac{1}{2} \sin 2 \psi_0  \right) d\omega \wedge d \varphi + \frac{\rho_c^2}{\rho^2} \sin \theta \cos \theta \, \omega d \omega \wedge d \varphi  ~ ,
\ee
\be
B_2 = - 2 \, \rho_c^2 \cos \theta \sin^2 \zeta \cos \theta_2 \, d \zeta \wedge d \phi_2 ~ ,
\ee
\begin{multline}
C_4 = - b_0^2 \sin \theta \, \frac{\rho_c^2}{\rho^2}  \, d x^0 \wedge d x^1 \wedge d x^2 \wedge dx^3 \\
- \frac{2 \, \rho_c^2 \cos \theta }{\sin^2 \psi_0} \left( \psi_0 - \frac{1}{2} \sin 2 \psi_0  \right) \sin^2 \zeta \, \omega  \cos \theta_2 \,  d \omega \wedge d \varphi \wedge d \zeta \wedge d \phi_2
\\
- 2 \, b_0^2 \,  \rho_c^2 \sin \theta \sin^2 \zeta  \, \omega  \cos \theta_2 \, d \omega \wedge d \varphi  \wedge d \zeta \wedge d \phi_2 ~ .
\end{multline}

\subsection{Overlap-zone $C_2$}
The overlap-zone $C_2$ gauge field can be decomposed as 
\be
C_2 = (C_{2})_0 + (C_{2})_\lambda + (C_{2})_{\rho_c^2} +(C_{2})_{\lambda \rho_c^2}
\ee
where
\begin{align}
(C_{2})_0 &= \frac{\omega}{\sin^2 \psi_0} \left( \psi_0 - \frac{1}{2} \sin 2 \psi_0  \right) d\omega \wedge d \varphi ~ ,\\
(C_{2})_\lambda &=  2 \lambda \, \rho  \cos \zeta  \, \omega d \omega \wedge d \varphi ~ ,\\
(C_{2})_{\rho_c^2} &=  \frac{\rho_c^2}{\rho^2} \sin \theta \cos \theta \, \omega d \omega \wedge d \varphi  ~ ,  \\
\label{C2over}
(C_{2})_{\lambda \rho_c^2} &= \lambda \rho_c^2 \ \Big( \text{Correction terms} \Big) ~ .
\end{align}
By requiring that the overlap-zone $C_2$ satisfies the relevant SUGRA equation to leading order in $\lambda$ and $\rho_c^2$, we can obtain the $\lambda \rho_c^2$ correction terms in the parenthesis of the RHS of Eq.\ \eqref{C2over}. 

The $C_2$ SUGRA equation is given by
\be
d \left( \star \tilde{F}_3 \right) + H_3 \wedge \star \tilde{F}_5 = 0 ~ .
\ee
As $C_0 = 0$, we have:
\be
d \Big( \star F_3 \Big) + H_3 \wedge \tilde{F}_5 = 0 ~.
\ee

To order $\lambda \rho_c^2$, we can write $\star F_3$ as
\begin{align}
\star F_3 &= \star \Big( (F_3)_0 +  (F_3)_{\rho_c^2}  + (F_3)_\lambda  + (F_3)_{\lambda \rho_c^2} \Big)\\
&= \star_{\lambda \rho_c^2} (F_3)_0 +  \star_{\lambda} (F_3)_{\rho_c^2} +  \star_{\rho_c^2} (F_3)_\lambda +  \star_{\text{flat}} (F_3)_{\lambda \rho_c^2}
\end{align}
where the subscript denotes the order we care about. For example, $\star_\lambda$ means the Hodge dual is performed with respect to the $\lambda$ corrected (KS) metric. With $(F_3)_0 =  d (C_2)_0 = 0$, $(H_3)_0 = 0$, $(\tilde{F}_5)_0 = 0$, and $(\tilde{F}_5)_\lambda = 0$, the $C_2$ SUGRA equation to leading order in $\lambda$ and $\rho_c^2$ can be written as 
\be
\label{1}
d \Big( \star_{\lambda} (F_3)_{\rho_c^2} +  \star_{\rho_c^2} (F_3)_\lambda +  \star_{\text{flat}} (F_3)_{\lambda \rho_c^2} \Big) + (H_3)_{\lambda} \wedge (\tilde{F}_5)_{\rho_c^2} = 0
\ee
with
\be
(H_3)_\lambda = - \frac{4}{\sqrt{3}} \lambda \, \rho^2 \sin^2 \zeta \sin \theta_2 \Big( \rho \cos \zeta \, d \zeta \wedge d \theta_2 \wedge d \phi_2 + \sin \zeta \, d \rho \wedge d \theta_2 \wedge d \phi_2 \Big)  ~,
\ee
\begin{multline}
(\tilde{F}_5)_{\rho_c^2} = - 2 \, b_0^2 \, \rho_c^2 \sin \theta \ \omega  \sin^2 \zeta  \sin \theta_2 \, d \zeta \wedge d \theta_2 \wedge d \varphi \wedge d \phi_2 \wedge d \omega \\
+ \frac{2 \, b_0^2 \, \rho_c^2 \sin \theta}{\rho^3} d x^0 \wedge d x^1 \wedge d x^2 \wedge dx^3 \wedge d \rho ~,
\end{multline}
\begin{multline}
d \Big( \star_\lambda (F_3)_{\rho_c^2} \Big) = 4 b_0^4 \, \lambda \rho_c^2 \sin^2 \zeta \cos \zeta  \sin \theta  \cos \theta \sin \theta_2 \cot \psi_0 \, dx^0 \wedge dx^1 \wedge dx^2 \wedge \\dx^3 \wedge d \zeta \wedge d \theta_2 \wedge d \rho \wedge d \phi_2 ~,
\end{multline}
and
\begin{multline}
d \Big( \star_{\rho_c^2} (F_3)_\lambda \Big) = 
- \frac{8 \, b_0^4}{\sqrt{3}} \, \lambda  \rho_c^2  \sin^2 \zeta  \cos \zeta  \cos (2 \theta)  \sin \theta_2 \, dx^0 \wedge dx^1 \wedge dx^2 \wedge \\
 dx^3 \wedge d \zeta \wedge d \theta_2 \wedge d \rho \wedge d \phi_2 ~.
\end{multline}

Bringing all known expressions to the RHS, the $C_2$ SUGRA equation becomes
\begin{align}
d \Big( \star_{\text{flat}} \Big( d \, (C)_{\lambda \rho_c^2} \Big) \Big) &= - d \Big( \star_{\lambda} (F_3)_{\rho_c^2} \Big) -  d \Big( \star_{\rho_c^2} (F_3)_\lambda  \Big) - (H_3)_{\lambda} \wedge (\tilde{F}_5)_{\rho_c^2} \\
\label{2}
&= - \mathcal{A} \, \lambda  \rho_c^2 \, \sin^2 \zeta \cos \zeta \sin \theta_2 \, d x^0 \wedge d x^1 \wedge dx^2 \wedge dx^3 \wedge d \zeta \wedge d \theta_2 \wedge d \rho \wedge d \phi_2
\end{align}
with the constant $\mathcal{A}$ given by
\be
\mathcal{A} = \frac{4}{3} b_0^2 \left( - 2 \sqrt{3}  + \sin \theta \left( 3 \, b_0^2 \cos \theta \cot \psi_0 + 2 \sqrt{3} \left(1 + 2 \sin \theta \right) \right) \right) ~.
\ee
Equation (\ref{2}) can be solved easily. In particular, plugging in the ansatz
\be
(C_2)_{\lambda \rho_c^2} = \lambda \rho_c^2 \, c_{\omega \varphi} (\rho, \zeta) \, \omega d \omega \wedge d \varphi 
\ee
yields the differential equation
\be
\mathcal{A} \sin \zeta  \cos \zeta + b_0^4 \, \rho \left(2 \cos \zeta c_{\omega \varphi}^{(0,1)} + \sin \zeta \left( c_{\omega \varphi}^{(0,2)} + \rho \left( 3 c_{\omega \varphi}^{(1,0)} + \rho c_{\omega \varphi}^{(2,0)} \right) \right) \right) = 0 ~.
\ee
Solving this differential equation yields $c_{\omega \varphi}$: 
\be
c_{\omega \varphi}  (\rho, \zeta) =  \frac{\mathcal{A}}{4 \, b_0^4} \, \frac{ \zeta + 1/2 \, \sin2 \zeta }{\rho \sin \zeta } + \frac{ \mathtt{c}_1 \zeta + \mathtt{c}_2 }{\rho \sin \zeta} 
\ee
where $\mathtt{c}_1$ and $\mathtt{c}_2$ are integration constants.

\paragraph{Regularity conditions \& Integration constants}
The regularity conditions of the $C_2$ gauge field at $\zeta = 0$ and $\zeta = \pi$ fix the integration constants $\mathtt{c}_1$ and $\mathtt{c}_2$. In particular, we obtain:
\begin{align}
&\mathtt{c}_1 = - \frac{\mathcal{A}}{4 \, b_0^4} ~,  & &\mathtt{c}_2 = 0 ~ .
\end{align}
This yields the final expression 
\be
c_{\omega \varphi} (\rho, \zeta) = \frac{\mathcal{A} \cos \zeta}{4 \, b_0^4 \, \rho} ~.
\ee

\subsection{Overlap-zone $B_2$} 
Let us consider the overlap-zone $B_2$ SUGRA equation:
\be
d \left( e^{- 2\phi} \star H_3 - \star \tilde{F}_3 \wedge C_0 - \frac{1}{2} \tilde{F}_5 \wedge C_2 + \frac{1}{2} C_4 \wedge F_3 - \frac{1}{2} B_2 \wedge C_2 \wedge F_3 \right) = 0 ~ .
\ee
As $C_0$ vanishes, the equation becomes
\be
d \left( E_\phi \star H_3  - \frac{1}{2} \left( d C_4 - H_3 \wedge C_2 \right) \wedge C_2 + \frac{1}{2} C_4 \wedge F_3 - \frac{1}{2} B_2 \wedge C_2 \wedge F_3 \right) = 0 
\ee
where, for convenience, we have defined $E_\phi \equiv e^{- 2\phi}$. We can decompose $E_{\phi}$ as
\be
E_{\phi} = 1 + (E_{\phi})_{\rho_c^2} + (E_{\phi})_{\lambda \rho_c^2}
\ee
and $B_2$ as
\be
B_2 = (B_{2})_\lambda + (B_{2})_{\rho_c^2} + (B_{2})_{\lambda \rho_c^2} ~.
\ee

Analogous to the $C_2$ case, bringing all the known expressions to the RHS, we find that the $B_2$ SUGRA equation to leading order in $\lambda$ and $\rho_c^2$ becomes:
\be
\label{mae41}
d \, \left( \star_{\text{flat}} \Big( d \, \Big( (B_2)_{\lambda \rho_c^2} \Big) \Big) \right) = - \mathcal{B} \, \frac{\lambda \rho_c^2 \, \omega \sin \zeta}{\rho^2} \,  d x^0 \wedge ...  \wedge d x^3 \wedge d \omega \wedge d \varphi \wedge d \rho \wedge d \zeta   
\ee
with the constant $\mathcal{B}$ given by
\be
\mathcal{B} = 4 b_0^4 \cos \theta \cot \psi_0 - \frac{8 b_0^2 \left( 1 + \cos^2 \theta + \sin \theta \right)}{\sqrt{3}} ~.
\ee

Using the ansatz 
\be
(B_2)_{\lambda \rho_c^2} = \lambda \rho_c^2 \, b_{\theta_2 \phi_2} (\rho,\zeta) \sin \theta_2 d \theta_2 \wedge d \phi_2  ~,
\ee
we can easily solve equation \eqref{mae41} for the $b_{\theta_2 \phi_2} (\rho,\zeta)$ unknown function:
\be
b_{\theta_2 \phi_2} (\rho,\zeta) = \rho \left[ \mathtt{c}_3 \cos \zeta + \mathtt{c}_4 \left( \left(\zeta - \frac{\pi}{2} \right) \cos \zeta - \sin \zeta \right) + \frac{\hat{\mathcal{B}} \sin^3 \zeta}{4 \, b_0^4} \right]  
\ee
where $\mathtt{c}_3$ and $\mathtt{c}_4$ are integration constants. 

\paragraph{Regularity conditions \& Integration constants}
Interestingly, the integration constants $\mathtt{c}_3$ and $\mathtt{c}_4$ of the overlap-zone $B_2$ gauge field are fixed by regularity of the next order (i.e. order $\lambda \rho_c^4$) overlap-zone solution. In particular, we can solve for the next order overlap-zone solution while still leaving the integration constants $\mathtt{c}_3$ and $\mathtt{c}_4$ in the description of $(B_2)_{\lambda \rho_c^2}$. We then see that regularity conditions for the overlap-zone profile at order $\lambda \rho_c^4$ forces the integration constants $\mathtt{c}_3$ and $\mathtt{c}_4$ to vanish. Thus, we have the final expression for the $b_{\theta_2 \phi_2} (\rho,\zeta)$ unknown function:
\be
b_{\theta_2 \phi_2} (\rho,\zeta) =  \frac{\hat{\mathcal{B}} \, \rho \sin^3 \zeta}{4 \, b_0^4}  ~.
\ee

\subsection{Overlap-zone $C_4$}
The $C_4$ SUGRA equation to leading order in $\lambda$ and $\rho_c^2$ is trivial. Instead, the determination of the overlap-zone $C_4$ gauge field is obtained through the duality condition $\tilde{F}_5 = \star \tilde{F}_5$. Note 
that, as $\tilde{F}_5 = d C_4 - H_3 \wedge C_2$ and $(C_2)_0 \neq 0$, the duality condition $\tilde{F}_5 = \star \tilde{F}_5$ contains both $(B_2)_{\lambda \rho_c^2}$ and $(C_4)_{\lambda \rho_c^2}$ correction terms. However, we can write $(C_4)_{\lambda \rho_c^2} = (\hat{C}_4)_{\lambda \rho_c^2} + (B_2)_{\lambda \rho_c^2} \wedge (C_2)_0$ and try to find $(\hat{C}_4)_{\lambda \rho_c^2}$ instead of $(C_4)_{\lambda \rho_c^2}$. By doing this, we can automatically get rid of the $(B_2)_{\lambda \rho_c^2}$ corrections in the duality equation, making our computations simpler.  

Analogous to the previous analysis, by explicitly decomposing the fluxes and Hodge star operations into $\lambda$, $\rho_c^2$, and $\lambda \rho_c^2$ terms, we can express the duality condition to leading order $\lambda$ and $\rho_c^2$ as a set of differential equations. By choosing a suitable ansatz for the $C_4$ corrections terms, we can solve these equations to obtain $(C_4)_{\lambda \rho_c^2}$:
\begin{multline}
(C_4)_{\lambda \rho_c^2} =  \lambda \rho_c^2 \,  \frac{\mathcal{C}_1 \cos \zeta}{\rho}  d x^0 \wedge d x^1 \wedge d x^2 \wedge d x^3 + \lambda \rho_c^2 \, \mathcal{C}_2 \, \rho \sin^3 \zeta  \, \omega \sin \theta_2 \, d \omega \wedge d \varphi \wedge d \theta_2 \wedge d \phi_2 \\
+ (B_2)_{\lambda \rho_c^2} \wedge (C_2)_{0} 
\end{multline}
where
\begin{align}
\mathcal{C}_1 &= \frac{2}{\sqrt{3}} \cos \theta \Big(1 - \sin \theta \Big) + b_0^2 \cot \psi_0 \sin \theta ~, \\
\mathcal{C}_2 &= - \frac{2}{\sqrt{3}} \cos \theta \Big(  1 +  \sin \theta \Big) - b_0^2 \cot \psi_0 \sin \theta  ~.
\end{align}
Of course, the general solution obtained from solving the duality condition contains some integration constants. However, these integration constants are fixed by requiring that the $C_4$ gauge field is regular at $\zeta = 0$ and $\zeta = \pi$.

\subsection{Overlap-zone dilaton and metric}
Contrary to our approach for the fluxes, where it is more convenient to postpone the specification of an ansatz, for the analysis of the metric and dilaton, it is easier to set an ansatz right away. Based on the symmetries of our overlap-zone profile, a simple choice for the ansatz is as follows:
\be
\label{mae42}
\phi = \frac
{\rho_c^2 \cos^2 \theta}{ 2 \, b_0^2 \rho^2} + \lambda \rho_c^2 \, \Phi (\rho, \zeta) ~,
\ee
and
\begin{multline}
\label{mae43}
g_{\mu \nu} dx^\mu dx^\nu = b_0^2 \left( 1 - \frac{\rho_c^2 \sin^2 \theta}{2 \, b_0^2 \rho^2}  \right) \Big( - (d x^0)^2 + (d x^1)^2 + (d x^2)^2 + (d x^3)^2 \Big) \\
+  b_0^2 \left( 1 + \frac{\rho_c^2 \sin^2 \theta}{2 \, b_0^2 \rho^2} + 2 \, \lambda \rho \cos \zeta \cot \psi_0  \right) \Big(d \omega^2 + \omega^2 d \varphi^2 \Big) \\
+ b_0^2 \left( 1 + \frac{\rho_c^2}{ b_0^2 \rho^2} - \frac{\rho_c^2 \sin^2 \theta}{2 \, b_0^2  \rho^2}  \right) \left( d \rho^2 + \rho^2 \left( d \zeta^2 + \sin^2 \zeta \left( d \theta_2^2 +  \sin^2 \theta_2 \, d \phi_2^2  \right) \right) \right) \\
+ \lambda \rho_c^2 \Bigg[ f_M(\rho, \zeta) \, \Big( - (d x^0)^2 + (d x^1)^2 + (d x^2)^2 + (d x^3)^2 \Big) + f_{\omega \omega} (\rho, \zeta) \, \Big( d \omega^2 + \omega^2 d \varphi^2 \Big) \\
+ f_{\rho \rho} (\rho, \zeta) \, \Big( d \rho^2 + \rho^2 \left( d \zeta^2 + \sin^2 \zeta \left( d \theta_2^2 + \sin^2 \theta_2 d \phi_2^2  \right) \right) \Big) \Bigg] ~. 
\end{multline}
Here, $\Phi (\rho, \zeta)$, $f_{M} (\rho, \zeta)$, $f_{\omega \omega} (\rho, \zeta)$, and $f_{\rho \rho} (\rho, \zeta)$ are unknown correction functions. Our task is to find an expression for these functions such that the dilaton and metric satisfy the relevant SUGRA equations to leading order in $\lambda$ and $\rho_c^2$. Plugging \eqref{mae42}-\eqref{mae43} into the dilaton and metric SUGRA equations \eqref{mae6} and \eqref{mae7}, we obtain a set of coupled partial differential equations of the unknown functions $\Phi$, $f_{M}$, $f_{\omega \omega}$, and $f_{\rho \rho}$. By noting that the $\rho$ dependence of these unknown functions can be determined purely from dimensional analysis, we can solve these differential equations and obtain regular expressions for the unknown functions:
\begin{align}
\Phi (\rho, \zeta) &= \frac{\cos^2 \theta \cot \psi_0}{2\, b_0^2} \, \frac{\cos \zeta}{\rho} ~ , \\
f_M (\rho, \zeta) &=  \left( \frac{1}{2} \cot \psi_0 \sin^2 \theta + \frac{\sin 2 \theta}{\sqrt{3} \, b_0^2}  \right) \frac{\cos \zeta}{\rho} ~ , \\
f_{\omega \omega} (\rho, \zeta) &=  \left( \frac{1}{2} \cot \psi_0 \sin^2 \theta - \frac{\sin 2 \theta}{\sqrt{3} \, b_0^2}  \right) \frac{\cos \zeta}{\rho} ~, \\
f_{\rho \rho} (\rho, \zeta) &= \left( \frac{1}{12} \Big( 1 + 7 \cos 2 \theta  \Big) \cot \psi_0 + \frac{2 \cos \theta \left( 1 - 2 \sin \theta \right)}{3 \sqrt{3} \, b_0^2}\right) \frac{\cos \zeta}{\rho} ~ .
\end{align}

We note that the overlap-zone dilaton and metric SUGRA equations enforce a relationship between the constants $\psi_0$ and $\theta$:
\be
\label{mae44}
\cot \psi_0 = \frac{2}{\sqrt{3}} \left( \frac{1}{b_0^2} \sec \theta + \frac{1}{b_0^2} \tan \theta \right) ~.
\ee
This equation is independent of the correction terms.

\subsection{Summary of the overlap-zone solution}
We can now collect the overlap-zone solution. The metric is 
\begin{multline}
g_{\mu \nu} dx^\mu dx^\nu = b_0^2 \left( 1 - \frac{\rho_c^2 \sin^2 \theta}{2 \, b_0^2 \rho^2}  \right) \Big( - (d x^0)^2 + (d x^1)^2 + (d x^2)^2 + (d x^3)^2 \Big) \\
+  b_0^2 \left( 1 + \frac{\rho_c^2 \sin^2 \theta}{2 \, b_0^2 \rho^2} + 2 \, \lambda \rho \cos \zeta \cot \psi_0  \right) \Big(d \omega^2 + \omega^2 d \varphi^2 \Big) \\
+ b_0^2 \left( 1 + \frac{\rho_c^2}{ b_0^2 \rho^2} - \frac{\rho_c^2 \sin^2 \theta}{2 \, b_0^2  \rho^2}  \right) \left( d \rho^2 + \rho^2 \left( d \zeta^2 + \sin^2 \zeta \left( d \theta_2^2 +  \sin^2 \theta_2 \, d \phi_2^2  \right) \right) \right) \\
+ \lambda \rho_c^2 \Bigg[ f_M(\rho, \zeta) \, \Big( - (d x^0)^2 + (d x^1)^2 + (d x^2)^2 + (d x^3)^2 \Big) + f_{\omega \omega} (\rho, \zeta) \, \Big( d \omega^2 + \omega^2 d \varphi^2 \Big) \\
+ f_{\rho \rho} (\rho, \zeta) \, \Big( d \rho^2 + \rho^2 \left( d \zeta^2 + \sin^2 \zeta \left( d \theta_2^2 + \sin^2 \theta_2 d \phi_2^2  \right) \right) \Big) \Bigg] ~. 
\end{multline}
with
\begin{align}
f_M (\rho, \zeta) &=  \left( \frac{1}{2} \cot \psi_0 \sin^2 \theta + \frac{\sin 2 \theta}{\sqrt{3} \, b_0^2}  \right) \frac{\cos \zeta}{\rho} ~ , \\
f_{\omega \omega} (\rho, \zeta) &=  \left( \frac{1}{2} \cot \psi_0 \sin^2 \theta - \frac{\sin 2 \theta}{\sqrt{3} \, b_0^2}  \right) \frac{\cos \zeta}{\rho} ~, \\
f_{\rho \rho} (\rho, \zeta) &= \left( \frac{1}{12} \Big( 1 + 7 \cos 2 \theta  \Big) \cot \psi_0 + \frac{2 \cos \theta \left( 1 - 2 \sin \theta \right)}{3 \sqrt{3} \, b_0^2}\right) \frac{\cos \zeta}{\rho} ~ .
\end{align}
The dilaton is
\be
\phi = \frac
{\rho_c^2 \cos^2 \theta}{ 2 \, b_0^2 \rho^2} + \lambda \rho_c^2 \, \Phi (\rho, \zeta)
\ee
with 
\be
\Phi (\rho, \zeta) = \frac{\cos^2 \theta \cot \psi_0}{2\, b_0^2} \, \frac{\cos \zeta}{\rho} ~ .
\ee
The $C_2$ gauge field is 
\begin{multline}
C_2 = \frac{\omega}{\sin^2 \psi_0} \left( \psi_0 - \frac{1}{2} \sin 2 \psi_0  \right) d\omega \wedge d \varphi + 2 \lambda \, \rho  \cos \zeta  \, \omega d \omega \wedge d \varphi \\
+ \frac{\rho_c^2}{\rho^2} \sin \theta \cos \theta \, \omega d \omega \wedge d \varphi + \lambda \rho_c^2 \frac{\mathcal{A} \cos \zeta}{4 \, b_0^4 \, \rho} \, \omega d \omega \wedge d \varphi
\end{multline}
with 
\be
\mathcal{A} = \frac{4}{3} b_0^2 \left( - 2 \sqrt{3}  + \sin \theta \left( 3 \, b_0^2 \cos \theta \cot \psi_0 + 2 \sqrt{3} \left(1 + 2 \sin \theta \right) \right) \right) ~.
\ee
The $B_2$ gauge field is given by
\begin{multline}
B_2 = - \frac{4}{\sqrt{3}} \lambda \rho^3 \sin^2 \zeta \cos \zeta \cos \theta_2 \, d \zeta \wedge d \phi_2 - \frac{4}{\sqrt{3}} \lambda \rho^2 \sin^3 \zeta \cos \theta_2 \, d \rho \wedge d \phi_2 \\
- 2 \, \rho_c^2 \cos \theta \sin^2 \zeta \cos \theta_2 \, d \zeta \wedge d \phi_2 \\
+ \lambda \rho_c^2 \, \frac{\mathcal{B} \, \rho \sin^3 \zeta}{4 \, b_0^4} \sin \theta_2 d \theta_2 \wedge d \phi_2 
\end{multline}
with
\be
\mathcal{B} = 4 b_0^4 \cos \theta \cot \psi_0 - \frac{8 b_0^2 \left( 1 + \cos^2 \theta + \sin \theta \right)}{\sqrt{3}} ~.
\ee

The $C_4$ gauge field is given by
\begin{multline}
C_4 = - b_0^2 \sin \theta \, \frac{\rho_c^2}{\rho^2}  \, d x^0 \wedge d x^1 \wedge d x^2 \wedge dx^3 - 2 \, b_0^2 \,  \rho_c^2 \sin \theta \sin^2 \zeta  \, \omega  \cos \theta_2 \, d \omega \wedge d \varphi  \wedge d \zeta \wedge d \phi_2 \\ 
+ \lambda \rho_c^2 \,  \frac{\mathcal{C}_1 \cos \zeta}{\rho}  d x^0 \wedge d x^1 \wedge d x^2 \wedge d x^3 + \lambda \rho_c^2 \, \mathcal{C}_2 \, \rho \sin^3 \zeta  \, \omega \sin \theta_2 \, d \omega \wedge d \varphi \wedge d \theta_2 \wedge d \phi_2 \\
+ B_2 \wedge (C_2)_{0} 
\end{multline}
with 
\be
(C_2)_0 = \frac{\omega}{\sin^2 \psi_0} \left( \psi_0 - \frac{1}{2} \sin 2 \psi_0  \right) d\omega \wedge d \varphi ~,
\ee
and
\begin{align}
\mathcal{C}_1 &= \frac{2}{\sqrt{3}} \cos \theta \Big(1 - \sin \theta \Big) + b_0^2 \cot \psi_0 \sin \theta ~, \\
\mathcal{C}_2 &= - \frac{2}{\sqrt{3}} \cos \theta \Big(  1 +  \sin \theta \Big) - b_0^2 \cot \psi_0 \sin \theta  ~.
\end{align}
The constants $\psi_0$ and $\theta$ are related by
\be
\cot \psi_0 = \frac{2}{\sqrt{3}} \left( \frac{1}{b_0^2} \sec \theta + \frac{1}{b_0^2} \tan \theta \right) ~.
\ee
Through direct substitution, one can easily check that the above profile satisfies all SUGRA equations to leading order in $\lambda$ and $\rho_c^2$.  

\end{appendices}

\bibliographystyle{JHEP}
\bibliography{ref}

\providecommand{\href}[2]{#2}\begingroup\raggedright\begin{thebibliography}{10}

\bibitem{Maldacena:2001pb}
J.~M. Maldacena and H.~S. Nastase, {\itshape {The Supergravity dual of a theory
  with dynamical supersymmetry breaking}},  {\em JHEP} {\bfseries 09} (2001)
  024, [\href{http://arxiv.org/abs/hep-th/0105049}{{\ttfamily
  hep-th/0105049}}].

\bibitem{Kachru:2002gs}
S.~Kachru, J.~Pearson, and H.~L. Verlinde, {\itshape {Brane / flux annihilation
  and the string dual of a nonsupersymmetric field theory}},  {\em JHEP}
  {\bfseries 06} (2002) 021,
  [\href{http://arxiv.org/abs/hep-th/0112197}{{\ttfamily hep-th/0112197}}].

\bibitem{Klebanov:2010qs}
I.~R. Klebanov and S.~S. Pufu, {\itshape {M-Branes and Metastable States}},
  {\em JHEP} {\bfseries 08} (2011) 035,
  [\href{http://arxiv.org/abs/1006.3587}{{\ttfamily arXiv:1006.3587}}].

\bibitem{Kachru:2003aw}
S.~Kachru, R.~Kallosh, A.~D. Linde, and S.~P. Trivedi, {\itshape {De Sitter
  vacua in string theory}},  {\em Phys. Rev. D} {\bfseries 68} (2003) 046005,
  [\href{http://arxiv.org/abs/hep-th/0301240}{{\ttfamily hep-th/0301240}}].

\bibitem{Kachru:2003sx}
S.~Kachru, R.~Kallosh, A.~D. Linde, J.~M. Maldacena, L.~P. McAllister, and
  S.~P. Trivedi, {\itshape {Towards inflation in string theory}},  {\em JCAP}
  {\bfseries 10} (2003) 013,
  [\href{http://arxiv.org/abs/hep-th/0308055}{{\ttfamily hep-th/0308055}}].

\bibitem{Bena:2012zi}
I.~Bena, A.~Puhm, and B.~Vercnocke, {\itshape {Non-extremal Black Hole
  Microstates: Fuzzballs of Fire or Fuzzballs of Fuzz ?}},  {\em JHEP}
  {\bfseries 12} (2012) 014, [\href{http://arxiv.org/abs/1208.3468}{{\ttfamily
  arXiv:1208.3468}}].

\bibitem{Klebanov:2000hb}
I.~R. Klebanov and M.~J. Strassler, {\itshape {Supergravity and a confining
  gauge theory: Duality cascades and chi SB resolution of naked
  singularities}},  {\em JHEP} {\bfseries 08} (2000) 052,
  [\href{http://arxiv.org/abs/hep-th/0007191}{{\ttfamily hep-th/0007191}}].

\bibitem{Myers:1999ps}
R.~C. Myers, {\itshape {Dielectric branes}},  {\em JHEP} {\bfseries 12} (1999)
  022, [\href{http://arxiv.org/abs/hep-th/9910053}{{\ttfamily
  hep-th/9910053}}].

\bibitem{Bena:2014jaa}
I.~Bena, M.~Gra\~na, S.~Kuperstein, and S.~Massai, {\itshape {Giant Tachyons in
  the Landscape}},  {\em JHEP} {\bfseries 02} (2015) 146,
  [\href{http://arxiv.org/abs/1410.7776}{{\ttfamily arXiv:1410.7776}}].

\bibitem{Nguyen:2019syc}
N.~Nguyen, {\itshape {Comments on the stability of the KPV state}},  {\em JHEP}
  {\bfseries 11} (2020) 055, [\href{http://arxiv.org/abs/1912.04646}{{\ttfamily
  arXiv:1912.04646}}].

\bibitem{Bena:2009xk}
I.~Bena, M.~Grana, and N.~Halmagyi, {\itshape {On the Existence of Meta-stable
  Vacua in Klebanov-Strassler}},  {\em JHEP} {\bfseries 09} (2010) 087,
  [\href{http://arxiv.org/abs/0912.3519}{{\ttfamily arXiv:0912.3519}}].

\bibitem{Gubser:2000nd}
S.~S. Gubser, {\itshape {Curvature singularities: The Good, the bad, and the
  naked}},  {\em Adv. Theor. Math. Phys.} {\bfseries 4} (2000) 679--745,
  [\href{http://arxiv.org/abs/hep-th/0002160}{{\ttfamily hep-th/0002160}}].

\bibitem{Blaback:2014tfa}
J.~Bl\r{a}b\"ack, U.~H. Danielsson, D.~Junghans, T.~Van~Riet, and S.~C. Vargas,
  {\itshape {Localised anti-branes in non-compact throats at zero and finite
  $T$}},  {\em JHEP} {\bfseries 02} (2015) 018,
  [\href{http://arxiv.org/abs/1409.0534}{{\ttfamily arXiv:1409.0534}}].

\bibitem{Massai:2012jn}
S.~Massai, {\itshape {A Comment on anti-brane singularities in warped
  throats}},  \href{http://arxiv.org/abs/1202.3789}{{\ttfamily
  arXiv:1202.3789}}.

\bibitem{Bena:2012bk}
I.~Bena, M.~Grana, S.~Kuperstein, and S.~Massai, {\itshape {Anti-D3 Branes:
  Singular to the bitter end}},  {\em Phys. Rev. D} {\bfseries 87} (2013),
  no.~10 106010, [\href{http://arxiv.org/abs/1206.6369}{{\ttfamily
  arXiv:1206.6369}}].

\bibitem{Gautason:2013zw}
F.~F. Gautason, D.~Junghans, and M.~Zagermann, {\itshape {Cosmological
  Constant, Near Brane Behavior and Singularities}},  {\em JHEP} {\bfseries 09}
  (2013) 123, [\href{http://arxiv.org/abs/1301.5647}{{\ttfamily
  arXiv:1301.5647}}].

\bibitem{Cohen-Maldonado:2015ssa}
D.~Cohen-Maldonado, J.~Diaz, T.~van Riet, and B.~Vercnocke, {\itshape
  {Observations on fluxes near anti-branes}},  {\em JHEP} {\bfseries 01} (2016)
  126, [\href{http://arxiv.org/abs/1507.01022}{{\ttfamily arXiv:1507.01022}}].

\bibitem{Cohen-Maldonado:2016cjh}
D.~Cohen-Maldonado, J.~Diaz, and F.~F. Gautason, {\itshape {Polarised
  antibranes from Smarr relations}},  {\em JHEP} {\bfseries 05} (2016) 175,
  [\href{http://arxiv.org/abs/1603.05678}{{\ttfamily arXiv:1603.05678}}].

\bibitem{Emparan:2009cs}
R.~Emparan, T.~Harmark, V.~Niarchos, and N.~A. Obers, {\itshape {World-Volume
  Effective Theory for Higher-Dimensional Black Holes}},  {\em Phys. Rev.
  Lett.} {\bfseries 102} (2009) 191301,
  [\href{http://arxiv.org/abs/0902.0427}{{\ttfamily arXiv:0902.0427}}].

\bibitem{Emparan:2009at}
R.~Emparan, T.~Harmark, V.~Niarchos, and N.~A. Obers, {\itshape {Essentials of
  Blackfold Dynamics}},  {\em JHEP} {\bfseries 03} (2010) 063,
  [\href{http://arxiv.org/abs/0910.1601}{{\ttfamily arXiv:0910.1601}}].

\bibitem{Armas:2016mes}
J.~Armas, J.~Gath, V.~Niarchos, N.~A. Obers, and A.~V. Pedersen, {\itshape
  {Forced Fluid Dynamics from Blackfolds in General Supergravity Backgrounds}},
   {\em JHEP} {\bfseries 10} (2016) 154,
  [\href{http://arxiv.org/abs/1606.09644}{{\ttfamily arXiv:1606.09644}}].

\bibitem{Armas:2018rsy}
J.~Armas, N.~Nguyen, V.~Niarchos, N.~A. Obers, and T.~Van~Riet, {\itshape
  {Metastable Nonextremal Antibranes}},  {\em Phys. Rev. Lett.} {\bfseries 122}
  (2019), no.~18 181601, [\href{http://arxiv.org/abs/1812.01067}{{\ttfamily
  arXiv:1812.01067}}].

\bibitem{Michel:2014lva}
B.~Michel, E.~Mintun, J.~Polchinski, A.~Puhm, and P.~Saad, {\itshape {Remarks
  on brane and antibrane dynamics}},  {\em JHEP} {\bfseries 09} (2015) 021,
  [\href{http://arxiv.org/abs/1412.5702}{{\ttfamily arXiv:1412.5702}}].

\bibitem{Bena:2012ek}
I.~Bena, A.~Buchel, and O.~J.~C. Dias, {\itshape {Horizons cannot save the
  Landscape}},  {\em Phys. Rev. D} {\bfseries 87} (2013), no.~6 063012,
  [\href{http://arxiv.org/abs/1212.5162}{{\ttfamily arXiv:1212.5162}}].

\bibitem{Bena:2013hr}
I.~Bena, J.~Blaback, U.~H. Danielsson, and T.~Van~Riet, {\itshape {Antibranes
  cannot become black}},  {\em Phys. Rev. D} {\bfseries 87} (2013), no.~10
  104023, [\href{http://arxiv.org/abs/1301.7071}{{\ttfamily arXiv:1301.7071}}].

\bibitem{Hartnett:2015oda}
G.~S. Hartnett, {\itshape {Localised Anti-Branes in Flux Backgrounds}},  {\em
  JHEP} {\bfseries 06} (2015) 007,
  [\href{http://arxiv.org/abs/1501.06568}{{\ttfamily arXiv:1501.06568}}].

\bibitem{Blaback:2019ucp}
J.~Bl\r{a}b\"ack, F.~F. Gautason, A.~Ruip\'erez, and T.~Van~Riet, {\itshape
  {Anti-brane singularities as red herrings}},  {\em JHEP} {\bfseries 12}
  (2019) 125, [\href{http://arxiv.org/abs/1907.05295}{{\ttfamily
  arXiv:1907.05295}}].

\bibitem{Bena:2010gs}
I.~Bena, G.~Giecold, and N.~Halmagyi, {\itshape {The Backreaction of Anti-M2
  Branes on a Warped Stenzel Space}},  {\em JHEP} {\bfseries 04} (2011) 120,
  [\href{http://arxiv.org/abs/1011.2195}{{\ttfamily arXiv:1011.2195}}].

\bibitem{M2M5brane}
J.~Armas, N.~Nguyen, V.~Niarchos, and N.~A. Obers, {\itshape {Thermal
  transitions of metastable M-branes}},  {\em JHEP} {\bfseries 08} (2019) 128,
  [\href{http://arxiv.org/abs/1904.13283}{{\ttfamily arXiv:1904.13283}}].

\bibitem{Bhattacharyya:2007vjd}
S.~Bhattacharyya, V.~E. Hubeny, S.~Minwalla, and M.~Rangamani, {\itshape
  {Nonlinear Fluid Dynamics from Gravity}},  {\em JHEP} {\bfseries 02} (2008)
  045, [\href{http://arxiv.org/abs/0712.2456}{{\ttfamily arXiv:0712.2456}}].

\bibitem{Camps:2012hw}
J.~Camps and R.~Emparan, {\itshape {Derivation of the blackfold effective
  theory}},  {\em JHEP} {\bfseries 03} (2012) 038,
  [\href{http://arxiv.org/abs/1201.3506}{{\ttfamily arXiv:1201.3506}}].
  [Erratum: JHEP 06, 155 (2012)].

\bibitem{Gorbonos:2004uc}
D.~Gorbonos and B.~Kol, {\itshape {A Dialogue of multipoles: Matched asymptotic
  expansion for caged black holes}},  {\em JHEP} {\bfseries 06} (2004) 053,
  [\href{http://arxiv.org/abs/hep-th/0406002}{{\ttfamily hep-th/0406002}}].

\bibitem{Harmark:2003yz}
T.~Harmark, {\itshape {Small black holes on cylinders}},  {\em Phys. Rev. D}
  {\bfseries 69} (2004) 104015,
  [\href{http://arxiv.org/abs/hep-th/0310259}{{\ttfamily hep-th/0310259}}].

\bibitem{Cvetic:2000db}
M.~Cvetic, G.~W. Gibbons, H.~Lu, and C.~N. Pope, {\itshape {Ricci flat metrics,
  harmonic forms and brane resolutions}},  {\em Commun. Math. Phys.} {\bfseries
  232} (2003) 457--500, [\href{http://arxiv.org/abs/hep-th/0012011}{{\ttfamily
  hep-th/0012011}}].

\bibitem{Bena:2015kia}
I.~Bena and S.~Kuperstein, {\itshape {Brane polarization is no cure for
  tachyons}},  {\em JHEP} {\bfseries 09} (2015) 112,
  [\href{http://arxiv.org/abs/1504.00656}{{\ttfamily arXiv:1504.00656}}].

\bibitem{Gregory:1993vy}
R.~Gregory and R.~Laflamme, {\itshape {Black strings and p-branes are
  unstable}},  {\em Phys. Rev. Lett.} {\bfseries 70} (1993) 2837--2840,
  [\href{http://arxiv.org/abs/hep-th/9301052}{{\ttfamily hep-th/9301052}}].

\bibitem{Hovdebo:2006jy}
J.~L. Hovdebo and R.~C. Myers, {\itshape {Black rings, boosted strings and
  Gregory-Laflamme}},  {\em Phys. Rev. D} {\bfseries 73} (2006) 084013,
  [\href{http://arxiv.org/abs/hep-th/0601079}{{\ttfamily hep-th/0601079}}].

\bibitem{Santos:2015iua}
J.~E. Santos and B.~Way, {\itshape {Neutral Black Rings in Five Dimensions are
  Unstable}},  {\em Phys. Rev. Lett.} {\bfseries 114} (2015) 221101,
  [\href{http://arxiv.org/abs/1503.00721}{{\ttfamily arXiv:1503.00721}}].

\bibitem{Lehner:2010pn}
L.~Lehner and F.~Pretorius, {\itshape {Black Strings, Low Viscosity Fluids, and
  Violation of Cosmic Censorship}},  {\em Phys. Rev. Lett.} {\bfseries 105}
  (2010) 101102, [\href{http://arxiv.org/abs/1006.5960}{{\ttfamily
  arXiv:1006.5960}}].

\bibitem{Figueras:2015hkb}
P.~Figueras, M.~Kunesch, and S.~Tunyasuvunakool, {\itshape {End Point of Black
  Ring Instabilities and the Weak Cosmic Censorship Conjecture}},  {\em Phys.
  Rev. Lett.} {\bfseries 116} (2016), no.~7 071102,
  [\href{http://arxiv.org/abs/1512.04532}{{\ttfamily arXiv:1512.04532}}].

\bibitem{Armas:2019iqs}
J.~Armas and E.~Parisini, {\itshape {Instabilities of Thin Black Rings: Closing
  the Gap}},  {\em JHEP} {\bfseries 04} (2019) 169,
  [\href{http://arxiv.org/abs/1901.09369}{{\ttfamily arXiv:1901.09369}}].

\bibitem{Polchinski:1998rr}
J.~Polchinski, {\em {String theory. Vol. 2: Superstring theory and beyond}}.
\newblock Cambridge Monographs on Mathematical Physics. Cambridge University
  Press, 12, 2007.

\bibitem{Herzog:2001xk}
C.~P. Herzog, I.~R. Klebanov, and P.~Ouyang, {\itshape {Remarks on the warped
  deformed conifold}},  in {\em {Modern Trends in String Theory: 2nd Lisbon
  School on g Theory Superstrings Lisbon, Portugal, July 13-17, 2001}}, 2001.
\newblock \href{http://arxiv.org/abs/hep-th/0108101}{{\ttfamily
  hep-th/0108101}}.

\bibitem{Minasian:1999tt}
R.~Minasian and D.~Tsimpis, {\itshape {On the geometry of nontrivially embedded
  branes}},  {\em Nucl. Phys. B} {\bfseries 572} (2000) 499--513,
  [\href{http://arxiv.org/abs/hep-th/9911042}{{\ttfamily hep-th/9911042}}].

\end{thebibliography}\endgroup

\end{document}